\documentclass[showpacs,superscriptaddress,floatfix,amsmath,amssymb,twocolumn,aps,longbibliography]{revtex4-1}
\usepackage{bm}% bold math
\usepackage{graphicx}
\usepackage{color}
\usepackage[normalem]{ulem}
\usepackage{epstopdf}
\usepackage{soul}
\usepackage{url}
\usepackage[normalem]{ulem}
\usepackage{verbatim}
\def\ep{\varepsilon}

\newcommand{\Ket}[1]{\vert #1 \rangle}
\newcommand{\Bra}[1]{\langle #1 \vert}

\def\Im{{\rm Im}}
\def\n{{\bm\nabla}}
\def \sgn{{\rm sgn}}
\def\Jo{{\cal J}^{\rm out}}
\def\Jin{{\cal J}^{\rm in}}
\usepackage{hyperref}
\hypersetup{pdftex,colorlinks=true,allcolors=blue,bookmarksnumbered=true}
\usepackage{hypcap}

\begin{document}
\title{Nonlocal random walk over Floquet states of a dissipative nonlinear oscillator}
\author{Yaxing Zhang}
\affiliation{Department of Physics, Yale University, New Haven, Connecticut 06511, USA}
\author{M. I. Dykman}
\affiliation{Department of Physics and Astronomy, Michigan State University, East Lansing, Michigan 48824, USA}
\date{\today}
\begin{abstract}
We study transitions between the Floquet states of a periodically driven oscillator caused by the coupling of the oscillator to a thermal reservoir.  The analysis refers to the oscillator that is driven close to triple its eigenfrequency and displays resonant period tripling. The interstate transitions  result in a random ``walk'' over the states. We find the transition rates and show that the walk is nonlocal in the state space: the stationary distribution over the states is formed by the transitions between remote  states. This is to be contrasted with systems in thermal equilibrium, where the distribution is usually formed by transitions  between nearby states. The analysis of period tripling allows us to explore the features of the multi-state Floquet dynamics including those missing in the previously explored models of driven oscillators such as the absence of detailed balance for low temperatures. We use the results to study switching between the period-3 states of the oscillator due to quantum fluctuations and find the scaling of the switching rates with the parameters.
\end{abstract}
\maketitle

\section{Introduction}

The distribution over the eigenstates of a quantum system weakly coupled to a thermal reservoir is formed by the coupling-induced interlevel transitions. In thermal equilibrium, the transitions are balanced, the rates of transitions $m\to n$ and $n\to m$ weighted with the state populations are equal for any states $m,n$ \cite{Einstein1917}. 
This detailed balance condition does not hold away from thermal equilibrium. When the distribution over the states of a nonequilibrium system is stationary, it means that the overall probability to come to a state from all other states is equal to the overall probability to leave this state. In this case, if the distribution over the states is steep, the transitions to a given state from remote states may become important. This is despite the fact that the rates of such transitions, which are determined by the overlapping of the corresponding wave functions, are exponentially small. 

One can think of the interstate transitions due to the coupling to a reservoir as a random walk over the states. Where the transitions between remote states are important, the walk becomes nonlocal, in the state space.

An important group of nonequilibrium system where the locality of the walk over the states can be explored is provided by periodically driven systems. The interest in such system has surged in recent years in various areas of physics and in various contexts, cf. \cite{Peano2006,Martin2017,Weinberg2017,Desbuquois2017,Schmidt2018a,Seetharam2018a,Lohse2018} and references therein. Periodically driven systems have  well-defined quantum states, the Floquet or quasienergy states ~\cite{shirley1965,zeldovich1967,ritus1967}, which form a complete set. It has been appreciated \cite{kohn2001} that, when a periodically driven system is brought in contact with a thermal reservoir, there is no detailed balance, and the resulting stationary distribution over the Floquet states is complicated. 

A remarkable and important exception to the lack of detailed balance in Floquet systems is provided by a nonlinear oscillator driven close to its eigenfrequency or parametrically modulated close to twice the eigenfrequency \cite{Drummond1980,Kryuchkyan1996}.  For the basic model of such an oscillator, that includes the quartic nonlinearity of the potential and a simple dissipation mechanism, the oscillator has detailed balance, if the temperature of the thermal reservoir is zero.  The detailed balance breaks down for nonzero temperature \cite{Dykman1979,Dykman1988a,Marthaler2006}. However, the very occurrence of it for $T=0$ is fascinating. Since it happens for two major models, one may ask whether this is a generic feature of periodically driven dissipative oscillators. 

In this paper, we study the rates of transitions between the Floquet states and the  stationary distribution over these states for a nonlinear oscillator driven close to triple its eigenfrequency $\omega_0$. The quantum coherent behavior of such an oscillator displays peculiar features such as the nontrivial geometric phase and the oscillations of the Floquet wave functions in the classically inaccessible region \cite{Guo2013a,Zhang2017}. An observation of period tripling in the quantum regime with a coupled-modes superconducting cavity has been already reported \cite{Svensson2017a}.
 Here we explore the dissipative dynamics of this system and the associated quantum fluctuations. It turns out that the analytical tools developed for the previously explored oscillator models \cite{Dykman1988a,Marthaler2006,Peano2014} do not apply in the present case. Moreover, even for $T=0$ the oscillator does not have detailed balance. This suggests that the detailed balance for resonant and parametric driving could be an artifact. 

Another important feature of the oscillators driven close to the eigenfrequency or modulated close to twice the eigenfrequency is that their stationary distribution over the Floquet states is formed  locally, i.e., by transitions between a few close states. This is the case both at zero temperature and above a certain very low temperature set by $\hbar$ weighted with a combination of the nonlinearity and driving  parameters \cite{Guo2013,Peano2014}. This again poses the question of whether the locality is a generic property of the $T=0$- stationary distribution of a driven oscillator. Answering this question for period tripling is one of the central topics of the present paper.

Classically, an oscillator that displays period tripling has four stable states: three states of period-three vibrations, with the same amplitude but the phases that differ by $2\pi/3$, see Fig.~\ref{fig:quasienergy_surface}~(a), and also the quiet state with no vibrations excited. Quantum fluctuations lead to switching between these states. For the driving at frequency close to $\omega_0$ and $2\omega_0$ it was found that, unless the dissipation rate is exponentially small, the switching occurs via transitions over the quasienergy barrier even for $T=0$. The mechanism was called quantum activation \cite{Marthaler2006}. The period-3 vibrations are different in several respects. One has to find how the oscillator switches between the period-3 states and between these states and the quiet state. In particular, the oscillator can possibly switch directly between the period-3 states or first go to the quiet state and then switch from this state.

 In Sec.~II below, for the oscillator that displays resonant period tripling, we provide the Hamiltonian in the rotating wave approximation (RWA) and discuss the Floquet eigenstates and the Wannier-type states, which are relevant where the wells of the RWA Hamiltonian have many bound states. In Sec.~III we develop the WKB theory of the rates of dissipation-induced transitions between the intrawell states and demonstrate the breaking of the detailed balance even for $T=0$. In Sec.~IV we study the distribution over the states using an eikonal-type approximation. In this approximation the distribution is formed by transitions between nearby states. Sec.~V is one of the central parts of the paper. Here we analyze the breaking of the local approximation for $T=0$. We also show that, already for small Planck numbers of the oscillator, the locality is restored. The results are compared with numerical analysis based on calculating the transitions rates using the Wannier-type wave functions. In Sec.~VI we study switching between different wells of the RWA Hamiltonian in the regime of weak damping and also in the vicinity of the bifurcation points where the period-3 states of the oscillator emerge. Sec.~VII contains conclusions. The Appendices describe technical details of the calculations.

\section{The model and the rotating wave approximation}
\label{sec:model}

Period tripling generically occurs in strongly driven nonlinear systems. However, for weakly damped oscillators it may emerge already for a relatively weak drive provided the frequency of the driving force $\omega_F$ is close to triple the oscillator eigenfrequency $\omega_0$ \cite{Nayfeh2004}. Several aspects of the quantum formulation of the problem in the absence of dissipation were discussed previously \cite{Zhang2017}. To set the scene, here we briefly summarize and generalize the formulation.

A simple Hamiltonian of the oscillator that displays resonant period tripling is  
\begin{align}
\label{eq:hamiltonian}
H&=\frac{1}{2}(p^2 + \omega_0^2 q^2) +\frac{1}{4} \gamma q^4 -\frac{1}{3}F_0q^3\cos\omega_F t,
\end{align}
where $p$ and $q$ are the momentum and coordinate of the oscillator. The conditions of resonant driving and comparatively weak nonlinearity are 
\[|\delta\omega|, |\gamma|\langle q^2\rangle \ll \omega_0, \quad \delta\omega = \frac{1}{3}\omega_F - \omega_0.\]
Without loss of generality, we set $\gamma>0$ and $F_0>0$. The analysis of \cite{Zhang2017} referred to the case $\delta\omega>0$. An alternative scaling that allows one to consider the case $\delta\omega=0$ is considered in Appendix~\ref{sec:fixed_point_analysis}. 

To study the oscillator dynamics near its eigenfrequency, we switch to the rotating frame at one third of the driving frequency via the unitary transformation $U(t)=\exp(-i\omega_Ft a^\dagger a/3)$, with $a=[(\omega_F q/3)+ip](3/2\hbar\omega_F)^{1/2}$. We then introduce slowly varying in time coordinate $Q$ and momentum $P$ of the oscillator in the rotating frame,
\begin{align}
\label{eq:new_variables}
&U^\dagger(t )q U(t) = C(Q\cos\phi +P\sin\phi), \\
& U^\dagger (t) p U(t) =- C\frac{\omega_F}{3}(Q\sin\phi -P\cos\phi),\qquad \phi = \omega_Ft/3,\nonumber
\end{align}
where $C=(8\omega_F|\delta\omega|/9\gamma)^{1/2}$. The operators $Q,P$ satisfy the commutation relation 
\begin{align}
\label{eq:commutation}
[Q,P]=i\lambda, \quad \lambda= \frac{27\gamma\hbar}{8\omega_F^2|\delta\omega|}.
\end{align}
Parameter $\lambda$ plays the role of the Planck constant in the rotating frame. It is dimensionless, and in what follows we use the WKB approximation, which applies for $\lambda \ll 1$. In terms of the raising and lowering operators, $Q=(\lambda/2)^{1/2}(a+a^\dagger)$ and $P=-i (\lambda/2)^{1/2}(a-a^\dagger)$.

\begin{figure}[t]
\includegraphics[width = 4.1 cm]{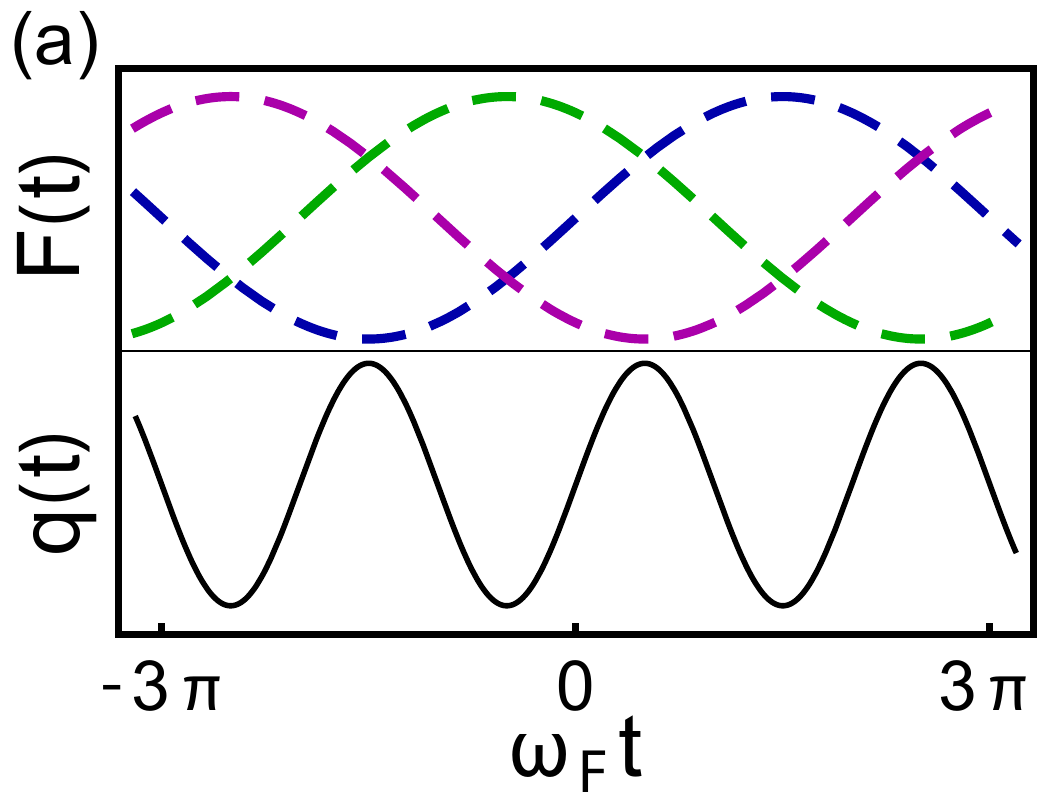} \hfill
\includegraphics[width = 4.1 cm]{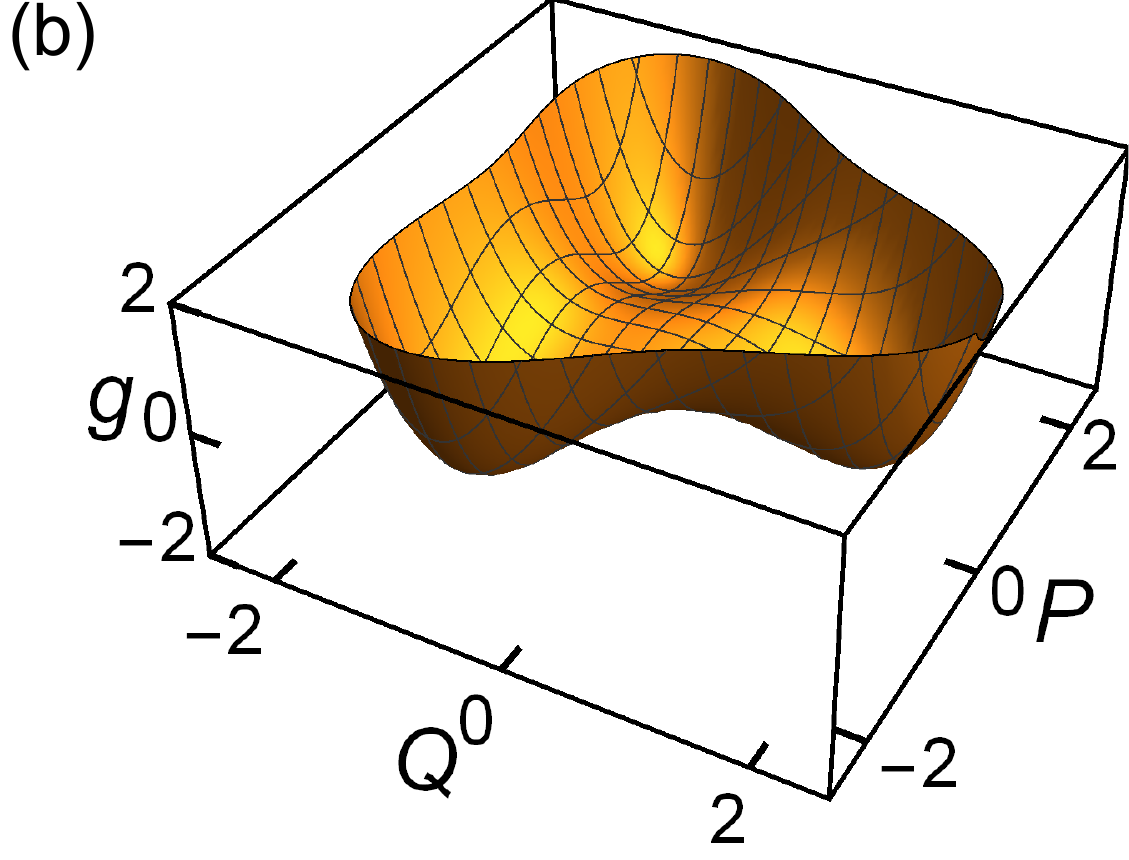} \hfill
\includegraphics[width = 4.2 cm]{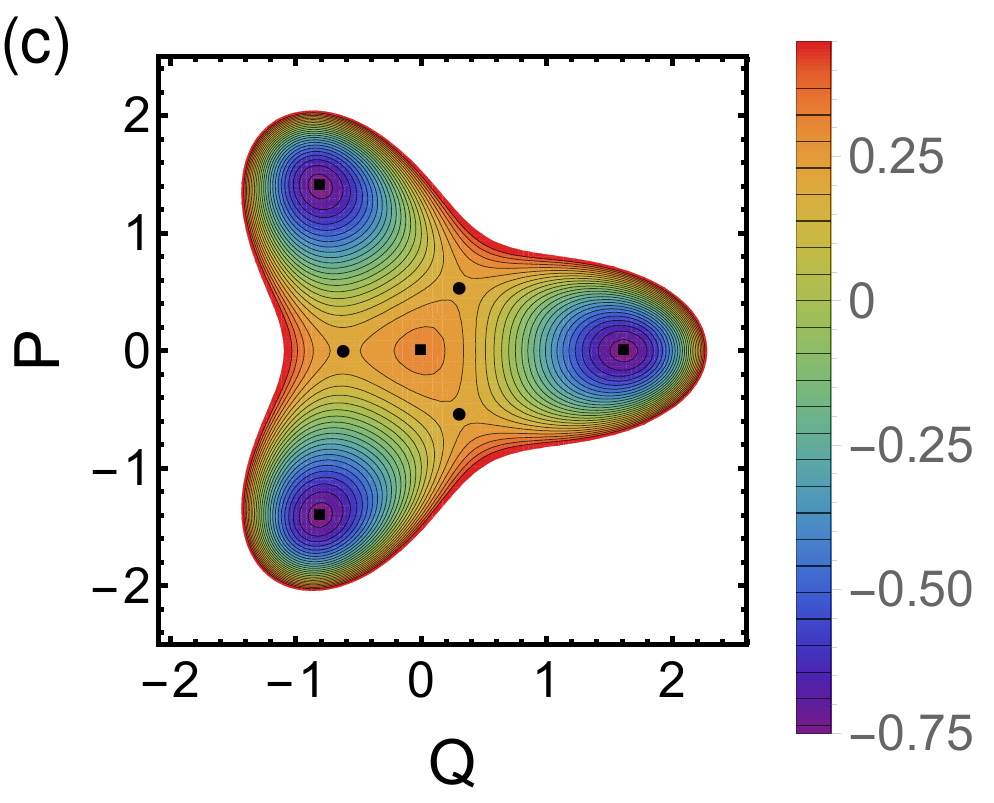} \hfill
\includegraphics[width = 4.1 cm]{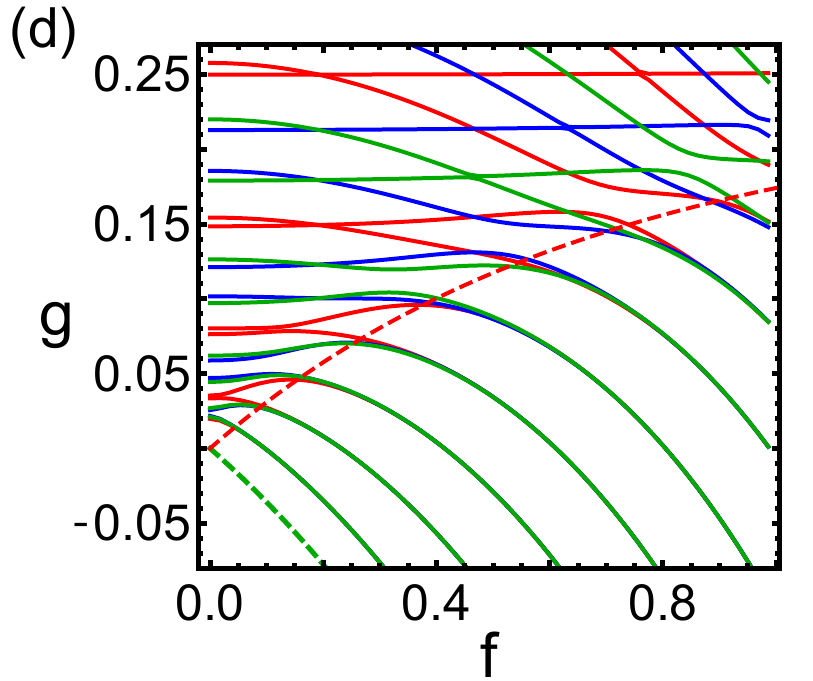}
\caption{(a) The oscillator coordinate $q(t)$ in the three period-3 vibrational states with the phases differing by $2\pi/3$ and the drive $F(t)=F_0\cos\omega_Ft$. 
(b) The classical RWA Hamiltonian $g(Q,P)$, Eq.~(\ref{eq:Hamiltonian_RWA}), as a function of the coordinate and momentum in the rotating frame for the scaled amplitude of the driving field $f=1$ and $\delta\omega>0$. (c) The classical orbits of the Hamiltonian $g$. The local maximum and the three local minima are marked by the squares, and the three saddle points are marked by the filled circles. (d) Evolution of the scaled quasienergies (calculated modulo $\hbar\omega_F/3$) with the varying $f$. The scaled Planck constant is $\lambda = 0.04$. The solid lines show the eigenvalues $g_n^{(k)}$  for $k=0$ (red), 1 (blue), and 2 (green). The dashed lines show the value of the function $g(Q,P)$ at the local minima (green) and the saddle (red); at its extremum at the origin $g(0,0)=1/4$.}
\label{fig:quasienergy_surface}
\end{figure}

In the rotating wave approximation (RWA), the motion in the rotating frame  is described by the canonically transformed Hamiltonian $U^\dagger H U-i\hbar U^\dagger\dot U =  (\hbar|\delta\omega|/\lambda)g(Q,-i\lambda\partial_Q)$, where
\begin{align}
\label{eq:Hamiltonian_RWA}
&g(Q,P) = \frac{1}{4}[Q^2+P^2 -\sgn(\delta\omega)]^2 -\frac{1}{3} f(Q^3 - 3PQP) \nonumber\\
&f=F_0/(8\omega_F \gamma |\delta\omega|)^{1/2}.
\end{align}
Here, $g(Q,P)$ as a function of the operators $Q,P$ is the scaled dimensionless Hamiltonian of the oscillator in the rotating frame calculated in the RWA. It depends on a single parameter, the scaled field amplitude $f$, and does not depend on time.

The Hamiltonian $g(Q,P) $  as a function of the classical coordinate and momentum is plotted in Fig.~\ref{fig:quasienergy_surface}~(b). In the case shown in the figure, $g(Q,P)$ has three local minima at a nonzero $Q^2+P^2$. For the frequency detuning  $\delta\omega>0$ the onset of the minima has no threshold in the scaled field amplitude $f$, whereas for $\delta\omega<0$ the minima emerge for $f>2$, see Appendix~\ref{sec:fixed_point_analysis}. In the classical limit and in the presence of weak dissipation, these minima correspond to the three period-3 vibrational states in the laboratory frame, cf. Fig.~\ref{fig:quasienergy_surface}~(a). 

The Hamiltonian $g(Q,P)$ also has a local maximum or minimum at $Q=P=0$ depending on whether $\delta\omega >0$ or $\delta\omega<0$. It corresponds to a classically stable quiet state with no vibrations. The dissipative classical dynamics is outlined in Appendix~\ref{sec:Langevin}.% In this paper, unless otherwise indicated, we consider the case $\delta\omega>0$. 

\subsection{The quasienergy spectrum}
\label{subsec:quasienergy_spectrum}

The Schr\"odinger equation in the rotating frame in terms of the scaled RWA Hamiltonian $g$ reads
\begin{align}
i\lambda \frac{\partial}{\partial \tau}|\phi\rangle = g(Q, -i\lambda\partial_Q) |\phi\rangle.
\end{align}
We have introduced here the dimensionless time $\tau =  |\delta \omega| t$. 
The Hamiltonian $g$ is invariant with respect to a rotation in the $Q,P$ phase plane by $2\pi/3$. Such rotation is described by the operator $\hat N_3 = \exp(-2\pi ia^+a /3)$, which commutes with $g$.
Therefore the eigenfunctions $\phi_n^{(k)}$ of $g$ are also eigenfunctions of $\hat N_3$,
\begin{align}
\label{eq:N_3}
\hat N_3\phi_n^{(k)} = \exp(-2\pi i k/3) \phi_n^{(k)}.
\end{align}
The superscript $k=0,1,2$ here determines the eigenvalue of the rotation operator $\hat N_3$. The subscript $n$ is the other number that enumerates the eigenstates of the Hamiltonian $g(Q,-i\lambda\partial_Q)$; this Hamiltonian has an infinite countable set of eigenfunctions and eigenvalues.

With the account taken of the interrelation between the oscillator Hamiltonian in the laboratory frame (\ref{eq:hamiltonian}) and $g$, the eigenvalues $g_n^{(k)}$ of $g$ determine  the quasienergies $\ep_n^{(k)}$ of the driven oscillator in the RWA \cite{Zhang2017}, 
\[\ep_n^{(k)} = \frac{\hbar |\delta\omega|}{\lambda}g_n^{(k)} +\frac{2\pi k}{3}\hbar\omega_F.\]
As seen from this expression, the eigenvalues $g_n^{(k)}$ give the the scaled quaisenergies calculated in the reduced Brillouin zone $0\leq \ep < \hbar\omega_F/3$. In what follows we call these eigenvalues scaled RWA energies, or just RWA energies, for brevity.

To calculate the eigenvalues and eigenfunctions of the scaled RWA Hamiltonian $g$, it is convenient to rewrite $g$ in terms of the ladder operators $a$ and $a^\dagger$,
\begin{align}
\label{eq:g_in_ladder_ops}
g& = \lambda \Bigl[-\sgn(\delta\omega)( a^\dagger a+1/2) + \lambda a^\dagger a(a^\dagger a +1) \nonumber\\
&- \frac{f}{3}\sqrt{2\lambda}(a^{\dagger 3}+a^3)\Bigr] + \frac{1}{4}(1+\lambda^2).
\end{align}
Several low-lying eigenvalues of $g$ as  functions of the scaled driving amplitude $f$ are shown in Fig.~\ref{fig:quasienergy_surface}(d). It is clear from Eq.~(\ref{eq:N_3}) and can be also directly seen from Eq.~(\ref{eq:g_in_ladder_ops}) that the eigenvalues $g_n^{(k)}$ form groups with different values of $k=0,1,2$. The eigenvalues with different $k$ can cross with the  varying $f$, as they correspond to different eigenvalues of $\hat N_3$. The eigenvalues with the same $k$ anticross. 

An important qualitative feature of the evolution of the spectrum is that, with the increasing $f$, the eigenvalues with different $k$ cluster into triples. This can be understood from the form of the function $g(Q,P)$ in Fig.~\ref{fig:quasienergy_surface}~(b) and (c). For nonzero $f$ and $\delta\omega>0$, this function has three minima separated by three saddle points. The clustering occurs once the values of $g_n^{(k)}$ become smaller than the saddle-point value $g_s$ of $g(Q,P)$, which is shown as the red dashed line in Fig.~\ref{fig:quasienergy_surface}. The states with $g_n^{(k)}<g_s $ correspond to linear combinations of the intrawell states of $g(Q,P)$ that have the same RWA energy in the neglect of interwell tunneling. The tunneling between the three wells leads to the level splitting within a triple of the eigenstates. It sharply falls off as the states go deeper into the wells, see Sec.~\ref{subsec:intrawell}. 
As the driving strength increases, the wells of $g(Q,P)$ become deeper whereas the saddle points get closer to the local maximum of $g(Q,P)$ at $Q=P=0$. Thus, the number of intrawell states increases, whereas the number of the states localized near $Q=P=0$ decreases.

\subsection{Intrawell states}
\label{subsec:intrawell}

Of central interest for this paper is the regime of a relatively strong drive where the three wells of the Hamiltonian function $g(Q,P)$ are well-separated in phase space compared to the typical quantum scale of the phase-space area $\lambda$. One can then think of the states localized mostly inside the wells. The number of such states is $\propto 1/\lambda$. We will be interested in the case where this number is large.

We will denote the intrawell states by $\Ket{n}_\nu\equiv |\psi_\nu(Q; n)\rangle$, where $\nu=0,1,2$ enumerates the wells. In what follows we choose the $\nu=0$ well to have a minimum on the $P=0$-axis, see Fig.~\ref{fig:quasienergy_surface}. The wells with $\nu=1,2$ are symmetrically located with respect to the $P=0$ axis, with their minima at the same $Q$. We use $({}Q_\nu,{}P_\nu)$ to denote the positions of the minima of the wells. These points are vortices of an equilateral triangle on the $(Q,P)$ plane, see Fig.~\ref{fig:quasienergy_surface}. We denote the minimal value of $g$ as $g_{\min}\equiv g({}Q_\nu,{}P_\nu)$.   

The wave functions $\psi_\nu(Q;n)$ are approximate eigenstates of the Hamiltonian $g$. Near the bottom of a well $\nu$, functions $\psi_\nu(Q;n)$  have the form of the eigenfunctions of the harmonic oscillator in their central part. The Hamiltonian of this oscillator is obtained by expanding $g(Q,P)$ about $({}Q_\nu,{}P_\nu)$ to the second order in $Q-{}Q_\nu, P - {}P_\nu$.  The lowest state was described earlier \cite{Zhang2017}. The large-$n$ states can be obtained in the WKB approximation, 
\begin{align}
\label{eq:WKB_nu}
&\psi_\nu(Q;n)=C _n^{(\nu)}(\partial_P g)^{-1/2}\exp[iS^{(\nu)}(Q,g_n)/\lambda],\nonumber\\
&\partial_Q S^{(\nu)} = P^{(\nu)}(Q,g_n)
\end{align}
Here, $P^{(\nu)}(Q,g)$ is the solution of the equation $g(Q,P)=g$ that corresponds to the Hamiltonian orbit circling the $\nu$th minimum of $g(Q,P)$ with the classical RWA energy $g$. Parameter $C^{(\nu)}_n$ is the normalization constant.  For large $n$, one finds $g_n$ by calculating $S_\nu(Q,g)$ and applying the standard Bohr-Sommerfeld quantization condition \cite{landau1977}. By symmetry the values of $g_n$  are the same for all wells $\nu=0,1,2$. 

The intrawell wave functions of different wells overlap, which is the tunneling effect. However, for small $\lambda$ the overlapping  is exponentially small. As a result we have, to the leading order in the overlapping,
\begin{align}
\label{eq:g_matrix}
_\nu\!\Bra{n} g \Ket{n}_\nu = g_n,\qquad _\nu\!\Bra{n} g \Ket{n}_{\nu\pm 1} = J_{\nu\pm}(g_n).
\end{align} 
The hopping integral $J_{\nu\pm}(g_n)$ for the lowest intrawell states was found in Ref.~\onlinecite{Zhang2017}; for higher states it is discussed in Sec.~\ref{subsec:onset_quantum_activation}. The notion of intrawell states is meaningful if $|J_{\nu\pm}(g_n)|\ll |g_n-g_{n\pm 1}|$. Since the symmetry operator $\hat N_3$ can be thought of as rotating all wells together, $\nu\to \nu+1, \forall \nu$, we see that $J_{\nu\pm}$ is independent of $\nu$ and $J_{\nu+}\equiv J_{0+}=J_{\nu-}^*\equiv J_{0-}^*$.

To the lowest order in the overlap integrals, the eigenfunctions $\phi_\nu^{(k)}$ and eigenvalues $g_n^{(k)}$ ($k=0,1,2$) of the RWA Hamiltonian $g$ can be found in the same way as the Bloch wave functions and the energy levels in the tight-binding approximation in solid-state physics \cite{Ziman1979},
\begin{align}
\label{eq:nth_wave_functions}
&\phi_n^{(k)}(Q) \approx  3^{-1/2}\sum_\nu \psi_\nu(Q;n)  e^{-2\pi i \nu k/3},\nonumber\\
&g_n^{(k)} - g_n \approx 2{\rm Re}[ J_{0+}(g_n)\exp(-2\pi i k/3)]. 
\end{align}
From Eqs.~(\ref{eq:WKB_nu}) - (\ref{eq:nth_wave_functions}), the approximately calculated Bloch-type wave functions $\phi_n^{(k)}(Q)$ have maxima inside the wells of $g(Q,P)$ and are small outside the wells.  It is clear from qualitative arguments that this is also true for the wave functions $\phi_n^{(k)}(Q)$ calculated exactly, for example, by numerically  diagonalizing the Hamiltonian (\ref{eq:g_in_ladder_ops}). It is important that, since the operator $g(Q,-i\lambda\partial_Q)$ is real, its eigenfunctions can be also chosen to be real.

Using the exact Bloch-type wave functions $\phi_n^{(k)}(Q)$, one can construct Wannier-type wave functions 
\begin{align}
\label{eq:Wannier_defined}
\psi_{\nu; {\rm W}}(Q;n)= 3^{-1/2}\sum_k \phi_n^{(k)}(Q)\exp(2\pi i\nu k/3).
\end{align}
Since the exact wave functions $\phi_n^{(k)}(Q)$ are orthogonal and normalized, the Wannier functions (\ref{eq:Wannier_defined}) are also orthogonal and normalized, in contrast to $\psi_\nu(Q;n)$. However, one can make functions $\psi_\nu(Q;n)$ and $\psi_{\nu; {\rm W}}(Q;n)$ with the same $\nu$ and $n$ very close to each other inside the $\nu$th  well, simultaneously for all $\nu$. This requires choosing the exact functions $\phi_n^{(k)}(Q)$ to be real and properly choosing the signs of all $\phi_n^{(k)}(Q)$ with a given $n$. As Eq.~(\ref{eq:nth_wave_functions}) suggests, inside the well $\nu=0$, where only $\psi_0(Q;n)$ is large (in the absolute value), if we want it to be close to $\psi_{0; {\rm W}}(Q;n)$, all exact wave functions $\phi_n^{(k)}(Q) $ should be close to each other. This determines the choice of the otherwise arbitrary signs of the numerically calculated $\phi_n^{(k)}(Q) $ with different $k$.

\section{Dissipation-induced intrawell transitions}
\label{sec:intrawell_transitions}

Coupling of the oscillator to a thermal reservoir leads to dissipation and to transitions between the intrawell states $|n\rangle_\nu\equiv \Ket{\psi_\nu(Q;n}$ of the Hamiltonian $g(Q,-i\lambda\partial_Q)$. We consider the parameter range where the dissipation-induced widths of the intrawell levels $g_n$, which are determined by the dissipation  rate, are much larger than the tunnel splitting $\sim |J_{0\pm}(g_n)|$. In this case the statistical distribution over the intrawell states of the oscillator is formed much faster then the distribution over different wells. By symmetry arguments, the intrawell transition rates are the same for the different wells of $g(Q,P)$. For concreteness, we will consider transitions between the states in the well $\nu= 0$ and use a simplified notation $|n\rangle \equiv |n\rangle _{\nu=0}$. 

\begin{figure}[h]
\includegraphics[width = 7.5 cm]{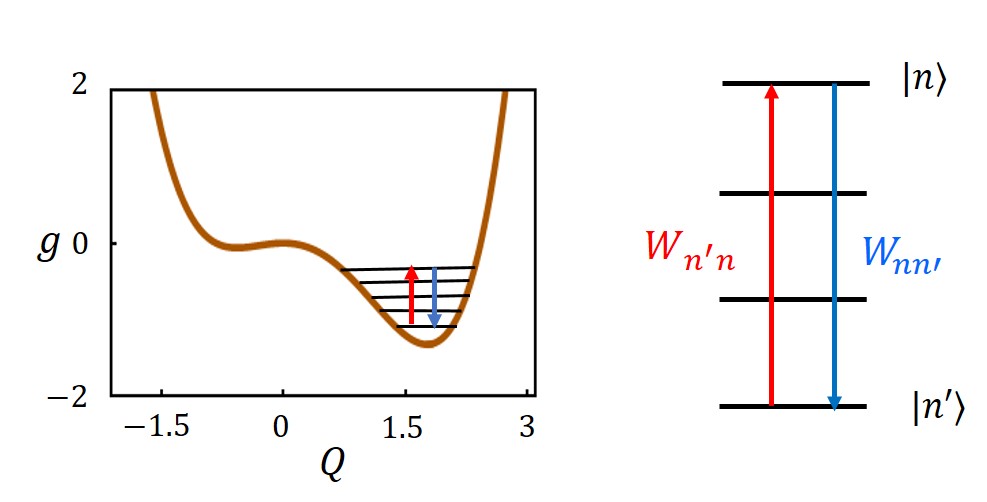} 
\caption{The cross-section of the Hamiltonian function $g(Q,P)$ by the plane $P=0$  and the sketch of dissipation-induced transitions between the intrawell states. Even at zero temperature there occur transitions both up and down the $g$-axis. The direction of more probable transitions determines which Floquet state is predominantly occupied in the presence of dissipation. In particular, inside the well of $g(Q,P)$, the transitions toward the minimum of the well (blue arrow) have higher rates than away from the minimum. For the oscillator driven at $\approx 3\omega_0$, the transition rates do not satisfy the detailed balance condition even for $T=0$, in contrast to the oscillator driven at $\approx \omega_0$ or $\approx 2\omega_0$.}
\label{fig:schematic_hoppings}
\end{figure}

As mentioned above, we enumerate the intrawell states in such a way that $g_n>g_m$ if $n>m$.  Even at zero temperature, where dissipation corresponds to emission of excitations into the thermal reservoir and thus the oscillator makes transitions only from its Fock states with higher (true) energy to the Fock states with lower energy, in terms of the intrawell states $\Ket{n}$ the transitions occur both toward larger and lower $n$. Formally, this is a consequence of the states $|n\rangle$ being linear combinations of the Fock states. The transitions between the states $|n\rangle$ are sketched in Fig.~\ref{fig:schematic_hoppings}.

The rate of a transition $|n\rangle \to |m\rangle$ is determined by the overlapping of the wave functions $|n\rangle,|m\rangle$. Therefore it falls off exponentially with the ``distance'' $|n-m|$. However, as we show in the next section, because the stationary state populations $\rho_n$ also fall off exponentially with increasing $n$, transitions from the remote small-$n$ states close to the bottom of the well can make a dominating contribution to the populations $\rho_n$ of large-$n$ states. This means strong non-locality of the distribution $\rho_n$ in the state-number space. Such effect is fairly general for systems far from equilibrium, whereas we are not aware of systems in thermal equilibrium where it would be known, unless there are selection rules that make the transition rates very different for different groups of states.

\subsection{Balance equation for the state populations}
\label{subsec:balance_equation}

To describe dissipation, we assume that the periodically driven nonlinear oscillator is weakly coupled to a thermal reservoir and that the coupling is linear in the oscillator dynamical variables. Then dissipation comes from the oscillator transitions between neighboring Fock states with emission or absorption of excitations in the reservoir. We further assume that the spectral density of these excitations weighted with the interaction is nearly constant around  the oscillator eigenfrequency within a band with width given by the detuning $|\delta \omega|$ and the nonlinear shift of the oscillator eigenfrequency $|\gamma| \langle q^2\rangle/\omega_0$. Then the oscillator dynamics in slow time $\tau = |\delta\omega|t$ is Markovian and is described by a quantum kinetic equation of the Landau-Lindblad form (cf. \cite{Landau1927,Walls2008}),
\begin{align}
\label{eq:master_general}
\frac{d}{d\tau} \rho =& (i/\lambda)[\rho, g] - \kappa{\cal D}[a]\rho,\nonumber\\
{\cal D}[a]\rho =& (\bar n+1)(a^\dagger a\rho - 2a\rho a^\dagger + \rho a^\dagger a)\nonumber\\
&+\bar n (a a^\dagger\rho -2 a^\dagger \rho a + \rho a a^\dagger).
\end{align}
Here, $\kappa$ is the dimensionless decay rate of the oscillator amplitude; it is related to the decay rate of the amplitude in the unscaled time (the coefficient of viscous friction, in the classical description) $\Gamma$ as $\kappa=\Gamma/\|\delta\omega|$; $\bar n$ is the Planck number, $\bar n = [\exp(\hbar\omega_0/k_B T)-1]^{-1}$.

Since the overlap integrals of the wave functions in different wells of $g(Q,P)$ are exponentially small, one can disregard mixing of the states in different wells by the relaxation superoperator ${\cal D}$. For $\kappa \gg |J_{0\pm}|$, dissipation-induced intrawell transitions occur much faster than interwell transitions.  For the dimensionless time $\tau$ small compared to the long time of the interwell relaxation, the interwell transitions can be disregarded; they are discussed below in Sec.~\ref{sec:tunneling}.

If the distance between the intrawell levels $g_{n+1} -g_n$ is large compared to the level broadening $\propto \lambda\kappa$, the off-diagonal matrix elements of the density matrix $\langle n|\rho|n'\rangle$  ($n\neq n'$) are small compared to the state populations $\rho_n \equiv \langle n|\rho|n\rangle$ (we recall that $|n\rangle$ refers to the intrawell states in the well $\nu=0$; for concreteness, we consider the matrix elements of $\rho$ for the states in this well). If we disregard the off-diagonal matrix elements, we obtain a balance equation for the state populations $\rho_n$,
\begin{align}
\label{eq:balance_equation}
\frac{d}{d\tau} \rho_n = -\sum_{n'} W_{nn'} \rho_n +  \sum_{n'} W_{n'n} \rho_{n'},
\end{align}
where $W_{nn'}$ are the rates of the dissipation-induced transition $|n\rangle \to |n'\rangle$. From Eq.~(\ref{eq:master_general}), 
\begin{align}
\label{eq:hopping_rates}
W_{nn'} = 2\kappa \left[ (\bar n +1)\left|\langle n'|a|n\rangle\right|^2 +  \bar n \left|\langle n|a|n'\rangle \right|^2\right]
\end{align}
We note that in a nonequilibrium system the ratio of the transition rates is not given by the Einstein relation for equilibrium systems,  $W_{nn'}/W_{n'n} \neq (\bar n + 1)/\bar n$~\cite{Kubo1957}.
%We also note that, for the coupling to a bath linear in $q,p$,  the coefficients $W_{nn'}$ can be expressed in terms of the coupling matrix elements using the Fermi golden rule, to the lowest order in the coupling.

\subsection{Transition rates in the semiclassical limit}
\label{sec:matrix_elements}

The rates of the dissipation-induced intrawell transitions $W_{n\, n\pm m}$ are determined by the matrix elements
\[\langle n+m|\hat a|n\rangle \equiv a_m(g_n).\]  
For small  $\lambda$ and for not too large $|m|\ll 1/\lambda$
%[the states $|n\rangle$ and $|n+m\rangle$ are localized well within the same well $\nu=0$ of $g(Q,P)$],
such matrix elements are given by the Fourier components of the vibrations of the corresponding classical dynamical variable with the scaled RWA energy $g_n$~\cite{landau1977},
\begin{align}
\label{eq:matrix_elements}
a_m(g) =\frac{\omega(g)}{2\pi}\int_{-\pi/\omega(g)}^{\pi/\omega(g)}d\tau \exp[-im\omega(g)\tau]a(\tau; g).
\end{align}
Here, $\omega(g)$ is the cyclic frequency of the periodic motion along the intrawell classical trajectory $g(Q,P)=g$, which is described by the Hamiltonian equations 
\begin{align}
\label{eq:classical_eom}
\frac{d}{d\tau} Q = \partial_P g, \qquad \frac{d}{d\tau} P = -\partial_Q g.
\end{align}
The function $a(\tau;g)$ is calculated as a function of time on this trajectory and is expressed in terms of the coordinate $Q$ and momentum $P$ on the trajectory as $a(\tau;g) = (2\lambda)^{-1/2}[Q(\tau;g)+iP(\tau;g)]$. Equations (\ref{eq:hopping_rates}) - (\ref{eq:classical_eom}) allow one to calculate the hopping rates numerically in the WKB approximation.

The WKB ansatz (\ref{eq:matrix_elements}) requires $n, n+m\gg 1$. For $n, n+m$ close to 1, the wave functions $\Ket{n}, \Ket{n+m}$ are close to the wave functions of a harmonic oscillator (see below) and the calculation of the matrix elements of $Q$ and $P$ is straightforward.

In terms of the matrix elements (\ref{eq:matrix_elements}), the transition rates become 
\begin{align}
\label{eq:semiclassical_hopping_rates}
W_{n \,n+m} = 2\kappa [(\bar n +1)|a_m(g_n)|^2 + \bar n |a_{-m}(g_n)|^2].
\end{align}
Here, we have neglected corrections $\sim \lambda$ and, for large $n$, set $a_{-m}(g_{n+m})\approx a_{-m}(g_n)$, in which case $W_{n\,n+m} \approx W_{n-m \,n}$.

The matrix elements $a_{m}(g_n)$ slowly vary with $n$ for large $n$, but rapidly change with $m$. The structure of the solution of the balance equation depends on the asymptotic behavior of $a_m$ for large $|m|$, such that $1\ll |m|\ll1/\lambda$. To find the large-$|m|$-behavior of $a_m(g)$, one can shift the integration contour in Eq.~(\ref{eq:matrix_elements})  into complex time. For positive and negative $m$ the contour is shifted into the lower and upper halfplane of the $\tau$-plane, respectively. The locations of the singularities of the function $a(\tau;g)$ in the complex time plane determine the exponential decay rate of the matrix element $a_m$.
The details of this calculation are in Appendix~\ref{sec:contour_integral}. For large negative $m$, we have

\begin{align}
\label{eq:a_m<0}
|a_{m<0}(g_n)|\approx &\frac{(3/2)^{1/6}\Gamma(1/3)}{2\pi \sqrt{\lambda}}\left[ \frac{\omega^2(g_n)}{|m|f}\right]^{1/3} \nonumber \\
&\times \exp[-|m|\omega(g_n)\tau_\infty(g_n)].
\end{align}
For large positive $m$, we have 
\begin{align}
\label{eq:a_m>0}
|a_{m>0}(g_n)|\approx &\frac{(2/3)^{1/6}\Gamma(2/3)}{2\pi\sqrt{\lambda}}\left[\frac{f\omega(g_n)}{m^2}\right]^{1/3} \nonumber \\
&\times \exp[-m\omega(g_n)\tau_\infty(g_n)].
\end{align}

\begin{figure}[ht]
\includegraphics[width = 7.5 cm]{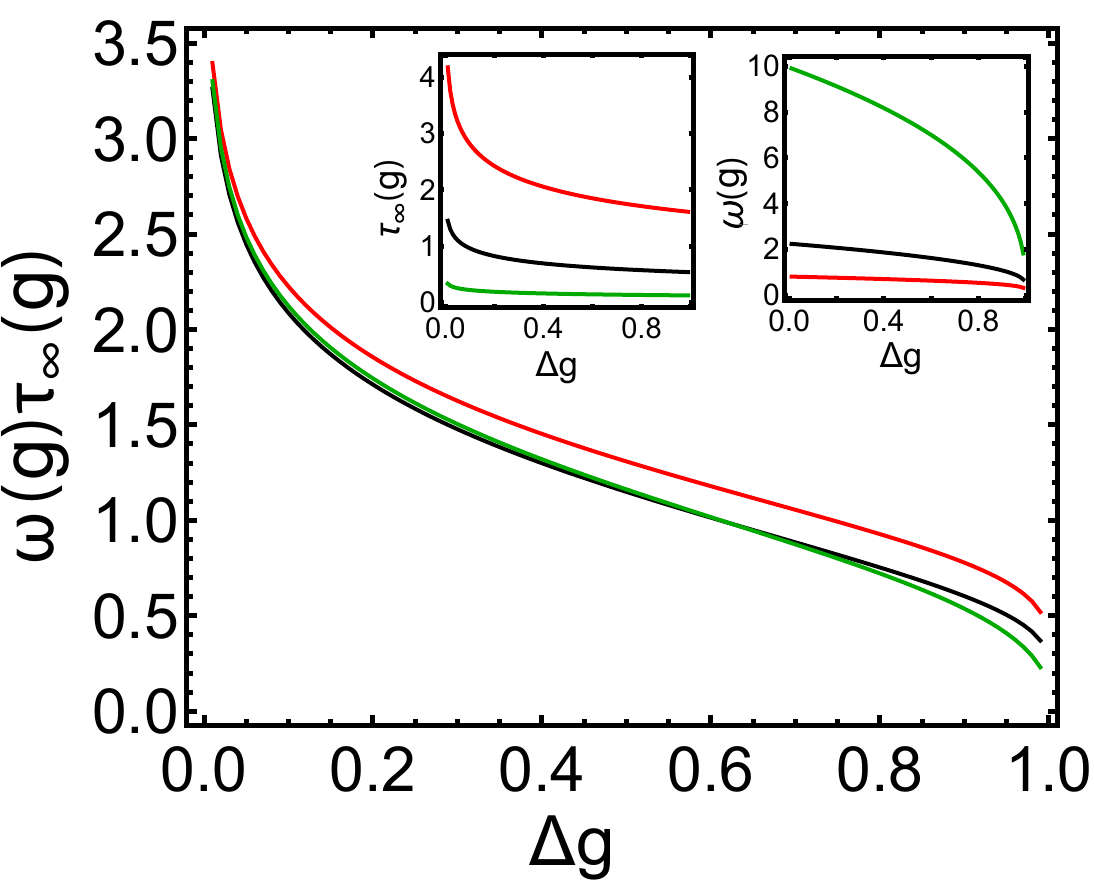}
\caption{The product $\omega(g)\tau_\infty(g)$ as a function of the distance to the bottom of the well  $\Delta g{} = (g-g_{\min})/(g_s-g_{\min})$; here, $\tau_\infty(g)$ is the minimal imaginary time  to go to infinity along the Hamiltonian trajectory (\ref{eq:classical_eom}), whereas $\omega(g)$ is the classical frequency of intrawell vibrations with a given $g$. The curves with different colors refer to the scaled drive amplitude $f$ = 0.1 (red), 0.5 (black),  and 2 (green); $\delta\omega>0$. The insets shows $\omega(g)$ and $\tau_\infty(g)$ separately for the same values of $f$. Note that the values $g_{\rm min}$ and $g_s$ of $g(Q,P)$ at the minimum and the saddle point depend on $f$. }
\label{fig:tau_infty}
\end{figure}

The parameter $\tau_\infty$ is the smallest distance along the imaginary time $\tau$ to the singularity of $Q(\tau;g)$ and $P(\tau;g)$. It is defined by Eq.~(\ref{eq:imaginary_time})  and is the same in the upper and the lower half-plane of the complex $\tau$ plane. It is shown in Fig.~\ref{fig:tau_infty}. 

Since the distance to the singularities is the same in the both half-planes, the matrix elements $|a_m(g_n)|^2$, and thus the transition rates,  fall off with $|m|$ exponentially with the same exponent proportional to $\tau_\infty>0$ for $m>0$ and $m<0$. This is qualitatively different from the rates of transitions between the Floquet  states of an oscillator driven by a force with frequency close to the eigenfrequency or  a parametrically modulated oscillator at frequency close to twice the oscillator  eigenfrequency \cite{Marthaler2006,Guo2013,Peano2014}. In the both latter cases, for $T=0$, the rates of transitions away from the stable state fall off exponentially faster with the interstate distance than the rates of transitions toward the stable state.

We show in Fig.~\ref{fig:tau_infty} the decrement $\omega(g)\tau_\infty(g)$ of the fall-off rate  of the matrix elements Eqs.~(\ref{eq:a_m<0}) and (\ref{eq:a_m>0}) with $|m|$. The decrement monotonically decreases with the increasing  $g$, and interestingly, for a given $\Delta g{}$, it depends only weakly on the scaled drive amplitude $f$. For $g$ close to $g_{\rm min}$, it diverges logarithmically 
\begin{align}
\label{eq:tao_infty_diverge}
\omega(g)\tau_\infty \approx -\frac{1}{2}\ln \Delta g{}, \qquad \Delta g{}\equiv \frac{g-g_{\rm min}}{g_s-g_{\rm min}}\ll 1,
\end{align}
due to the divergence of $\tau_\infty(g)$ seen in the inset of Fig.~\ref{fig:tau_infty}, and weakly depends on $f$, for a given $\Delta g$. This is  because classical vibrations near the minima of $g(Q,P)$ become almost harmonic. Therefore, higher-order Fourier components of the classical motion decay fast. For $g$ close to the saddle point value $g_s$, the period of the classical orbit diverges logarithmically and the motion becomes extremely anharmonic: very slow near the saddle points, but fast away from them. As a result, $\omega(g)\tau_\infty(g)$ approaches zero logarithmically for $g$ close to $g_s$.

Another distinctive and important feature of period tripling is that the singularities of $a(\tau;g)$ in the complex plane $\tau$ are branching points rather than  poles, as for resonant driving or parametric modulation. The branching points at $ i\tau_\infty$ and $-i\tau_\infty$ are different. Respectively, the prefactors in Eqs.~(\ref{eq:a_m<0}) and (\ref{eq:a_m>0}) show a power-law dependence on $m$, but with different exponents.

The parameter used to obtain the large-$|m|$ asymptotic behavior of the coefficients $a_m(g)$ is $|m|\omega(g)\tau_\infty\gg 1$. For $f\lesssim 1$ we have $\tau_\infty \sim 1$, see Fig.~\ref{fig:tau_infty}  ($\tau_\infty \gg 1$ for $g\to g_{\min}$). Then Eqs.~(\ref{eq:a_m<0}) and (\ref{eq:a_m>0}) apply provided $|m|\omega(g) \gg 1$. In this case for $\bar n =0$  the ratio of the rates of transitions away and toward $g_{\min}$ is $W_{n\,n+m}/W_{n\,n-m} \propto [f^2/m\omega(g)]^{2/3} \ll 1$ ($m>0$). This shows that, in the stationary regime, the system is mostly localized near the minimum of the well of $g(Q,P)$: it is much more probable to make a transition toward this minimum than away from it. From Eq.~(\ref{eq:Hamilton_eqns}), for  $f\gg 1$ the frequency $\omega(g)$ scales with $f$ as $f^2$, and therefore still $W_{n\,n+m}/W_{n\,n-m}\ll 1$ for $m>0$ as long as $m\omega(g)\tau_\infty(g)\gg 1$.

\subsection{Breakdown of the detailed balance}
\label{subsec:detailed balance}

A periodically driven oscillator is far from thermal equilibrium. Generally, it should not have detailed balance. The absence of detailed balance was shown for a resonantly driven oscillator in the classical limit \cite{Dykman1979}. However, as mentioned in the Introduction, in the deeply quantum regime of zero temperature, a resonantly driven Duffing oscillator with the relaxation described by the master equation (\ref{eq:master_general}) has detailed balance in the RWA \cite{Drummond1980}. Moreover, for the same type of relaxation, detailed balance for zero temperature was found also for a Duffing oscillator parametrically modulated close to twice its eigenfrequency \cite{Kryuchkyan1996}. The physical reason for the detailed balance, which exists only for $T=0$ and only for this relaxation mechanism is not known (it breaks down in the presence of dephasing \cite{Marthaler2006}). It is interesting and, in a way, illuminating, to find out whether detailed balance persists for $T=0$ and  relaxation described by Eq.~(\ref{eq:master_general}) in the case of an oscillator driven close to triple the eigenfrequency.

The condition of detailed balance is the relation between the rates of interstate transitions that shows that the ratio of the rates of direct transitions back and forth between the states is equal to the ratio of the rates of transitions via an intermediate state, 
\[ \frac{W_{nn'}}{ W_{n'n}} =  \frac{W_{nn''} }{W_{n'n''}} \frac{W_{n''n'}}{ W_{n''n}} .\]
For $T=0$ we have $W_{nn'} = 2\kappa |a_{n'-n}(g_n)|^2$. As seen from Eqs.~(\ref{eq:a_m<0}) and (\ref{eq:a_m>0}), for period tripling there is no detailed balance even for $T=0$. This shows that the detailed balance for modulation at frequencies $\approx \omega_0$ and $\approx 2\omega_0$  for $T=0$ is non-generic.

\section{Quasistationary distribution over intrawell states}
\label{sec:distribution}
Over the dimensionless relaxation time $\sim \kappa^{-1}$,  transitions between the intrawell states form a quasistationary distribution over these states. This distribution can be found from the stationary solution of the balance equation (\ref{eq:balance_equation}). An important property of the transition rates is
\[W_{n+m\,n} >W_{n\,n+m}, \quad m>0,\]
that is, the system is more likely to go from a higher-lying to a lower-lying state than in the opposite direction. This property is seen from Eqs.~(\ref{eq:a_m<0}) and (\ref{eq:a_m>0}) for large $m$ and from the numerical evaluation of the semiclassical rates $W_{nn'}$ based on Eq.~(\ref{eq:matrix_elements}) for $|n-n'|\sim 1$. It is clear then that the stationary distribution will be maximal near the bottom of the well of $g(Q,P)$. However, since the probabilities of transitions up along the $g_n$ axis are nonzero, the stationary distribution over the levels $g_n$ will have a nonzero width.

We are interested in the semiclassical parameter range $\lambda\ll 1$, where the number of intrawell bound states is large. 
The analysis of the distribution should be done differently in different ranges of the values of the scaled RWA energy $g$. We note that it does not apply in a narrow range of $g$ close to the saddle value $g_s$, where the interwell transitions have to be taken into account.

\subsection{The vicinity of the minima of the wells of $g(Q,P)$}
\label{sec:harmonic_approximation}

The distribution over intrawell states can be found analytically near the minima of the RWA Hamiltonian function $g(Q,P)$. In the rotating frame, the vibrations of $Q,P$ about a minimum are almost harmonic and, respectively, the levels $g_n$ are almost equidistant. To find the distribution for the well $\nu=0$, we expand $g(Q,P)$ about the $\nu=0$-minimum,
\begin{align}
\label{eq:g_expanded}
&g \approx  g_{\rm min} + \frac{1}{2}g_{PP} P^2 + \frac{1}{2}g_{QQ} (Q-Q_0)^2, \nonumber \\
& g_{PP} = 3fQ_0, \qquad g_{QQ} = fQ_0 +2\sgn(\delta\omega),  \nonumber\\ 
&Q_0 = \frac{1}{2}[f+\sqrt{f^2+4\sgn(\delta\omega)}].
\end{align}
Here $(Q_0,P_0=0)$ is the position of the minimum and $g_{\min}$ is the value of $g(Q,P)$ at the minimum, $ g_{\min}=fQ_0^2(3f-4Q_0)/12$. 

It is convenient to change from $P$ and $Q-Q_0$ to the raising and lowering operator $b^\dagger$ and $b$ using a squeezing transformation,
\begin{align}
\label{eq:g_harmonic}
& Q - Q_0 + iP = (2\lambda)^{1/2}(b \cosh\phi_* - b^\dagger \sinh\phi_*), \nonumber \\
& g \approx g_{\rm min} + \lambda \omega_{\rm min}(b^\dagger b + 1/2),\quad \omega_{\rm min} = \sqrt{g_{PP}g_{QQ}},
\end{align}
where $\omega_{\rm min}$ is the vibration frequency at the bottom of the well of $g(Q,P)$ and the squeezing parameter $\phi_*$ is given by the equation $\tanh\phi_* = (|g_{QQ}|^{1/2}-|g_{PP}|^{1/2})/(|g_{QQ}|^{1/2}+|g_{PP}|^{1/2})$. The intrawell states $\Ket{n}$ near the bottom of the well are well approximated by the eigenstates of the operator $b^\dagger b$.

From Eq.~(\ref{eq:g_harmonic}), the lowering operator of the oscillator $a= (2\lambda)^{-1/2}(Q+iP)$ is a linear combination of the operators $b$ and $b^\dagger$.  Therefore the transitions rates $W_{n\,n+m}$, Eq.~(\ref{eq:hopping_rates})  are nonzero for $m=\pm 1$, and this is the case even for $\bar n=0$, where the oscillator makes dissipative transitions only to a lower Fock state. The rates $W_{n\,n+m}$ are easy to find from  Eq.~(\ref{eq:g_harmonic}), taking into account that $\Bra{n}b\Ket{n+m} = (n+1)^{1/2}\delta_{m,1}$. Substituting the rates into the balance equation (\ref{eq:balance_equation}), we find that the stationary solution of this equation near the minimum of the well is  
\begin{align}
\label{eq:Boltzmann}
&\rho_n \propto \exp(-n\lambda \omega_{\rm min}/T_{\rm eff}), \nonumber \\
&\lambda \omega_{\rm min}/T_{\rm eff} = \log[(\bar n_{\rm eff} + 1)/\bar n_{\rm eff}]. 
\end{align}
where $n_{\rm eff}$ is the effective Planck number,
\begin{align}
\label{eq:effective_nbar}
\bar n_{\rm eff} = \bar n + (2\bar n +1) \sinh^2\phi_*.
\end{align}

Equation (\ref{eq:Boltzmann}) has the form of the Boltzmann distribution. As seen from Eq.~(\ref{eq:effective_nbar}), the effective temperature of the distribution over the intrawell states is nonzero even where the effective temperature of the thermal reservoir is equal to zero. This is the quantum heating effect \cite{Dykman1988a,Marthaler2006}. For a resonantly driven oscillator it was observed experimentally in Ref.~\onlinecite{Ong2013}.

\begin{figure}
\includegraphics[width = 5.5 cm]{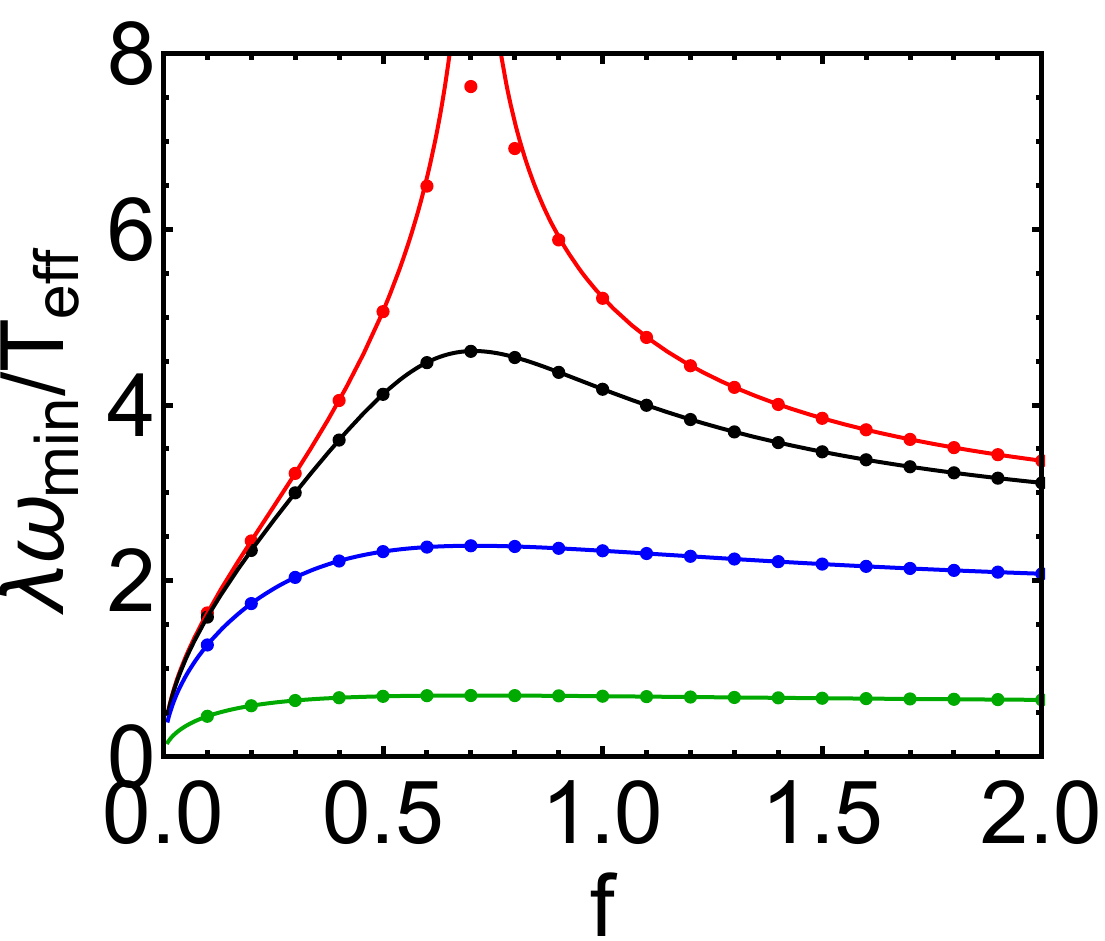}
\caption{The scaled inverse temperature $\lambda \omega_{\rm min}/T_{\rm eff}$ in the harmonic approximation as a function of the scaled drive amplitude $f$ for $\delta\omega>0$. The red, black, blue, and green lines are given by Eqs.~(\ref{eq:Boltzmann}) and (\ref{eq:effective_nbar}) with the equilibrium Planck number $\bar n =0, 0.01,0.1, 1$, respectively. The dots show the result of the eikonal approximation, Eq.~(\ref{eq:equation_for_xi}), for $g$ close to $g_{\rm min}$ ($\Delta g = 0.0001$). We define the effective temperature in this case from the expression $\lambda \omega_{\rm min}/T_{\rm eff}=[R'(g) \omega(g)]_{g\to g_{\min}} $. The difference between the dots and the curves is pronounced for $\bar n=0$ and small $|f-1/\sqrt{2}|$, where $T_{\rm eff} \to \infty$ in the harmonic approximation that gives Eqs.~(\ref{eq:Boltzmann}) and (\ref{eq:effective_nbar}). This difference decreases with the decreasing $\Delta g$. }
\label{fig:harmonic_R'}
\end{figure}

The dependence of $\bar n_{\rm eff}$ on $f$ is different for different signs of the detuning of the drive frequency $\delta\omega= (\omega_F/3)-\omega_0$. For $\delta\omega<0$, as mentioned previously, $g(Q,P)$ has a minimum at $Q=P=0$, whereas the wells at nonzero $Q^2+P^2$ appear only for $f>2$. In this case $\bar n_{\rm eff}$ monotonically decreases as $f$ increases. Near the threshold $f = 2$,  $\bar n_{\rm eff}$ diverges as $(f-2)^{-1/4}$. For large $f\gg 1$, $\bar n_{\rm eff}$ asymptotically approaches $ (2\bar n + 1)/\sqrt{3}-1/2$. 

For $\delta\omega>0$ function $g(Q,P)$ has the form of a Mexican hat, with the top at $Q=P=0$ and the rim at $Q^2+P^2 = 1$ for $f=0$. The wells $\nu=0,1,2$ emerge for $f>0$ with no threshold, as a modulation of the depth of the rim. The effective Planck number $\bar n_{\rm eff}$ is non-monotonic as a function of the scaled drive amplitude $f$. It diverges as $f^{-1/2}$ for $f\to 0$. As $f$ increases, $\bar n_{\rm eff}$ reaches a minimum at $f=1/\sqrt{2}$, where $\bar n_{\rm eff}=\bar n$. This corresponds to the disappearance of the squeezing, $\sinh\phi_* = 0 $ for $f=1/\sqrt{2}$. For $f>1/\sqrt{2}$,  $\bar n_{\rm eff}$ increases with the increasing $f$ and approaches the same asymptotic value as in the case $\delta\omega<0$. 

We show in Fig.~\ref{fig:harmonic_R'} the scaled effective inverse temperature $\lambda \omega_{\rm min}/T_{\rm eff}$ as a function of the scaled drive amplitude $f$ for $\delta\omega>0$. For $\bar n = 0$, the inverse temperature diverges at $f=1/\sqrt{2}$ as $\left|\log(f-1/\sqrt{2})^2\right|$.

For $f=1/\sqrt{2}$ and $\bar n=0$ the analysis of the distribution over the states $\Ket{n}$ has to go beyond the harmonic approximation (\ref{eq:g_harmonic}) and  take into account the terms of higher order in $P$ and $Q-Q_0$ in Eq.~(\ref{eq:g_expanded}). By perturbation theory in the parameter $\lambda$, the cubic terms in $P$ and $Q-Q_0$ lead to a finite transition rate $W_{02}\propto \lambda$, and more generally, $W_{n\,n+2}\propto \lambda$ for $n\sim 1$. The finite rates of the upward in $g$ transitions result in a nonzero population of the excited states; the distribution over the states falls off as a power series in $\lambda$.

\subsection{The eikonal approximation}
\label{subsec:eikonal}

To find the stationary solution of the balance equation (\ref{eq:balance_equation})  away from the minimum of $g(Q,P)$ it is convenient to seek this solution in the eikonal form,
\begin{align}
\label{eq:eikonal}
\rho_n = \exp[-R(g_n)/\lambda].
\end{align} 
As it stands, Eq.~(\ref{eq:eikonal}) just uses the fact that $\rho_n\geq 0$. It applies  near the minimum of the well of $g(Q,P)$ as well, as seen from Eq.~(\ref{eq:Boltzmann}), but is not limited to this range of $g$.

We will assume, and then check a posteriori,  that $R(g)$ is a smooth function of $g$. This is an analog of the eikonal approximation in optics or the WKB approximation in quantum mechanics. In this approximation, for $|n'-n|\ll 1/\lambda$ we have  
\begin{align}
\label{eq:local_approximation}
R(g_{n'}) &\approx R(g_n) + (g_{n'} - g_n) R'(g_n) \nonumber \\ 
& \approx R(g_n) +  \lambda \omega(g_n)R'(g_n) (n'-n),
\end{align}
where $R'(g)\equiv dR/dg$; we have used here the semiclassical expression for the quasienergy level spacing in terms of the classical vibration frequency $\omega(g)$.

The approximation (\ref{eq:local_approximation}) was used to find the stationary distribution of resonantly and parametrically driven quantum oscillators \cite{Dykman1988a,Marthaler2006} from the balance equation for the state populations (\ref{eq:balance_equation}). As we show, it allows one to solve the balance equation  for the considered here period tripling problem in a broad parameter range, but it can break down for $\bar n =0$. This is qualitatively different from the cases considered earlier. 

We emphasize that Eq.~(\ref{eq:local_approximation}) assumes that the {\em logarithm} of the distribution $\rho_n$ smoothly depends on $n$, whereas the distribution itself does not have to be a smooth function of $n$ and can vary by a factor ${\cal O}(1)$ when $n$ is incremented by 1. This is similar to the WKB approximation, where the wave function is oscillating fast, but the wavelength changes only a little on the distance given by the wavelength itself. 

From Eqs.~(\ref{eq:eikonal}) and (\ref{eq:local_approximation}),  we see that $\rho_{n+m}/\rho_n = \exp[-m\omega(g_n)R'(g_n)]$ for $|m|\ll 1/\lambda$. This means that $\omega(g_n)R'(g_n)$ can be interpreted as the $g_n$--dependent inverse temperature of the Boltzmann-like distribution over the RWA energy levels $g_n$. 

Equations (\ref{eq:balance_equation}) and (\ref{eq:local_approximation}) reduce the problem of finding the distribution over the excited intrawell states ($n\gg 1$) to an algebraic equation 
\begin{align}
\sum_m W_{n+m\,n} (\xi_n^m-1 )=0, \quad \xi_n = e^{-R'(g_n)\omega(g_n)}.
\label{eq:equation_for_xi}
\end{align} 
Here the hopping rates $W_{nn'}$ are given by the semiclassical expression (\ref{eq:semiclassical_hopping_rates}). As indicated above, for $|n-n'|\sim 1$ they can be calculated numerically from Eqs.~(\ref{eq:classical_eom}),  whereas for large $|n-n'|$ we have the explicit expressions (\ref{eq:semiclassical_hopping_rates}) - (\ref{eq:a_m>0}). In writing Eq.~(\ref{eq:equation_for_xi}) we used $W_{n \,n+m}=W_{n-m \, n}$ for large $n\gg 1$ and $|m|\ll n$. 

 A crucial feature of Eq.~(\ref{eq:equation_for_xi}) is that it is {\it local} in the level number $n$. It assumes that the stationary population $\rho_n$ is formed by the transitions from a few states $n+m$ surrounding a given state $n$, with $|m|\ll n, 1/\lambda$. Formally, it requires that the sum over $m$ in Eq.~(\ref{eq:equation_for_xi}) converges. If this is the case, Eq.~(\ref{eq:equation_for_xi}) defines a smooth function $R'(g)$ which depends on the single parameter, the scaled amplitude of the driving field $f$. However, the applicability of this approximation is not known a priori. We note that  Eq.~(\ref{eq:equation_for_xi}) does not contain the scaled Planck constant $\lambda$; in particular, $R'$ is independent of $\lambda$.

\subsubsection{The vicinity of the minima of the Hamiltonian function}

Typtically, the WKB approximation applies for highly excited states that correspond to large $n$ in Eq.~(\ref{eq:equation_for_xi}). In the considered case the situation is different. Because near the  minima of the quasienergy surface the dynamics in the rotating frame can be mapped onto the dynamics of an auxiliary weakly nonlinear oscillator, cf. Eq.~(\ref{eq:g_harmonic}), the semiclassical approximation works for $g$ approaching $g_{\rm min}$. One can show that for $g-g_{\min} \ll 1$, the semiclassical transition rates $W_{n+m\,n}$ in Eq.~(\ref{eq:semiclassical_hopping_rates})  scale as $W_{n+m\,n}\propto  (g-g_{\rm min})^{|m|}$. If one sets to zero the rates $W_{n+m\,n}$ with $|m|\geq 2$, Eq.~(\ref{eq:equation_for_xi}) gives the same result  for the distribution as in Eq.~(\ref{eq:Boltzmann}), with $R'(g_{\rm min}) \omega(g_{\rm min})\to \lambda \omega_{\rm min}/T_{\rm eff}$ for $g\to g_{\min}$. Clearly, the relation $W_{n\,n+m} = W_{n-m\,n}$ does not apply for small $n\sim 1$, but the range of small $n$ corresponds to a very narrow range of $g$ for small $\lambda$. 

In Fig.~\ref{fig:harmonic_R'} we show the result of solving Eq.~(\ref{eq:equation_for_xi}) for $g$  close to $g_{\rm min}$. It matches the solution (\ref{eq:Boltzmann}) obtained in the harmonic approximation well in a broad parameter range. However, it also shows that the transitions with $|m|>1$ lead to an occupation of the excited intrawell states for $\bar n=0$ even for the value of the driving amplitude $f$ where, in the harmonic approximation, such states remain unoccupied. This shows that $T_{\rm eff}$ becomes nonzero  if the transitions with $|m|>1$ are taken into account.

\subsubsection{Classical limit}
For large Planck numbers $\bar n$, Eq.~(\ref{eq:equation_for_xi}) greatly simplifies, since $\omega(g) R'(g) \ll 1$. One can then expand $\exp[-m\omega(g)R']$ in a series to second order in $R'$. Then, from Eqs.~(\ref{eq:semiclassical_hopping_rates}) and (\ref{eq:equation_for_xi}), 
\begin{align}
\label{eq:small_R_prime}
R' (g) &= \frac{2\sum_m |a_{-m}(g)|^2 m}{\omega(g)(2\bar n +1) \sum_m |a_{-m}(g)|^2 m^2}  
\end{align}
Using that the matrix elements $a_m(g)$ are the Fourier components of the dynamical variable $a(\tau;g)\propto Q(\tau;g)+iP(\tau;g)$  calculated for a classical orbit (\ref{eq:classical_eom}) with a given $g$, the sums in Eq.~(\ref{eq:small_R_prime}) can be written as integrals over the area enclosed by this orbit in the phase space~\cite{goldstein2001}. Then Eq.~(\ref{eq:small_R_prime}) can be rewritten as,
\begin{align}
\label{eq:small_R_prime_simplified}
&R'(g)= \frac{2 M(g)}{(2\bar n +1)N(g)},\quad M(g) =  \iint dQdP, \nonumber \\
& \,\, N(g) = \frac{1}{2}  \iint \n^2 g(Q,P) dQdP.
\end{align}
As indicated above, the integrals here are taken over the interior of the region $g(Q,P)=g$ within a well of $g(Q,P)$ and $\n^2 g(Q,P) \equiv \partial^2_Q g  + \partial^2_P g = 4(P^2+Q^2)-2\,{\rm sgn}(\delta\omega)$.

In the high temperature range $k_B T \gg \hbar \omega_0$, $R'$ is inversely proportional to $T$. The distribution over intrawell states goes over into the classical (but non-Boltzmann) distribution formed as a result of thermal-noise-induced diffusion over the RWA-energy.  

Equation~(\ref{eq:small_R_prime}) applies for an arbitrary temperature near the threshold of the period tripling,  but not too close to the threshold, so that the condition $\omega(g_{\min})\gg \kappa$ still holds. Being near the threshold means $f\ll 1$ for $\delta\omega>0$ and $f-2\ll 1$ for $\delta\omega<0$, to zeroth order in $\kappa$. Near the threshold, the vibration frequency $\omega (g) \ll 1$ for the whole range of the intrawell values of $g$. Indeed, near the minimum of $g(Q,P)$ we have $\omega_{\rm min} \approx(6f)^{1/2} \ll 1$ for $\delta\omega>0$ and $ \omega_{\rm min} \approx 2\sqrt{3}(f-2)^{1/4} \ll 1$ for $\delta\omega<0$. For larger $g$, $\omega(g)$ monotonically decreases and approaches zero as $g$ approaches $g_s$.

We compare in Appendix~\ref{sec:classical_compare} the result of the classical limit given by Eq.~(\ref{eq:small_R_prime}) with the semiclassical  calculation based on Eq.~(\ref{eq:equation_for_xi}). The classical results match well the semiclassical ones already for moderately small $\bar n \sim 1$.% as shown in Fig.~\ref{fig:Rprime_classical}.

\section{The breakdown of the local approximation}
\label{sec:non_locality}
The eikonal approximation relies on the stationary distribution  being formed locally by transitions between a few nearest states. The number of the states involved should be much smaller than $1/\lambda$, otherwise the expansion of the logarithm of the probability distribution (\ref{eq:local_approximation}) does not apply. As a consequence, the polynomial equation for $R'$ (\ref{eq:equation_for_xi}), which is a direct analog of the equation for the derivative of the logarithm of the wave function in the WKB approximation, does not apply either. It turns out that, for $\bar n =0$, the locality, and thus the eikonal approximation, can break down for the oscillator that displays period tripling. In this section, we study the condition of such a breakdown and show how to go beyond the local approximation. We focus on the case of a positive detuning $\delta\omega>0$.

\subsection{The locality condition and its breakdown for zero temperature}

The stationarity of the intrawell state populations $\rho_n$ means that the probability flux out of a state $n$ (the rate of leaving the state) and the probability flux into the state are equal. We can write the balance equation (\ref{eq:balance_equation}) in terms of the relative incoming and outgoing fluxes as
\begin{align}
\label{eq:fluxes_introduced}
&\frac{d}{d\tau} \log\rho_n = \sum_m\left[\Jin_m(g_n)-\Jo_m(g_n)\right], \nonumber\\
&\Jin_m(g_n)=W_{n+m\,n}\frac{\rho_{n+m}}{\rho_n}, \quad \Jo_m(g_n)=W_{n\,n+m}.
\end{align}
The exponential decay of the rate $W_{n\,n+m}\propto \exp[-2|m|\omega(g_n)\tau_\infty(g_n)]$ for large $|m|$ guarantees that the total relative outgoing flux $\sum_m\Jo_m$ is always finite. Since the stationary state populations $\rho_n$ decrease with the increasing $n$, that is, $\rho_{n+m}/\rho_n <1$ for $m>0$, the relative incoming flux from the states with larger RWA energy $\sum_{m>0}\Jin_m(g_n)$ is also finite. Moreover, it is local, only a few nearest states contribute to this flux.

In contrast, but for the same reason, the relative flux ${\cal J}^{\rm in}_{m<0}$ into a given state from the states with lower RWA energy  can be non-local. If the increase of the population $\rho_{n-|m|}$ with the increasing $|m|$ is faster than the decrease of the transition rate $W_{n-|m|\,n}$, the incoming relative flux ${\cal J}^{\rm in}_{m<0}$ increases with  the increasing distance between the states $|m|$. 

If we use the eikonal approximation for the state populations, Eq.~(\ref{eq:local_approximation}), we find that, for large $|m|$ (but $|m|\ll 1/\lambda$) and for a given $n$,  ${\cal J}^{\rm in}_{m<0}\propto  \exp[-|m|\omega (2\tau_\infty-R')]$. Thus, the locality condition for a given $n$  is
\begin{align}
\label{eq:locality_condition}
\lambda \ll \omega(g_n)[2\tau_\infty(g_n)-R'(g_n)].
\end{align}
The right-hand side of this inequality gives the fall-off rate of the relative influx with the distance to the lower-lying state, in the eikonal approximation. The left-hand side is proportional to the reciprocal number of states in the well of $g(Q,P)$, and thus puts a limit on the number of the lower-lying states. When the condition (\ref{eq:locality_condition}) is not met, the eikonal approximation (\ref{eq:local_approximation}) does not allow one to find the stationary probability distribution over the intrawell states. One needs to use the full  balance equation (\ref{eq:balance_equation})  (we remind that, in fact, we are discussing  the quasi-stationary distribution, as we have so far disregarded interwell transitions). Also, one should not use Eq.~(\ref{eq:semiclassical_hopping_rates}) to calculate the transition rates, but rather the full WKB wavefunctions or the Wannier-type functions to calculate the relevant matrix elements.

\begin{figure}[ht]
\includegraphics[width =5.5 cm]{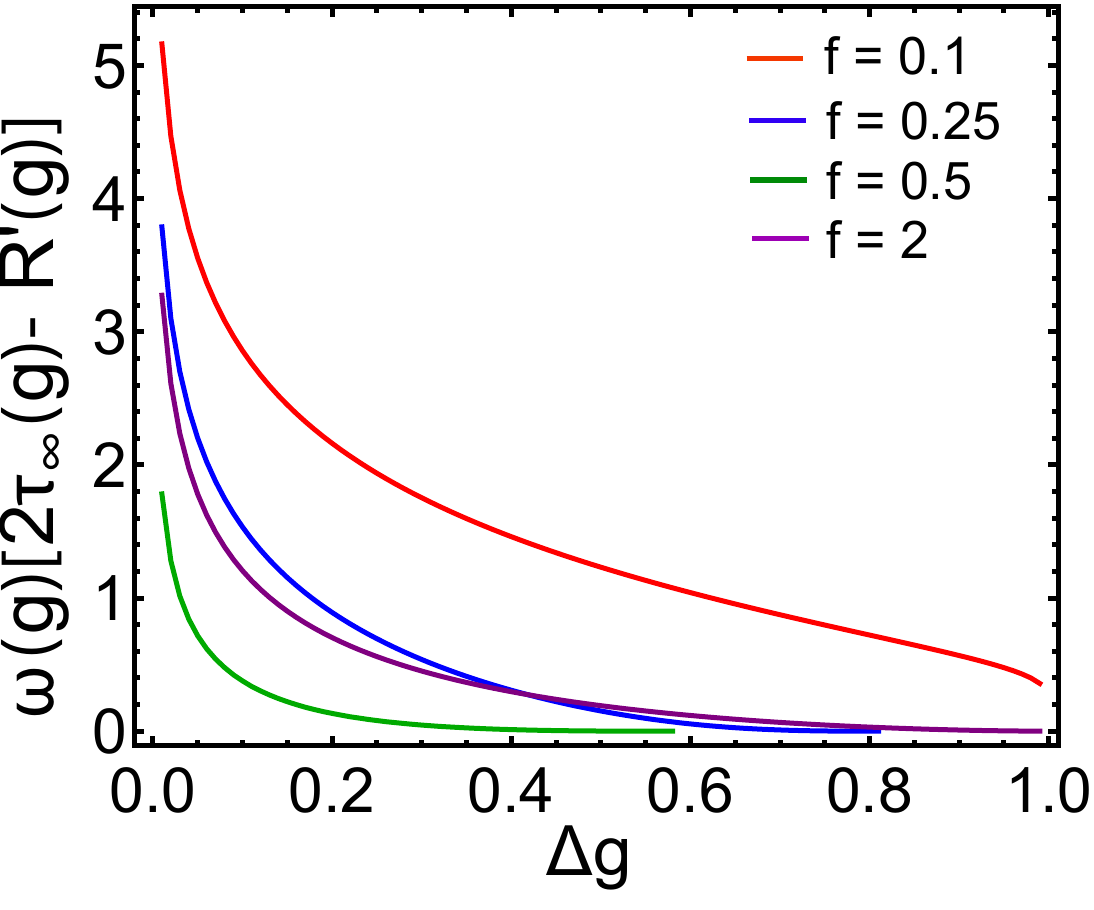} \hfill
\caption{The rate $\omega(g) [2\tau_\infty(g)-R'(g)]$ of the fall-off of the relative probability flux $\Jin_{-|m|}(g)$ [in Eq.~(\ref{eq:fluxes_introduced})] into the intrawell state with quasienergy $g$ from lower-lying states, for different $\Delta g = (g-g_{\min})/(g_s-g_{\min})$. The lines refer to the scaled driving amplitude $f$ = 0.1 (red), 0.25 (blue), $0.5$ (green), and 2 (purple).  As the RWA energy  $g$ approaches $g_{\rm min}$, one can show that $\omega(g) [2\tau_\infty(g)-R'(g)]$ diverges $\propto |\ln(g-g_{\rm min})|$.
}
\label{fig:breakdown_locality}
\end{figure}

We start with analyzing the locality breakdown in the semiclassical limit $\lambda \rightarrow 0$. The function $\omega(g) [2\tau_\infty(g)-R'(g)]$ calculated in the ``local'' approximation (\ref{eq:equation_for_xi}) is shown in Fig.~\ref{fig:breakdown_locality}. For relatively large (purple line) or small (red line) values of the scaled driving field $f$, function $2\tau_\infty(g)-R'(g)$ remains positive for all intrawell values of $g$, from $g = g_{\rm min}$ to $g= g_s$. Thus, the locality holds. However, for $f\sim 1$, function $2\tau_\infty(g)-R'(g)$ becomes equal to zero at some $g=g_{\rm NL}$ as seen from the data shown by the blue and green lines in Fig.~\ref{fig:breakdown_locality}. For $g>g_{\rm NL}$, Eq.~(\ref{eq:equation_for_xi}) does not have a nontrivial solution for $R'$, signaling the breakdown of the locality.

\begin{figure}[h]
\includegraphics[width =5.5 cm]{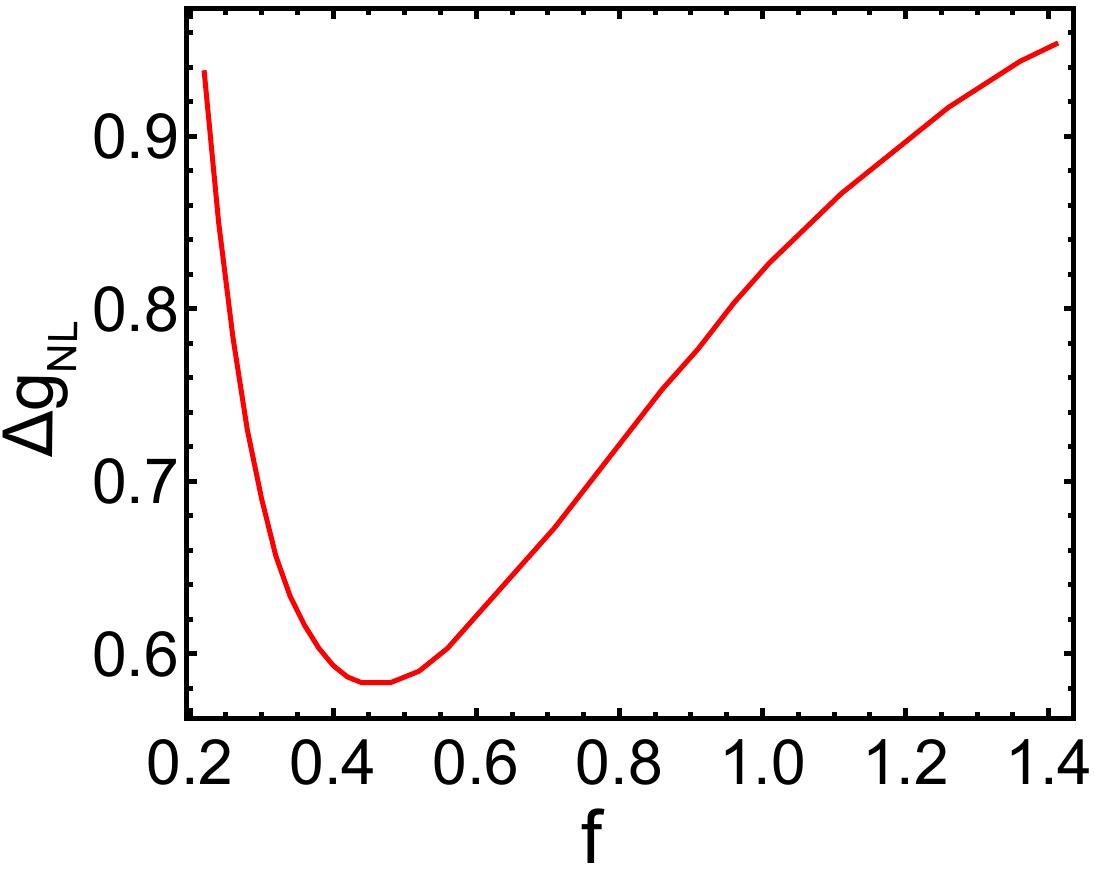}
\caption{The minimal  scaled RWA energy $\Delta g_{\rm NL}=(g_{\rm NL}-g_{\min})/(g_s-g_{\min})$ where the locality breaks down as a function of the scaled driving amplitude $f$ for $\delta\omega>0.$}
\label{fig:gNL}
\end{figure}

We denote the minimum value of $g$ where the locality breaks down by $g_{\rm NL}$. In Fig.~\ref{fig:gNL} we show the dependence of $g_{\rm NL}$ on the scaled driving amplitude $f$. As seen from the figure, $g_{\rm NL}$ first emerges with the increasing $f$ at the saddle point value $g_s$ of $g(Q,P)$ for $f\approx 0.2$. It then moves deeper into the well of $g(Q,P)$. The distance $\Delta g_{\rm NL}$  from $g_{\rm NL}$ to the bottom of the well scaled by the depth of the well $g_s-g_{\min}$ is minimal for $f \approx 0.5$. As $f$ further increases, $g_{\rm NL}$ goes back to $g_s$. For $f\gtrsim 1.4$ the nonlocality disappears.

To find accurately the location of $g_{\rm NL}$ from Eq.~(\ref{eq:equation_for_xi}), it is important to take into account the large-$|m|$ contribution to the influx ${\cal J}^{\rm in}_{m<0}$. This is because for $g$ close to $g_{\rm NL}$, ${\cal J}^{\rm in}_{m<0}(g)$ falls off with $|m|$ rather slowly, with the fall-off exponent $\omega(g) [2\tau_\infty(g)-R'(g)]$ close to zero as can be seen in Fig.~\ref{fig:breakdown_locality}. The large-$|m|$ contribution can be calculated using the asymptotic expression for the transition rates, Eqs.~(\ref{eq:hopping_rates}) and  (\ref{eq:a_m>0}). 

In the region of nonlocality we have $R'(g)>2\tau_\infty(g)$. The stationary probability distribution in this region is found numerically and is discussed in Sec.~\ref{subsec:numerical_Wannier}.

\subsection{Effect of nonzero temperature}
The very existence of the non-locality in the transitions among intrawell states depends sensitively on the interrelation between the temperature and the scaled Planck constant $\lambda$. In the limit $\lambda \rightarrow 0$ the locality is formally preserved at any nonzero temperature. This is seen from the expression  (\ref{eq:semiclassical_hopping_rates}) for the transition rates $W_{nn'}$. At nonzero temperatures, the second term in $W_{nn'}$, which comes from the absorption of energy from the thermal reservoir, becomes non-zero. Although this term is small for low temperatures, it can significantly change the distribution. We now provide analytical arguments to show that this is the case. The numerical results are given in the next section. 

The relative influx into the state with a given RWA energy $g_n$ can be written as a sum of the contribution from nearby states [small $|m|$ in Eq.~(\ref{eq:fluxes_introduced})] and that from the states that are further away (large $|m|$).  We will use that, where the locality holds, $\rho_{n+m}/\rho_n = \exp[-mR'(g_n)\omega(g_n)]$ and consider the ``most dangerous'' situation where $R'\approx 2\tau_\infty$ and therefore the contribution of the remote states is significant. Using Eqs.~(\ref{eq:a_m<0}) and (\ref{eq:a_m>0}) we find 
\begin{align}
\label{eq:influx_sum}
&\sum_{m<0} {\cal J}^{\rm in}_{m} \approx (\bar n+1) C_1 \Gamma(-1/3)[\omega(2\tau_\infty-R')]^{1/3} \nonumber\\
&+ \bar n C_2 \Gamma(1/3)[\omega(2\tau_\infty-R')]^{-1/3} + C_0.
\end{align}
Here $C_1,C_2$ are constants independent of $R'$, whereas the parameter $C_0$ comes from the small-$|m|$ terms in Eq.~(\ref{eq:fluxes_introduced})  and depends  on $2\tau_\infty-R'$ in a nonsingular way. 

As seen from Eq.~(\ref{eq:influx_sum}), for  $R'(g) \to 2\tau_\infty(g)$ the part of the relative  influx, which is proportional to $\bar n$, diverges. The other terms in Eq.~(\ref{eq:fluxes_introduced}) remain finite. Therefore, for any nonzero $\bar n$, the``local''  balance equation (\ref{eq:equation_for_xi}) has a nontrivial stationary solution for $R'(g)$ with $R'(g) < 2\tau_\infty(g)$, as the term $\propto \bar n$ can compensate other terms.  In other words, the locality is preserved in the limit $\lambda \rightarrow 0$ but $\bar n \neq 0$. However, as we have seen, the locality can break down for $\bar n=0$.

Where $R'(g_n)$ approaches $2\tau_\infty(g_n)$, the number of states that contribute to the influx into the state $n$ increases  $\propto [2\tau_\infty(g_n)-R'(g_n)]^{-1}$. For $\bar n\to 0$ this number diverges. Physically, the total number of states in a well of $g(Q,P)$ is always finite. It is $\propto 1/\lambda$.  Therefore for a nonzero $\lambda$, the transition from the non-locality at $\bar n=0$ to the locality at $\bar n \neq 0$ is smeared out. The ``non-locality'' means  that essentially all states with $g_m<g_n$ contribute to the stationary population  $\rho_n$, whereas the ``locality'' means that the number of such states is much smaller.  For a small nonzero $\bar n$, the number of the contributing states is given by the condition that the term $\propto \bar n [2\tau_\infty(g)-R'(g)]^{-1/3}$   in Eq.~(\ref{eq:influx_sum}) is of the same order as other terms in the balance equation. The number  of the contributing states $\sim  [2\tau_\infty(g)-R'(g)]^{-1}$ is thus $\propto \bar n^{-3}$. Therefore the locality holds where $\bar n^3 \gtrsim \lambda$.

We note that there is a profound difference between the non-locality for period tripling discussed here and that for an oscillator driven near its eigenfrequency~\cite{Guo2013,Peano2014} or parametrically modulated near its second overtone~\cite{Marthaler2006}. There, the locality holds for $\bar n =0$. The non-locality emerges only in a narrow range of nonzero $\bar n$ due to the asymmetry of the decay of the semiclassical matrix elements $a_{m>0}$ and $a_{m<0}$ with $|m|$. In contrast, in the period-tripling case, the semiclassical matrix elements $a_{m>0}$ and $a_{m<0}$ decay with $|m|$ with the same exponents, albeit with different prefactors. The non-locality then occurs at zero temperature. 
Another qualitative distinction is that, for period tripling, the logarithm of the probability distribution $\log\rho_n=-R(g_n)/\lambda$ remains a smooth function of $g_n$. The function $R(g)$ can still be expanded in a series, it is just that keeping the first term in this expansion is not sufficient for finding the probability distribution where the locality breaks down.

\subsection{Numerical analysis using the Wannier functions}
\label{subsec:numerical_Wannier}

To confirm the breakdown of the local approximation, we numerically find the Wannier-type intrawell wave functions defined by Eqs.~(\ref{eq:nth_wave_functions}) and (\ref{eq:Wannier_defined}), then calculate the transition rates using the matrix elements of the operators $a,a^\dagger$ on these wave functions, and then solve the balance equation~(\ref{eq:balance_equation}). This method does not require the local approximation, but it significantly relies on the scaled Planck constant $\lambda$ being nonzero. An alternative approach is to calculate the matrix elements using the Landau method \cite{landau1977} for states with significantly different quantum numbers. The Landau method was used \cite{Guo2013} in the problem of an oscillator driven close to its eigenfrequency, where the singularities of the classical trajectories (\ref{eq:classical_eom}) were simple poles periodically located on the complex-time plane. In the present case the trajectories  have multiple branching points, which significantly complicates the application of the method.

\begin{figure}[ht]
\includegraphics[width=5cm]{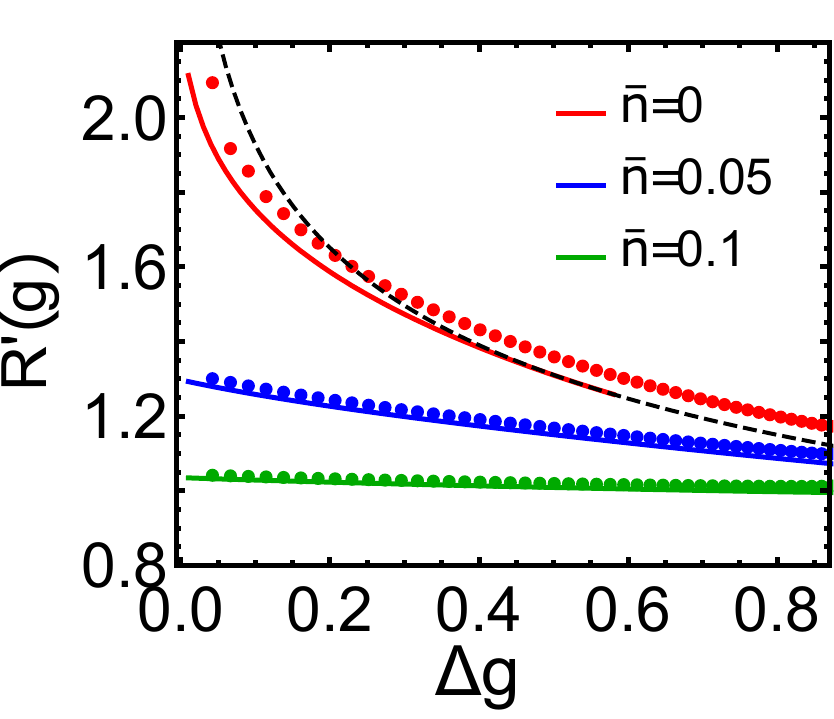}
\caption{The scaled inverse temperature $R'(g)$ of the intrawell probability distribution as a function of the RWA energy $g$, $\Delta g = (g-g_s)/(g_s-g_{\min})$. 
The scaled driving amplitude is $f=0.5$. The dots refer to $R'$ obtained by solving the balance equation~(\ref{eq:balance_equation}) using the Wannier functions, $R'(g_n) = -\lambda \log (\rho_{n+1}/\rho_{n-1})/(g_{n+1}-g_{n-1})$. The scaled Planck constant is $\lambda=0.004$. There are approximately 50 levels in each well. The solid lines refer to $R'$ obtained from Eq.~(\ref{eq:equation_for_xi}), which does not contain $\lambda$. The black dashed line shows $2\tau_{\infty}(g)$. At $\bar n = 0$, $R'(g)$ obtained using Eq.~(\ref{eq:equation_for_xi}) (red solid line) becomes equal to $2\tau_\infty(g)$ for $g = g_{\rm NL}$. For larger $g$, where $R'(g)>2\tau_\infty(g)$, the local approximation breaks down, Eq.~(\ref{eq:equation_for_xi}) no longer applies, and this is why the red line terminates at $g_{\rm NL}$. The difference between the dots and the solid lines of the same color is due to $\lambda$ being nonzero and is most significant when $R'$ becomes close to $2\tau_\infty$.% For $g\approx g_{\rm min}$, the values of $R'$ obtained using the Wannier-type wave functions are close to those shown in Fig.~\ref{fig:harmonic_R'}. 
}
\label{fig:compare_Wannier}
\end{figure}

We compare in Fig.~\ref{fig:compare_Wannier} the inverse temperature $R'(g)$ obtained using the Wannier-type functions with that obtained from Eq.~(\ref{eq:equation_for_xi}).
For a broad range of parameters, the inverse temperatures $R'(g_n)$ obtained using the two methods coincide. When $2\tau_\infty$ becomes close to $R'(g_n)$, the result obtained from Eq.~(\ref{eq:equation_for_xi}) deviates from the Wannier-functions method, indicating the breakdown of the local approximation that underlies Eq.~(\ref{eq:equation_for_xi}).

\begin{figure}[ht]
\includegraphics[width=4.2cm]{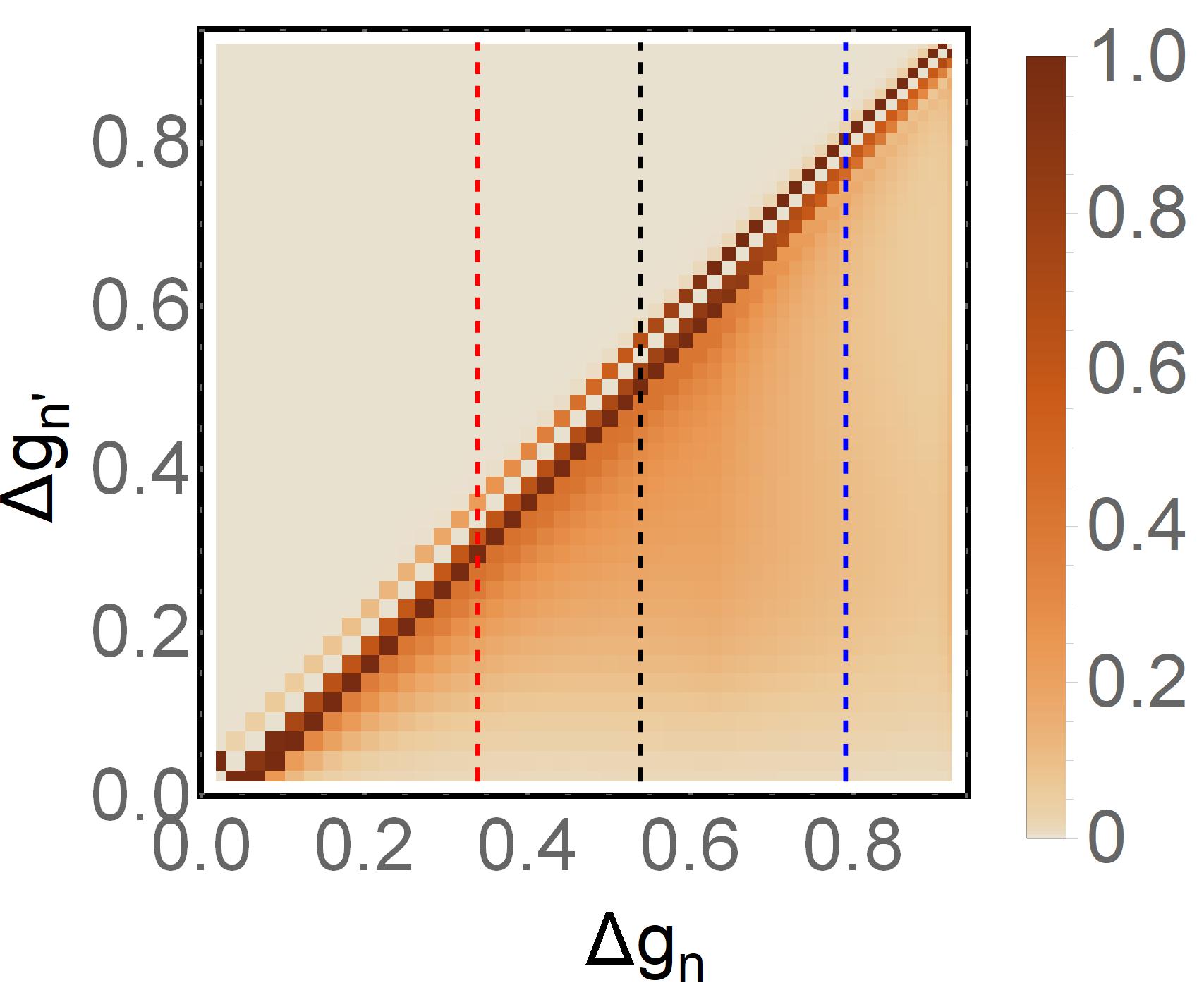}\hfill
\includegraphics[width=4.2cm]{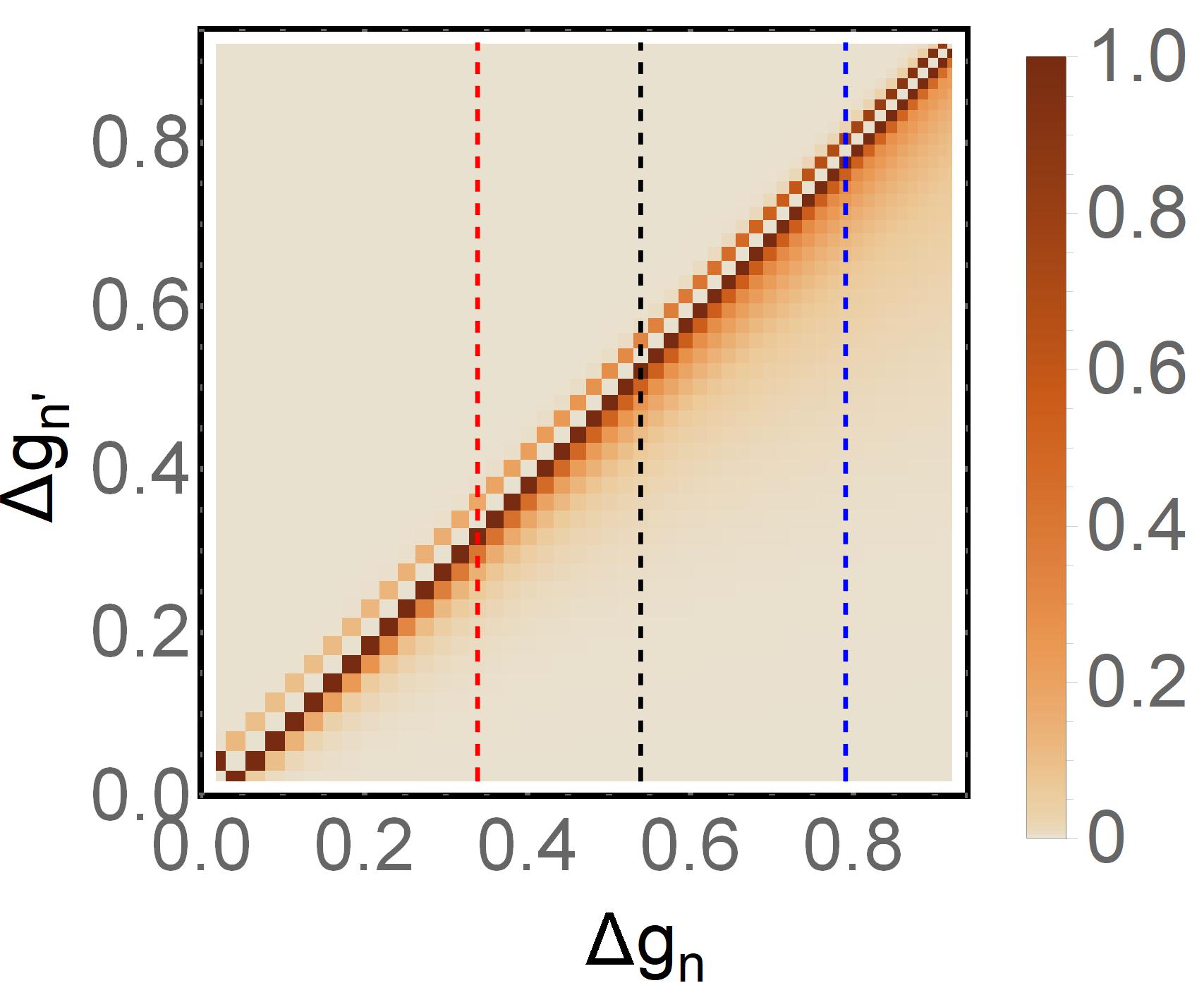} \\
\includegraphics[width=4cm]{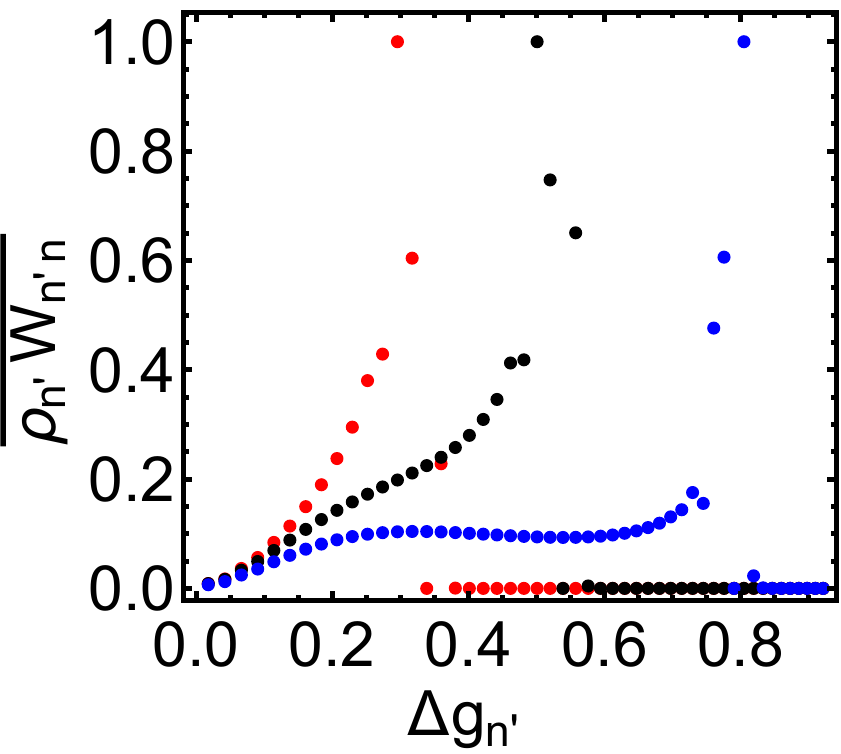} \hfill
\includegraphics[width=4cm]{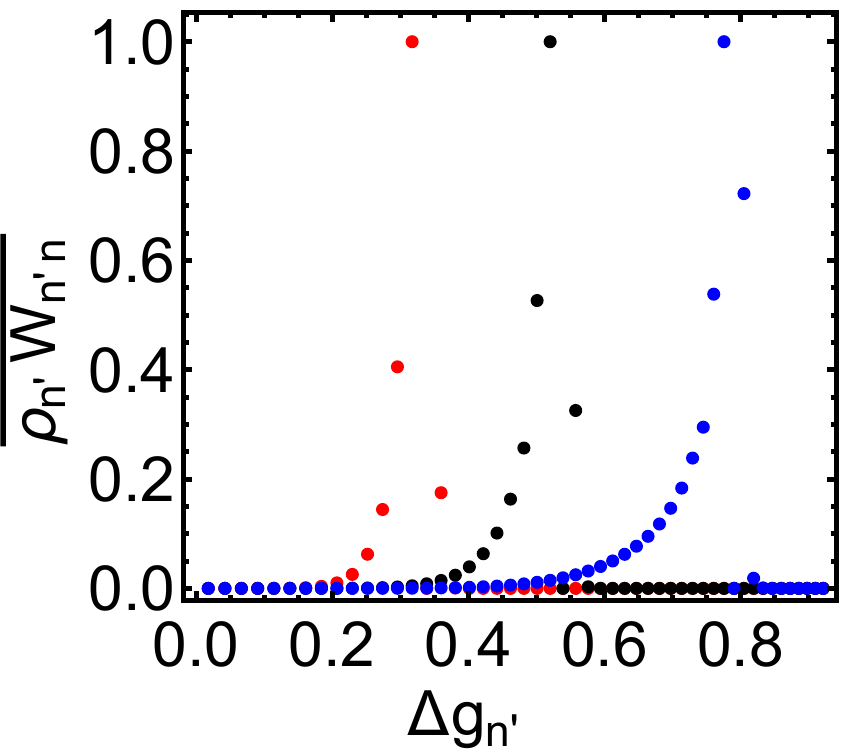}
\caption{Top panels: The normalized stationary probability flux $\Ket{n'}\to \Ket{n}$ given by the function $\overline{\rho_{n'} W_{n'n}}$, Eq.~(\ref{eq:normalized_flux}). $\Delta g_n = (g_n-g_s)/(g_s-g_{\rm min})$. The left and right panels refer to $\bar n = 0$ and $\bar n = 0.05$, respectively. The scaled drive amplitude  and the dimensionless Planck constant are $f=0.5$ and $\lambda = 0.004$ in the both panels, so that the number of levels in a well of $g(Q,P)$ is $\approx 50$. Bottom panels: the normalized flux $\overline{\rho_{n'} W_{n'n}}$ as a function of $n'$ for the three values of $n$ that correspond to the vertical red, black and blue cuts in the corresponding top panels. The dots are the cross-sections of the flux shown in the upper panels along the cuts, and the color coding of the dots corresponds to the color coding of the cuts. The left lower panel shows that, for $\bar n=0$, the decay of $\overline{\rho_{n'} W_{n'n}}$ with the increasing interstate distance slows down for larger $g_n$ and even becomes non-monotonic (see the blue dots). The right lower panel shows that, 
as the temperature increases, the decay of the flux $\overline{\rho_{n'} W_{n'n}}$ becomes exponential in $|g_n-g_{n'}|$,  indicating that the locality has been recovered.} 
\label{fig:influx}
\end{figure}

To visualize the nonlocality, we show in Fig.~\ref{fig:influx} the normalized probability flux 
\begin{align}
\label{eq:normalized_flux}
\overline{\rho_{n'} W_{n'n}}=\rho_{n'} W_{n'n}/\rm {max}_{n'} (\rho_{n'} W_{n'n})
\end{align}
from the state $\Ket{n'}$ to the state $\Ket{n}$ inside a well of $g(Q,P)$ in the stationary regime. The normalization factor is the maximal over $n'$ value of the  flux $\rho_{n'} W_{n'n}$. The flux is obtained  from Eq.~(\ref{eq:balance_equation}) using the transition rates calculated with the Wannier-type wave functions (\ref{eq:Wannier_defined}). At zero temperature (left panels), the fall-off of the  flux (\ref{eq:normalized_flux}) with the increasing interstate distance $n-n'$  becomes strongly non-exponential as the RWA energy of the state into which the transition occurs $g_n$ approaches $g_{\rm NL}$. For relatively large values of $g_n$, the flux $\rho_{n'} W_{n'n}$ becomes nonmonotonic in $n-n'$. This is a manifestation of the nonlocality: a significant portion of the probability flux into a high-lying state $\Ket{n}$ comes from the states much deeper in the well, because the increase of their population with $n-n'$ is faster than the fall-off of the transition rate  $W_{n'n}$. Moreover, the flux from the states that are further away can be larger than from at least some states closer to a given state $n$. At finite temperatures (right panels), such behavior goes away and $\rho_{n'} W_{n'n}$ falls off exponentially as $|n-n'|$ increases.

\section{Escape from the period-three states of a dissipative oscillator}
\label{sec:tunneling}

In the discussions in Secs.~\ref{sec:distribution} and~\ref{sec:non_locality}, we have disregarded tunneling between the states localized in different wells of $g(Q,P)$. As explained in Sec.~\ref{subsec:intrawell}, the tunneling causes splitting of the intrawell levels  into triplets with the level spacing $\sim |J_{0\pm}|$, see Eq.~(\ref{eq:nth_wave_functions}). If the scaled relaxation rate $\kappa$ is smaller than $|J_{0\pm}|/\lambda$, dissipation leads to transitions within the triplets and between different triplets, and ultimately there is formed a stationary distribution over the broadened tunnel-split Floquet states of the oscillator. However, for small $\lambda$ the level splitting $\sim |J_{0\pm}(g_n)|$ is exponentially small. Therefore such picture is relevant provided $\kappa$ is also exponentially small. 

In this section we focus on the parameter regime where the tunnel splitting is small compared to the dissipative level broadening, $ |J_{0\pm}|\ll \lambda\kappa $. In this case coherent tunneling is replaced by an interwell hopping (switching) with a rate $\propto |J_{0\pm}|^2/\lambda^2\kappa$, an analog of the quantum diffusion in solids \cite{Kagan1992}. Since the intrawell states in the rotating frame correspond to the vibrations at frequency $\omega_F/3$ with different phases, such interwell switching corresponds to  phase-flip transitions in the laboratory frame.

 If one disregarded the population of the excited intrawell states,  the switching would occur in the lowest state.  The dissipation-induced transitions between the intrawell states significantly complicate the picture. The distribution over the intrawell states is formed fast, over the dimensionless time $\kappa^{-1} \ll \lambda^2\kappa/|J_{0\pm}|^2$. The populations of the excited intrawell states $\rho_n$ fall off exponentially with the increasing $g_n$, whereas the tunneling matrix elements $|J_{0\pm}(g_n)|$ exponentially  increase with $g_n$. The rate of the interwell hopping is then determined by some optimal $g_n$ where $\rho_n|J_{0\pm}(g_n)|^2$ has a sharp maximum. Such competition of the exponential factors is similar to that in systems in thermal equilibrium, where the intrawell distribution is of the Boltzmann form \cite{Larkin1985}.

As we show below, in our case the optimal value of the RWA energy for interwell switching corresponds to the height of the barrier between the wells of $g(Q,P)$, i.e., the saddle-point value $g_s$. In other words, escape from a well of $g(Q,P)$  occurs via an over-barrier transition. This is an analog of thermal activation in systems in thermal equilibrium and is called quantum activation, since it is due to quantum fluctuations and occurs even for $T=0$. For quantum oscillators driven close to their eigenfrequency or parametrically modulated close to twice the eigenfrequency, quantum activation was found earlier \cite{Dykman1988a,Marthaler2006}. These systems have detailed balance for $T=0$ \cite{Drummond1980,Kryuchkyan1996}, whereas our system does not, and therefore the nature of the escape can be expected to be different.

We will focus on the case of positive detuning, $\delta\omega>0.$ For negative detuning, $\delta\omega<0$, along with interwell tunneling there also occurs tunneling between the intrawell states and the state localized near $Q=P=0$.

\subsection{The condition for the onset of quantum activation}
\label{subsec:onset_quantum_activation}

For small $\lambda$ the tunneling matrix element $J_{0\pm}(g_n)$ can be calculated in the WKB approximation. The problem is somewhat different from the problem of tunneling of a particle in a potential well, where the momentum as a function of coordinate and energy has only two values and is either real or imaginary. In the present case, the classical momentum $P(Q,g)$ defined by the equation $g(Q,P)=g$  has 4 branches as a function of the coordinate and is complex rather than purely imaginary in the classically inaccessible region. As a result the WKB wave function is not just decaying, but also oscillating in this region.  Nevertheless,  extending the analysis for the lowest intrawell state \cite{Zhang2017}, one can show that 
\begin{align}
\label{eq:tunnel_matrix_element}
&|J_{0\pm}(g)|\propto \omega(g) \exp[-S_{\rm tun}(g)/\lambda], \nonumber \\
&S_{\rm tun} (g) = \Im \int dQ P(Q,g).
\end{align}
Here, $S_{\rm tun}(g)$ is the imaginary part of the complex classical action calculated along the Hamiltonian trajectory (\ref{eq:classical_eom}) that goes in complex time from one well of $g(Q,P)$ to another well with the same RWA energy $g$. One has to choose the trajectory with the minimal $S_{\rm tun}$. 

To find the optimal value of the RWA energy $g$ for interwell hopping, we will use a quasicontinuous approximation for the intrawell distribution and replace $\rho_n=\exp[-R(g_n)/\lambda]$ with $\rho(g) = \exp[-R(g)/\lambda]$. In this approximation the rate of  interwell hopping with the RWA energy between $g$ and $g+dg$ is $\propto dg\,\exp\{-[R(g) + 2S_{\rm tun}(g)]/\lambda\}$. The optimal $g$ is determined by the extremum of the exponent.

If escape from a well of $g(Q,P)$ occurs via quantum activation, it means that the amplitude of the exponent decreases with the increasing $g$ for all values of $g$ inside the well of $g(Q,P)$, 
\begin{align}
\label{eq:q_activation_condition}
\partial_g R (g) + 2\partial_g S_{\rm tun}(g) < 0\quad  {\rm      for     }\quad  g_{\min} \leq g \leq g_s.
\end{align}

\begin{figure}[t]
\includegraphics[width =5.5 cm]{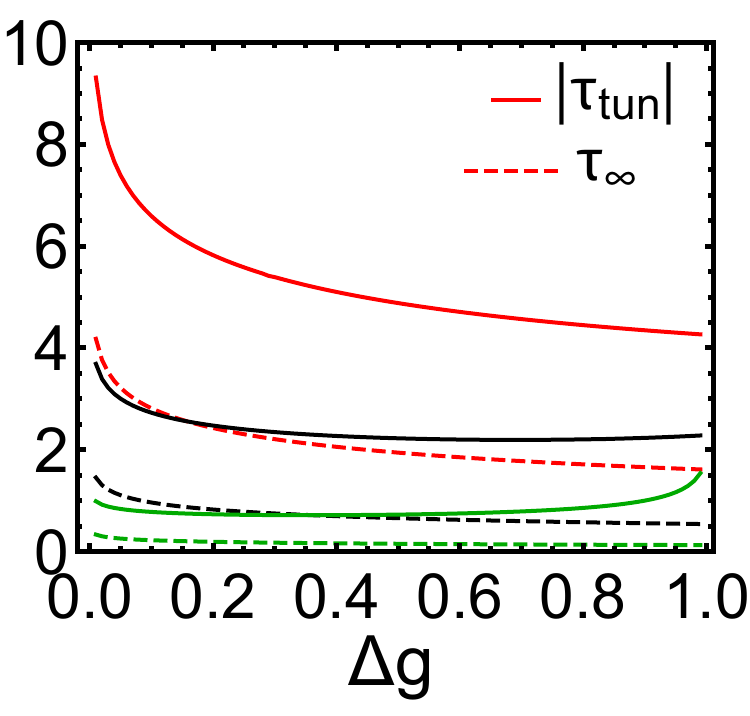} 
\caption{The imaginary tunneling time $|\tau_{\rm tun}(g)|$ (solid lines) and the limiting steepness of the exponent of the intrawell distribution $\tau_\infty(g)$ (dashed lines) as functions of $g$, $\Delta g=(g-g_{\min})/(g_s- g_{\min})$. The red, black, and green curves refer to the scaled driving amplitude $f =$ 0.1, 0.5, and 2, $\delta\omega>0.$ }
\label{fig:compare_tau_tun}
\end{figure}

We found in Sec.~\ref{sec:non_locality} that, where the intrawell distribution is  formed by transitions between comparatively close states (the locality condition), $R'(g_n)$ is limited by the decrement  $2\tau_\infty(g_n)$ of the decay of the intrawell transition rates $W_{nn'}$ with $|n-n'|$, i.e., $0<R'(g)\leq 2\tau_\infty(g)$. This limits the first term in Eq.~(\ref{eq:q_activation_condition}). 

The second term in Eq.~(\ref{eq:q_activation_condition}) is negative and is determined by the imaginary part of the time of moving along a Hamiltonian trajectory (\ref{eq:classical_eom}) through the classically inaccessible region between the wells,  
\begin{align}
\label{eq:tunnel_time_general}
\partial_g S_{\rm tun} = \tau_{\rm tun}(g) = {\rm Im}\,\int dQ[\partial_Pg(Q,P)]^{-1}.
\end{align}
The derivative $\partial_Pg$ is calculated here for $P=P(Q,g)$. The details of the calculation are given in Appendix~\ref{sec:imaginary_tunneling_time}. We note that the calculation is done differently in different regions of $g$ where the trajectories in the classically allowed region are ellipse-like or horse-shoe-like.

We show in Fig.~\ref{fig:compare_tau_tun} the imaginary tunneling time $\tau_{\rm tun}$ for various values of the driving amplitude. For $g$ close to $g_{\rm min}$, $\tau_{\rm tun}$ diverges logarithmically,
\[
\tau_{\rm tun}\sim \omega_{\min}^{-1}\log(g-g_{\rm min}).
\]
For $g$ close to $g_s$, one can show that $\tau_{\rm tun}$ is linear in $g_s-g$ and is finite at $g=g_s$; see Appendix~\ref{sec:imaginary_tunneling_time}. 

Also shown in Fig.~\ref{fig:compare_tau_tun} is the function $\tau_\infty(g)$, which limits $R'(g)/2$ in the region of locality. As seen from Fig.~\ref{fig:compare_tau_tun}, $|\tau_{\rm tun}|>\tau_\infty$, thus guaranteeing that the quantum-activation condition (\ref{eq:q_activation_condition}) holds true. Note that for $g$ close to $g_{\rm min}$, both $\tau_\infty$ and $\tau_{\rm tun}$ diverge logarithmically, but the prefactor of $\tau_{\rm tun}$ is twice as large as that of $\tau_\infty$ as can be seen from  Eqs.~(\ref{eq:tao_infty_diverge}) and (\ref{eq:tunnel_time_general}). 

In the range of the driving amplitude and the RWA energy where, for $\bar n=0$, the intrawell probability distribution is formed by transitions from remote states, we have $R'(g)>2\tau_\infty(g)$. We have found that the numerically calculated value of $R'$ in this range satisfies the condition (\ref{eq:q_activation_condition}). Therefore escape from a well of $g(Q,P)$ does occur via quantum activation.

\subsection{Quantum activation energy}

The rate of escaping from a well of $g(Q,P)$ in the rotating frame is the rate of escaping from a period-3 vibrational state in the laboratory frame. To logarithmic accuracy this rate is
\begin{align}
\label{eq:escape_rate}
W_{\rm esc}\sim \exp[-R_A/\lambda], \quad R_A = \int_{g_{\rm min}}^{g_s} dg R'(g).
\end{align}
The parameter $R_A$ is the quantum activation energy of escape. The quantum nature of the fluctuations leading to escape is clear from the fact that the escape rate displays activation dependence on $\hbar$,  $\log W_{\rm esc}\propto \lambda^{-1}\propto \hbar^{-1}$. Yet, as we mentioned, the escape occurs via an overbarrier transition, as if it were thermally activated, with $\hbar$ playing the role of temperature.

\begin{figure}[t!]
\includegraphics[width =5.5 cm]{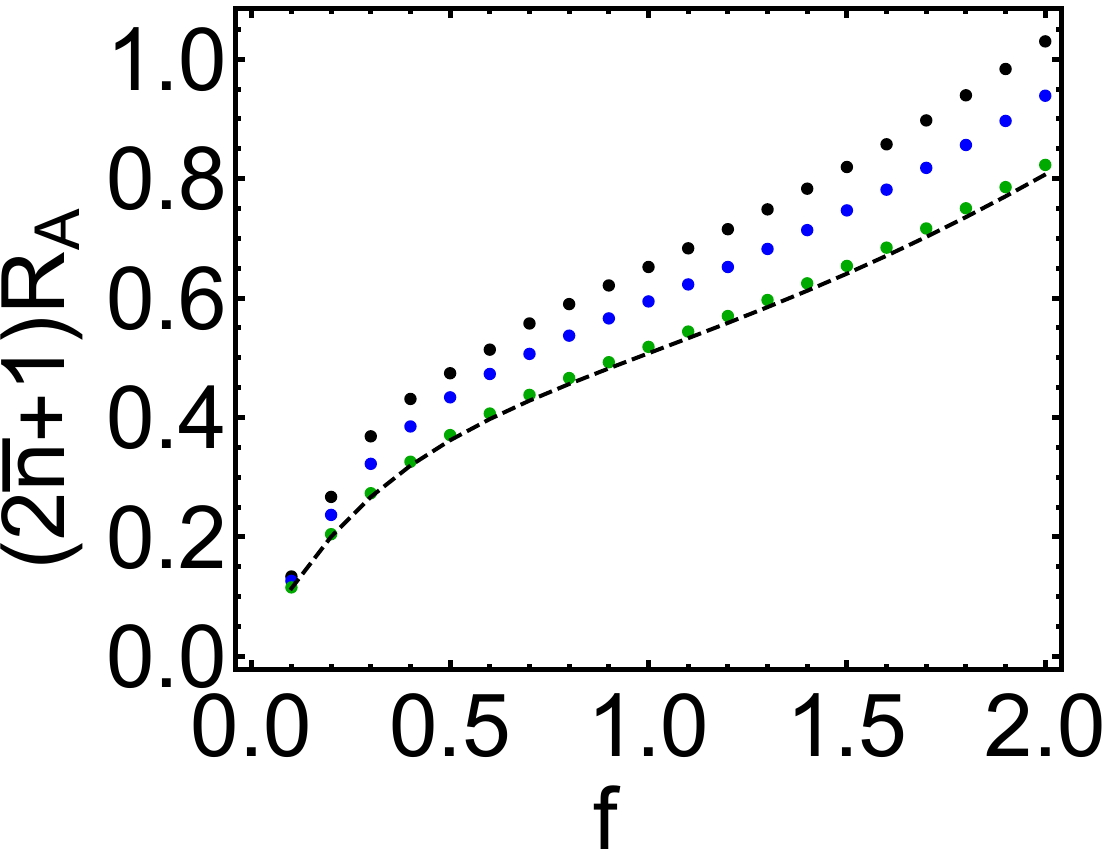} 
\caption{The semiclassical activation energy $R_A$ as a function of the scaled drive amplitude $f$. The black, blue, and green dots refer to $\bar n = 0.01, 0.1$ , and 1, respectively. The dashed line refers to the result of the classical limit obtained from Eq.~(\ref{eq:small_R_prime}). }
\label{fig:RA_semiclassical}
\end{figure}

We show in Fig.~\ref{fig:RA_semiclassical} the dependence of the activation energy $R_A$ on the scaled driving amplitude $f$. In contrast to the inverse effective temperature of the intrawell distribution near the minima of $g(Q,P)$, which is shown in Fig.~\ref{fig:harmonic_R'}, $R_A$ monotonically increases as $f$ increases. We present the results for $\bar n>\lambda^{1/3}$. In this case the intrawell distribution is formed by transitions $\Ket{n}\to \Ket{n'}$ with $|n-n'|\ll 1/\lambda$, and then $R_A$ is independent of $\lambda$. Already for the Planck number $\bar n\sim 1$, $R_A$ approaches the classical limit described by Eq.~(\ref{eq:small_R_prime}) and scales inversely proportional to $2\bar n +1$.

An important comment is due here. In classical systems, in a fluctuation-induced escape from a metastable state, with overwhelming probability the system goes over the saddle point on the boundary of the attraction basin to the metastable state, cf.  \cite{Freidlin1998}. In the case of the oscillator we are considering, as seen from the phase portrait in Appendix~\ref{sec:Langevin}, Fig.~\ref{fig:phase_portrait}, the classical switching occurs from a period-3 state to the zero-amplitude state, not to another period-3 state. 

A quantum oscillator, however, can directly switch between the period-3 states. This is because the condition of small tunnel splitting compared to the level broadening  can be violated near the saddle point, as the tunneling exponent $S_{\rm tun}$ falls off linearly with $g$ near the saddle-point value $g_s$, cf. Eq.~(\ref{eq:tunnel_time_general}) and Fig.~\ref{fig:compare_tau_tun}. If the tunnel splitting becomes comparable to the level broadening, the oscillator will tunnel between the wells once it has reached the values of $g$ close to  $g_s$. The overall rate of interwell switching will still be close to that given by Eq.~(\ref{eq:escape_rate}), as this equation gives the rate of reaching $g_s$, to the logarithmic accuracy.

\subsection{Escape near the bifurcation point where the period-3 state emerges}
\label{subsec:bifurcation}

In the previous sections we considered the weak damping limit, where the intrawell level spacing is much larger than the level widths. If the level broadening is disregarded, the onset of the period-3 vibrational states has no threshold in the amplitude of the driving field for $\delta\omega>0$. In the presence of dissipation, the driving field must have a finite amplitude for the period-3 states to emerge.  In terms of the theory of classical dynamical systems, the onset of period-3 vibrations is a bifurcation where the system acquires new stable states, in this case, period-3 states.

Generically,  one of the dynamical variables of the system becomes slow near a bifurcation point (the bifurcation parameter values). This variable is an analog of a soft mode. It controls the overall dynamics, as other dynamical variables follow it adiabatically \cite{Guckenheimer1997}. In the case of a driven quantum oscillator this significantly simplifies the problem of fluctuations \cite{Dykman2007,*Dykman2012}. Indeed, since the oscillator dynamics in the rotating frame is Markovian, and since it is controlled by a single dynamical variable, the operator nature of this variable becomes irrelevant. The variable commutes with itself, moreover, it commutes with itself at different instants of time. The fluctuations are still quantum, their intensity is nonzero even for $T\to 0$, but otherwise the dynamics is fully classical.

Classical dynamics of an oscillator resonantly driven close to triple its eigenfrequency is well understood \cite{Nayfeh2004,jordan2007}.  This dynamics is briefly summarized and the way to incorporate fluctuations consistent with the kinetic equation (\ref{eq:master_general}) is described in Appendix~\ref{sec:Langevin}. The important feature for the discussion here is that the period-3 states emerge as a result of three simultaneous saddle-node bifurcations occurring at three points $(Q_B^{(i)},P_B^{(i)})$ $(i=1,2,3)$ located at the vortices of an equilateral triangle on the $(Q,P)$-plane. At each point $(Q_B^{(i)},P_B^{(i)})$  there merge a stable and an unstable stationary state in the rotating frame, which correspond to the appropriate stable and unstable period-3 states in the laboratory frame. 

If the parameter $\kappa$ is close to its bifurcational value $\kappa_B \equiv \kappa_B(f)$, one can expand the classical equations of motion near a point $(Q_B^{(i)},P_B^{(i)})$ in $\delta Q^{(i)}=Q-Q_B^{(i)},\;  \delta P^{(i)}=P-P_B^{(i)}$ and rotate the variables so as to single out the slow variable $z=\delta Q^{(i)}\cos\phi_B^{(i)} + \delta P^{(i)}\sin\phi_B^{(i)}$. In the quantum Langevin equation for this variable one should keep the leading-order terms in $z, \kappa-\kappa_B$. This equation generically has the form \cite{Dykman2012}
\begin{align}
\label{eq:Langevin_soft_mode}
\frac{d}{d\tau}z = a_Bz^2 -b_B(\kappa-\kappa_B)+\xi_z(\tau), 
\end{align}
where $\langle \xi_z(\tau)\xi_z(\tau')\rangle = \lambda\kappa(2\bar n+1)\delta(\tau-\tau')$ is the quantum noise, which is $\delta$-correlated in the slow time $\tau$ (such structure of the noise corresponds to the Markovian dynamics in slow time). The explicit expressions for $Q_B^{(i)}, P_B^{(i)}, \phi_B^{(i)}$, and the parameters $a_B, b_B$ are given in Appendix~\ref{sec:Langevin}. Here we note that $a_Bb_B<0$. 

\begin{figure}
\includegraphics[width = 5.5cm]{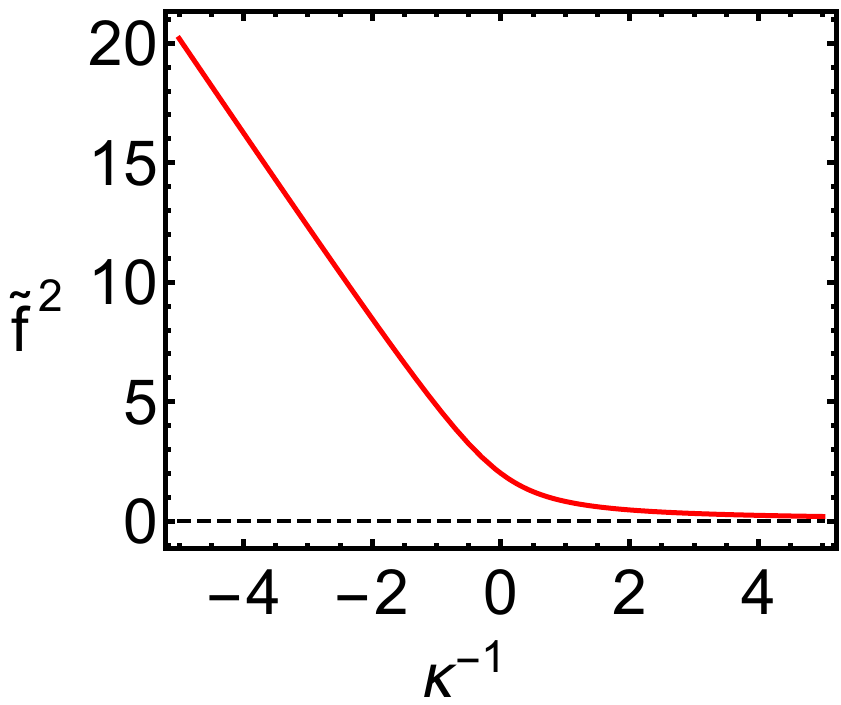}
\caption{The interrelation between the amplitude and frequency of the drive where the period tripling emerges. The abscissa shows the detuning of the driving field frequency $\delta\omega$ divided by the oscillator relaxation rate in the unscaled time, $\kappa^{-1}=\delta\omega/\Gamma$. The ordinate shows the scaled squared amplitude of the driving field $\tilde f^2 = F_0^2/24\omega_0^2\gamma\Gamma$; the scaling is independent of the frequency detuning.  Above the bifurcation curve the oscillator has three stable period-3 states and the stable zero-amplitude state; below this curve, only the zero-amplitude state is stable. The plot refers to $\gamma\,\delta\omega>0$. %For large positive detuning $\delta\omega >0$ (we assumed $\gamma>0$), the threshold in the drive power for the bifurcation approaches zero; for large negative detuning, the threhold approaches $\tilde f^2 > -4 \kappa^{-1}$ .}
}
\label{fig:bifurcation}
\end{figure}

For $\kappa <\kappa_B(f)$ the system (\ref{eq:Langevin_soft_mode}) has a stable and an unstable stationary state, if we disregard fluctuations. They are located at $\mp z_{\rm st}\,{\rm sgn} \,a_B$, where 
\[z_{\rm st} = [b_B(\kappa-\kappa_B)/a_B]^{1/2}.\] 
At $\kappa=\kappa_B(f)$ these states merge together, and for the scaled decay rate $\kappa>\kappa_B(f)$ the oscillator does not have period-3 states in the laboratory frame. Equivalently, period-3 states exist in the range of the scaled field amplitude where
\begin{align}
\label{eq:f_B_kappa}
f^2 >f_B^2(\kappa) = 2[(1+\kappa^2)^{1/2}-\sgn(\delta\omega)]
\end{align}
(see Apendix~\ref{sec:Langevin}). In Fig.~\ref{fig:bifurcation} we plot $\tilde f_B^2 = f_B^2/\kappa$; the parameter $\tilde f^2 = f^2/\kappa = F_0^2/24\omega_0\gamma\Gamma$ is advantageous as it does not depend on the frequency detuning of the drive from $3\omega_0$. 

Fluctuations caused by the noise $\xi_z(t)$ in Eq.~(\ref{eq:Langevin_soft_mode}) can lead to escape from the dynamically stable state. Even though our system is far away from thermal equilibrium, the escape rate $W_{\rm esc}$ near the bifurcation point can be found in the same way as for a Brownian particle in a potential well. The dynamics of such a particle in the case where inertial effects can be disregarded (the overdamped limit) is described by the Langevin equation that has the same form as Eq.~(\ref{eq:Langevin_soft_mode}), except that the noise is thermal. Using the Kramers solution of the thermal-equilibrium problem \cite{Kramers1940} and taking into account the explicit form of the intensity of the quantum noise $\xi_z(t)$, we obtain
\begin{align}
\label{eq:escape_bif}
\ln W_{\rm esc}\approx -\frac{2}{3}\frac{|b_B (\kappa-\kappa_B)|^{3/2}}
{|a_B|^{1/2}\kappa_B\lambda(2\bar n+1)}.
\end{align} 
The exponent of the escape rate scales with the distance to the bifurcation point as $|\kappa-\kappa_B|^{3/2}$ and scales with temperature as $(2\bar n+1)^{-1}$. In terms of the parameters of the driving field, near the bifurcation point we have $\kappa-\kappa_B \approx \Gamma[(\delta\omega)_B - \delta\omega]/(\delta\omega_B)^2\approx (f-f_B)/(df_B/d\kappa)$, where $(\delta\omega)_B$ is the bifurcational value of the frequency detuning and $f_B(\kappa)$ is given by Eq.~(\ref{eq:f_B_kappa}). The scaling (\ref{eq:escape_bif}) with the parameters of the driving field and with temperature holds also for a resonantly driven oscillator near a bifurcation point \cite{Dykman2007,*Dykman2012} and has been seen for such an oscillator in the experiment with Josephson junction oscillators \cite{Vijay2009,Vijay2012}.

We note that, even though the zero-amplitude state of the oscillator that displays period tripling is stable, for sufficiently strong driving quantum fluctuations lead to escape from this state into period-3 states. For a strong field, $f^2\gtrsim (1+\kappa^2)/\lambda(2\bar n+1)$, one can show that quantum fluctuations essentially wash away the zero-amplitude state. The full analysis of escape from the zero-amplitude state is beyond the scope of this paper.

%%%%%%%%%%%%%%%%%%%%%%%%%%%%%

\section{Summary and outlook}

This paper describes the dissipative quantum dynamics of a nonlinear oscillator resonantly driven into period tripling. The analysis is based on a minimalistic model.  The scaled RWA Hamiltonian in the rotating frame has one parameter,  the scaled amplitude of the drive $f$, the dissipation is described by one parameter, the scaled linear friction coefficient $\kappa$, and the only other parameter is the scaled Planck constant $\lambda\propto \hbar$. The model has allowed revealing generic features of the dissipative Floquet dynamics of a multi-state system. Of particular importance is that it revealed the features that remained hidden in the previously explored models.

%In contrast to the standard Floquet analysis, in the period-tripling problem the analysis of the quantum dynamics is  most conveniently done in terms of the ``Brillouin zone" of the quasienergies with the width $\hbar\omega_F/3$ rather than $\hbar\omega_F$ ($\omega_F$ is the driving frequency). This is reminiscent of the folding of the Brillouin zone in solids at the Peierls transition \cite{Ziman1979}. The eigenvalues of the scaled RWA Hamiltonian are naturally ordered in such a nomenclature. 

For period tripling, the RWA Hamiltonian as a function of the phase-space variables $g(Q,P)$ has three symmetrically located wells. In the most interesting regime, the wells contain many bound states. The number of the states is $\propto 1/\lambda \gg 1$. Because of the three-fold symmetry, the intrawell states would be degenerate if there were no interwell tunneling. The tunneling splits the eigenvalues of $g$ into triplets. 

Coupling to a thermal reservoir leads to transitions between the states. Where the transition rates  ($\propto \kappa$) exceed the exponentially small tunnel splitting (divided by $\hbar$), the distribution over the states is formed in two stages. First, over the dimensionless time $\sim 1/\kappa$ there is formed a distribution over the intrawell states. Then over an exponentially longer time, there is formed a distribution over the wells.

Arguably, the most unexpected feature of the dynamics is that the intrawell distribution is formed by transitions between remote states within the well, i.e., by a nonlocal walk over the intrawell states. This is despite the corresponding transition rates being exponentially small. The effect is a consequence of a very steep decay of the distribution with the increasing RWA energy.  

A broader implication of this effect is related to the fact that, even though the distribution is very steep, its logarithm is a smooth function of the number $n$ of an intrawell state [or, equivalently, of the eigenvalue $g_n$ of the RWA Hamiltonian]. A standard approach to the analysis of the distribution in this case is based on the eikonal approximation. Here, the logarithm of the distribution is expanded in the state number and the balance equation for the state populations is reduced to the equation for the derivative of the logarithm of the distribution over $g_n$, cf. \cite{Dykman1988a}. Such approach is a complete analog of the WKB approximation where the Schr\"odinger equation is reduced to the equation for the derivative of the logarithm  of the wave function. It has been used to analyze numerous problems of chemical kinetics and population dynamics \cite{Kamenev2011,Assaf2017}, in which case $n$ stands for the number of species and often has several components that correspond to different species.

Usually, the WKB approximation in quantum mechanics breaks down locally near singularities of the wave front, cf.~\cite{landau1977,Berry1989}. A similar behavior was found for the distribution of the resonantly driven oscillator \cite{Guo2013,Peano2014}. In contrast, in the present case, the logarithm of the probability distribution remains smooth. However, the local approximation, that underlies the eikonal method, breaks down. Such behavior may be expected also in other types of physical systems away from thermal equilibrium as well as in the problems of chemical kinetics and population dynamics.

The other generic feature of the dynamics of period tripling is the lack of detailed balance. This contrasts the previous results \cite{Drummond1980,Kryuchkyan1996,Marthaler2006} for the oscillator driven close to its eigenfrequency or twice the eigenfrequency, where the detailed balance holds at $T=0$. At the technical level, the semiclassical dultiple singularities of the classical Hamiltonian trajectories in the complex time and the branching at the singularities. This significantly complicates the analysis, but also broadens the applicability of the results. 

In common with the previously considered driven oscillators, the switching between the period-3 states occurs via transitions over the barrier of $g(Q,P)$ rather than via interwell tunneling, if the decay rate exceeds the tunnel splitting divided by $\hbar$. We have also found the scaling behavior of the switching rate with the distance to the bifurcation point where the period-3 states disappear. 

We have not provided the results for the switching from the quiet state. 
The rate of switching from this state is a peculiar and important problem on its own, starting with the very possibility to switch to any of the period-3 states. The switching rate can be made comparatively large in this case, while still remaining much smaller than the decay rate. A larger switching rate should grossly simplify the observation of the switching in the experiment, thus providing the means to study quantum-fluctuations induced switching where the detailed balance condition is strongly violated. Such an observation should be contrasted with the important  experiment on the switching of a resonantly driven quantum oscillator in the regime where the detailed balance was effectively present \cite{Vijay2009}. A detailed analysis of the switching from the quiet state is left for the future work.

In conclusion, our results show that the dissipative dynamics of resonant period tripling provides new insights into the effects of dissipation and quantum fluctuations in Floquet systems. Experimentally, these effects can be studied with various nonlinear microwave and optical cavities and also with nanomechanical systems, where quantum regime has been already recently reached \cite{Satzinger2018,Chu2018}. 

\acknowledgments

We are grateful to J. Ankerhold, S. M. Girvin and J. Gosner, who participated in the work at the early stage.  We also acknowledge valuable discussions with C.~Bruder, N.~L\"orch and V.~Peano. Y.Z. was supported by the National Science Foundation (DMR-1609326). M.I.D. acknowledges partial support from  the National Science Foundation (Grants No. DMR-1806473 and and CMMI-1661618).

%%%%%%%%%%%%%%%%%%%%%%%%%%%%%%%%%%%%%
%%%%%%%%%%%%%%%%%%%%%%%%%%%%%%%%%%%%%

\appendix

\section{Fixed points of the Hamiltonian $g$}
\label{sec:fixed_point_analysis}
The Hamiltonian function $g(Q,P)$, Eq.~(\ref{eq:Hamiltonian_RWA}), which describes the motion in the rotating frame  in the rotating-wave approximation, has three minima located at 
\begin{align}
\label{eq:minima_of_g}
(Q,P) = (Q_0,0), (-Q_0/2,\pm\sqrt{3}Q_0/2).
\end{align}
Here $Q_0$ and the minimal value of $g$ are given by the expressions
\begin{align}
\label{eq:min_g}
&Q_0 = (f+\sqrt{f^2+4\sgn(\delta\omega)})/2, \nonumber\\
&g = g_{\rm min} = -fQ_0[Q_0^2+3\sgn(\delta\omega)]/12.
\end{align}
For the detuning of the drive frequency $\delta\omega\equiv \frac{1}{3}\omega_F-\omega_0 >0$, function $g(Q,P)$ has a local maximum at $Q=P=0$, whereas for $\delta\omega<0$ it has a local minimum at $Q=P=0$. For $\delta\omega<0$, as seen from Eq.~(\ref{eq:min_g}), there is a threshold in the drive amplitude for the local minima to emerge. 

Function $g(Q,P)$ also has  three saddle points  located at 
\[
(Q,P) = (Q_s,0),(-Q_s/2,\pm\sqrt{3}Q_s/2).
\]
Here $Q_s$ and the saddle-point value of $g$ are given by the expressions
\begin{align}
\label{eq:saddle_point_g}
& Q_s = (f-\sqrt{f^2+4\sgn(\delta\omega)})/2,\nonumber\\
& g_s=-fQ_s[Q_s^2+3\sgn(\delta\omega)]/12.
\end{align}

For $\delta\omega>0$, the saddle points lie on the boundaries between the wells of $g(Q,P)$. For $\delta\omega<0$ the saddle points lie at the boundaries between the wells of $g(Q,P)$ located at nonzero $Q^2+P^2$, see Eq.~(\ref{eq:minima_of_g}), and the well at $Q=P=0$.

In the presence of weak dissipation, the three minima of $g(Q,P)$ given by Eq.~(\ref{eq:minima_of_g}) become stable period-three vibrational states, whereas the local maximum or minimum at the origin becomes a stable zero-amplitude state, see Appendix~\ref{sec:Langevin}.

The scaling used in Sec.~\ref{sec:model} does not allow one to consider the driving on exact resonance. An alternative scaling of the dynamical variables in the rotating frame is 
\begin{align}
\label{eq:Yaxing_scaling_QP}
&Q^{\prime}{} = (\lambda^{\prime}{}/2)^{1/2}(a+a^\dagger),\quad P^{\prime}{}=-i(\lambda^{\prime}{}/2)^{1/2}(a-a^\dagger),\nonumber\\
&\lambda^{\prime}{}= 9\hbar\gamma^2/4\omega_0F^2 
\end{align}
In these variables, the  Hamiltonian in the rotating frame in the RWA becomes $U^\dagger H U - i\hbar U^\dagger \dot U =(2F^4/27\gamma^3)g^{\prime}{}(Q^{\prime}{},P^{\prime}{})$ with
\begin{align}
\label{eq:Yaxing_Hamiltonian}
&g^{\prime}{}(Q^{\prime}{},P^{\prime}{}) = \frac{1}{4}(Q^{\prime}{}^2 + P^{\prime}{}^2)^2 - \frac{1}{8}\zeta^{\prime}{}(Q^{\prime}{}^2+P^{\prime}{}^2)\nonumber\\
&\qquad -(Q^{\prime}{}^3-3P^{\prime}{}Q^{\prime}{}P^{\prime}{})/6+ {\rm const}, \nonumber\\
&\qquad \zeta^{\prime}{} = 24\omega_0 \gamma\delta\omega/F^2.
\end{align}
Clearly, the dynamics is still described by the single parameter, in this case $\zeta^{\prime}{}=\sgn (\delta\omega)/f^2$, and by the dimensionless Planck constant $\lambda^{\prime}{}$, which is independent of the frequency detuning $\delta\omega$. The tristability requires $\zeta^{\prime}{}>-1/4$.

\section{Classical dissipative dynamics}
\label{sec:Langevin}

%%%%%%%%%%%%%%%%%%%%%%%%%%%

The Markov quantum kinetic equation in ``slow'' time (\ref{eq:master_general}) corresponds to the quantum Langevin equations of motion of variables $Q,P$ of the form 
\begin{align}
\label{eq:eom}
\dot Q =  -i\lambda^{-1}[Q,g] -\kappa Q +\xi_Q(\tau),\nonumber\\
\dot P= -i\lambda^{-1}[P,g] -\kappa P +\xi_P(\tau), %\nonumber\\
%&\Omega = \delta\omega/\Gamma.
\end{align}
where $\xi_Q(\tau)$ and $\xi_P(\tau)$ are Gaussian quantum noises, which are $\delta$-correlated in slow time, $\langle \xi_Q(\tau)\xi_Q(\tau')\rangle =  \langle \xi_P(\tau)\xi_P(\tau')\rangle =\lambda\kappa(2\bar n +1)\delta(\tau-\tau')$, and satisfy the commutation relation $\langle [\xi_Q(\tau),\xi_P(\tau')]\rangle =2i\lambda\kappa \delta(\tau-\tau')$, cf. \cite{Dykman2007,*Dykman2012}.

In the classical limit  the non-commutativity of the noise components can be disregarded and the noise correlators become independent of $\hbar$, $\langle \xi_Q(\tau)\xi_Q(\tau')\rangle =  \langle \xi_P(\tau)\xi_P(\tau')\rangle =(3\gamma\kappa k_BT/4\omega_0^3|\delta\omega|)\delta(\tau-\tau')$. The classical equations of motion read
\begin{align}
\label{eq:Langevin_classical}
\dot Q = \partial_P g -\kappa Q +\xi_Q(\tau),\nonumber\\
\dot P= -\partial_Q g -\kappa P +\xi_P(\tau), %\nonumber\\
%&\Omega = \delta\omega/\Gamma.
\end{align}

The dynamics described by Eq.~(\ref{eq:Langevin_classical}) in the absence of noise is well understood \cite{Nayfeh2004, jordan2007}. For a sufficiently strong driving (or a small decay rate) these equations can have  7 stationary states. One state is trivial,  $Q_{\rm st}=P_{\rm st} = 0$. This is a stable state, a focus on the phase plane $(Q,P)$, with the eigenvalues of the linearized motion equal to $-\kappa\pm i$. It corresponds to the local maximum (for $\delta\omega>0$) or minimum (for $\delta\omega<0$) of $g(Q,P)$.

Six other roots have the form $Q^{(j)}_{\rm st}=r^{(j)}_{\rm st}\cos\phi_{\rm st}^{(j)}$, $P^{j}_{\rm st}=r^{(j)}_{\rm st}\sin\phi_{\rm st}^{(j)}$ ($j=0,...,5$). The radii $r^{(j)}_{\rm st}$ take on one of the two values,
\begin{align}
\label{eq:stationary}
r_{\pm} =& \left\{\sgn(\delta\omega)+ \frac{1}{2}f^2 \right.\nonumber\\
&\left.\pm \left[f^2\sgn(\delta\omega) +(f^4/4)-\kappa^{2}\right]^{1/2}\right\}^{1/2},
\end{align}
whereas  the phases are given by the equation
\begin{align}
\label{eq:steady_positions}  
\exp[3i\phi_{\rm st}^{(j)}] = fr^{(j)}_{\rm st}/[(r_{\rm st}^{(j)})^2 -\sgn(\delta\omega) - i\kappa].
\end{align}
We set $j=0,1,2$ for the stationary states  $r^{(j)}_{\rm st}=r_+$ and $j=3,4,5$ for the states $r^{(j)}_{\rm st}=r_-$. 
A straightforward calculation shows that the eigenvalues of the linearized motion for $j=0,1,2$ have negative real parts, which shows that these states are stable. For the states $j=3,4,5$ the eigenvalues are real and have opposite signs, which means that these states are saddle points on the $(Q,P)$-plane.   

\begin{figure}[h]
\includegraphics[width = 5.5 cm]{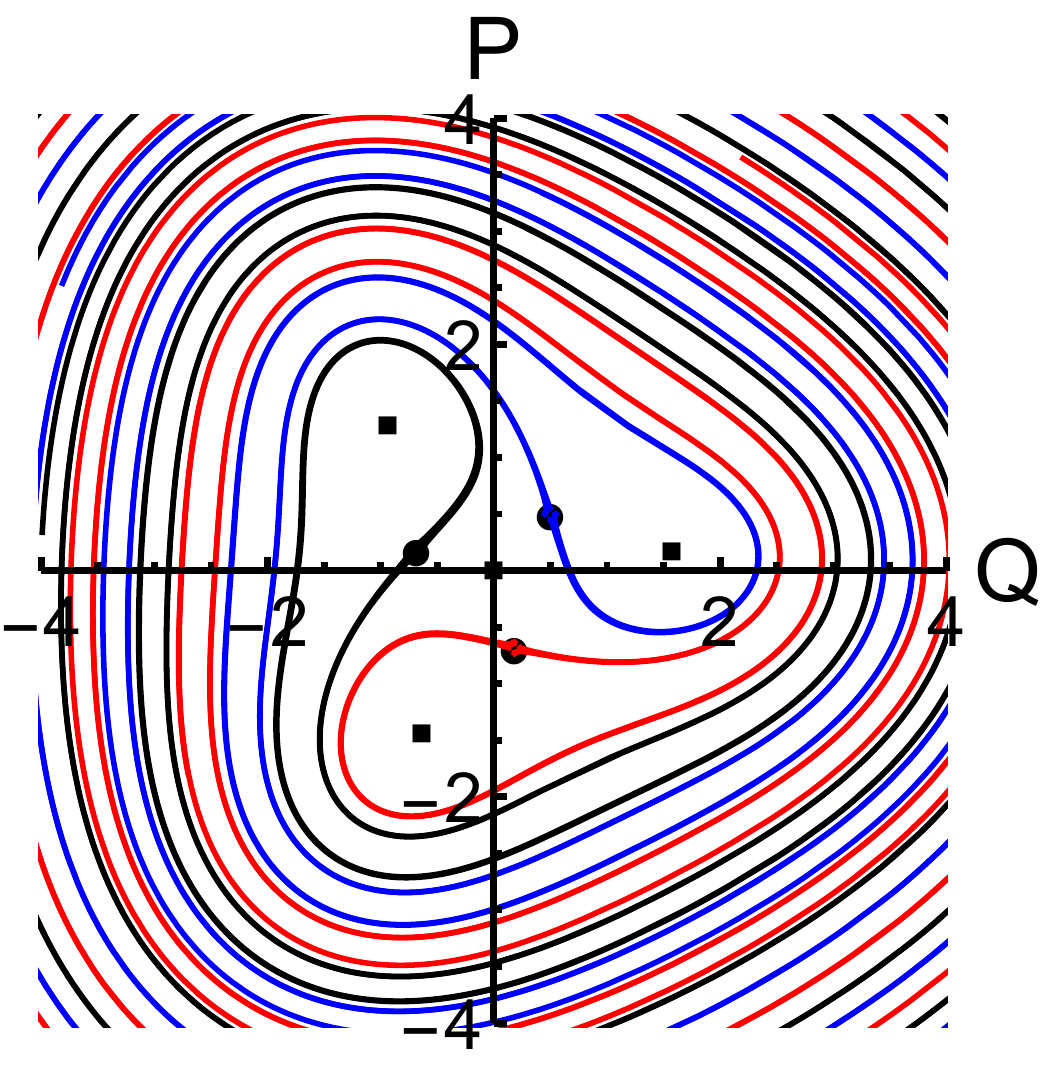} 
\caption{The phase portrait of a classical oscillator driven close to three times its eigenfrequency; $Q$ and $P$ are the scaled coordinate and momentum in the rotating frame.  The scaled amplitude of the driving field is $f=1$, the scaled decay rate is $\kappa =0.5$, and $\delta\omega>0$. The squares show the positions of the stable period-three states, whereas the circles show the positions of the saddle points. The solid lines show the separatrices that separate the basins of attraction to different stable period-three states and to the zero-amplitude state at the origin. As $\kappa\to 0$, the squares and the saddles rotate and shift and approach their position in Fig.~\ref{fig:quasienergy_surface}~(c).  }
\label{fig:phase_portrait}
\end{figure}

From Eq.~(\ref{eq:steady_positions}), the triples of the stable states and of the saddle points are located at the vertices of equilateral triangles on the $(Q,P)$ plane . The stable states are further away from the origin $Q=P=0$ , as $r_+>r_-$. The phase portrait is illustrated in Fig.~\ref{fig:phase_portrait}. For comparatively weak damping, the separatrices form spirals that loop around stable states. In the absence of noise, the oscillator will move inside the spiral of the separatrices of the same color toward the stable state inside the corresponding loop. If initially the oscillator is between the loops of different colors, it will move toward the zero-amplitude state. 

As seen from Eqs.~(\ref{eq:stationary}) and (\ref{eq:steady_positions}),  the stable and unstable states approach each other pairwise with the increasing $\kappa$ or decreasing $f$.  In Fig.~\ref{fig:phase_portrait}, the loops of the separatrices ``squeeze'' so that the squares inside the loops merge with circles on the loops. The merging of a stable state and a saddle is a saddle-node bifurcation. The bifurcational value of the scaled damping rate is given by the condition $r_+=r_-$, 
\begin{align}
\label{eq:saddle_node_position} 
\kappa_B\equiv \kappa_B(f)  =[ f^2\sgn(\delta\omega) + (f^4/4)]^{1/2}.
\end{align}
Equivalently, one can express the scaled bifurcational field amplitude $f_B$ in terms of $\kappa$, see Eq.~(\ref{eq:f_B_kappa}). It is important that the amplitude of the vibrations $r_B=(1+\frac{1}{2}f^2)^{1/2}$ is nonzero at the bifurcation point, i.e., even though there is a threshold in terms of the driving field strength, when the period-3 vibrations emerge in the laboratory frame, they have a nonzero amplitude. 

The analysis of the dynamics near the bifurcation point is the same for the three period-3 states. We will choose one of them and introduce  a scaled complex vibration amplitude at the bifurcation point $x_B= (Q_B + iP_B)$ with $Q_B,P_B$ given by one of the solutions of Eq.~(\ref{eq:steady_positions}) for $\kappa = \kappa_B$; by construction, $Q_B^2+P_B^2 = \sgn(\delta\omega) + f^2/2$. If we now set in the equations of motion (\ref{eq:Langevin_classical}) $Q=Q_B+\delta Q, \,P=P_B+\delta P$, expand 
the right hand sides in $\delta Q, \delta P$, and write the equation of motion for the combination
\begin{align}
\label{eq:slow_fast} 
z = \delta Q\cos\phi_B + \delta P\sin \phi_B,
\end{align}
we find that, if we choose the rotation angle as 
\begin{align}
\label{eq:phi_B}
\exp(2i\phi_B) = -i\frac{x_B^2 - 2fx_B^*}{\kappa_B - i(2|x_B|^2 -1)},
\end{align}
the linear in $\delta Q, \,\delta P$ terms drop out from this equation.

The combination of the variables (\ref{eq:slow_fast}) is the slow variable of the system. The linearly independent combination $y=-\delta Q\sin\phi_B + \delta P\cos\phi_B$ decays with the rate $2\kappa_B$ for $\kappa=\kappa_B$. 
Over the dimensionless time $\sim (2\kappa_B)^{-1}$, this variable approaches its quasi-stationary (adiabatic) value $y_{\rm ad}\equiv y_{\rm ad}(z)$ for a given $z$. One can show that, to the leading order, 
\begin{align*}
y_{\rm ad}= k_{\rm ad} z, \qquad k_{\rm ad}= -(1+f^2)/\kappa_B.
\end{align*}
The resulting equation for the slow variable $z$, to the leading order, has the form (\ref{eq:Langevin_soft_mode}) with the coefficients
\begin{align}
\label{eq:bifurcation_parameters}
&a_B = {\rm Im}\,\left[(1+ik_{\rm ad})^2\left(X_B^* + fe^{3i\phi_B}\right)\right] \nonumber\\
&\quad + 2(1+k_{\rm ad}^2){\rm Im}\,X_B, \qquad
%&(3y_B - f\sin 3\phi_c) k_{\rm ad}^2 + (2z_B +2f\cos3\phi_c)k_{\rm ad}  \nonumber \\
%&+ y_B + f\sin 3\phi_c, \nonumber \\
 b_B={\rm Re} X_B, 
\end{align}
where $ X_B=x_B \exp(-i\phi_B)$. It is important that the parameters $a_B$ and $b_B$ are the same for all three period-three states. Indeed, changing from one of the three stable states to another corresponds to the rotation $x_B\to x_B\exp(\pm 2\pi i/3)$. As seen from Eq.~(\ref{eq:phi_B}), such rotation leads to the angle $\phi_B$ being incremented by $\pm 2\pi/3$, as expected from the symmetry arguments. Therefore $X_B$ is the same for all three states as is also $\exp(3i\phi_B)$.

\section{Asymptotic behavior of the semiclassical matrix elements}
\label{sec:contour_integral}

Finding the asymptotic behavior of the matrix elements $a_m(g_n)$ in Eq.~(\ref{eq:matrix_elements}) for large $|m|$ requires understanding the analytical properties of the function $Q(\tau;g)+iP(\tau;g)\equiv (2\lambda)^{1/2}a(\tau;g)$. This function is determined by the classical Hamiltonian dynamics with the Hamiltonian $g(Q,P)$.  In the explicit form, the Hamiltonian equations (\ref{eq:classical_eom}) for $Q(\tau;g)$ and $P(\tau;g)$ read
\begin{align}
\label{eq:Hamilton_eqns}
&\dot Q   = P(Q^2 + P^2 -1) +2fQP,\nonumber\\
&\dot P = -Q(Q^2 + P^2 -1) +f(Q^2 - P^2).
\end{align}
Here and in what follows we consider the case $\delta\omega>0$.

For real time, Im~$\tau = 0$, the functions $Q(\tau;g)$ and $P(\tau;g)$ describe smooth Hamiltonian trajectories, which in the range of $g$ we are interested in, $g_{\min}<g<g_s$, are shown in Fig.~\ref{fig:quasienergy_surface}~(c).  However, $Q(\tau;g)$ and $P(\tau;g)$ generally become complex for complex time and have singularities where $|Q|$ and $|P|$ go to infinity. Because $g(Q,P)$ is quartic in $Q,P$, these singularities can be reached  in a complex time $\tau_{\rm sing}$ with finite real and imaginary parts. From Eq.~(\ref{eq:matrix_elements}), for large $|m|$ the Fourier components $a_{m> 0}$ and $a_{m<0}$ are given by  the values of $\tau_{\rm sing}$ with the smallest in the absolute value negative and positive imaginary part, respectively.  

\begin{figure}
\includegraphics[width = 4 cm]{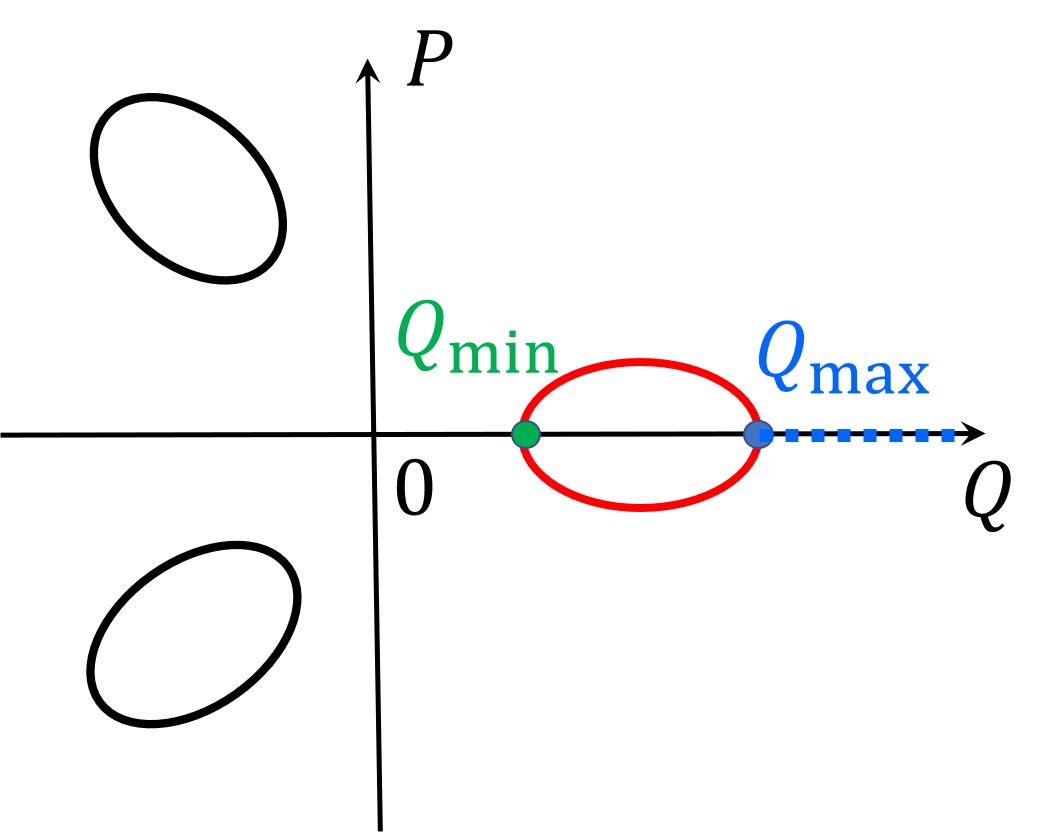} \hfill
\includegraphics[width = 4 cm]{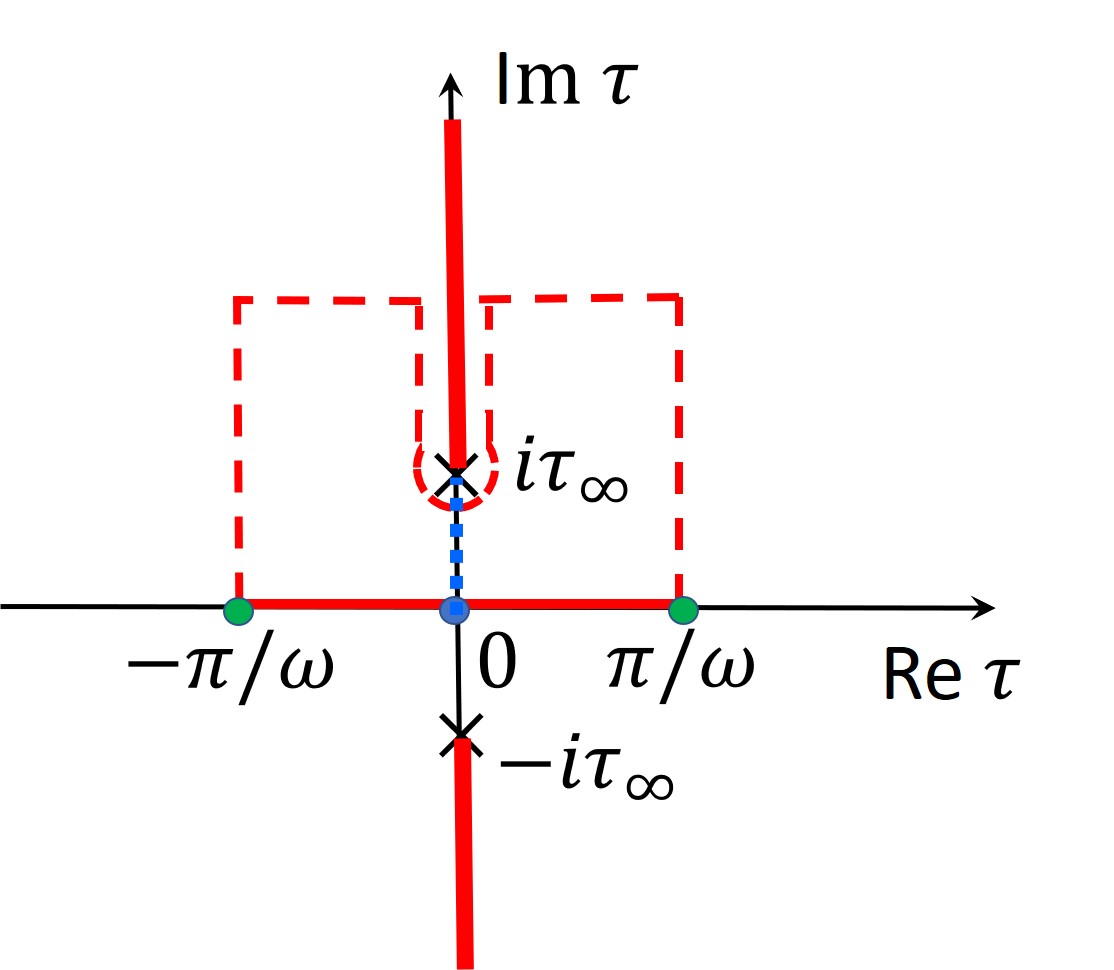}
\caption{Left panel: a sketch of the classical intrawell orbits with the same RWA energy $g$ such that $g_{\min}<g<g_s$; the general pattern of such orbits is shown in Fig.~\ref{fig:quasienergy_surface}~(c). Right panel: the contour on the $\tau$-plane for calculating the Fourier components $a_m(g)$. The thin red line on the real time axis from $\tau=-\pi/\omega$ to $\pi/\omega$ corresponds to the real trajectory shown by the red line in the left panel. The dashed red line is the deformed contour which goes to large $\Im~\tau$ where $\exp[-m\omega(g){\rm Im}~\tau]\ll \exp[-m\omega(g)\tau_\infty(g)])$ for large $m$. The contour goes around the singularity at $\tau = i\tau_\infty$. The thick red line along the imaginary-$\tau$ axis is the branch cut. The dotted blue line from $\tau = 0$ to $\tau = i\tau_\infty$ corresponds to the trajectory along the real-$Q$ axis from $Q=Q_{\rm max}$ to $Q \to\infty$ in the left panel.}
\label{fig:schematic_contour_integral}
\end{figure}

By symmetry, the values of Im~$\tau_{\rm sing}$ are the same for all three wells of $g(Q,P)$. It is most convenient to find them by considering the classical trajectory for the $\nu=0$-well of $g(Q,P)$, which is centered on the axis $P=0$ in Figs.~\ref{fig:quasienergy_surface}~(b) and (c). This trajectory  in real time is shown by the red line in Fig.~\ref{fig:schematic_contour_integral}.  For a given $g$, the momentum $P$ as a function of $Q$ on this trajectory has the form 
\begin{align}
\label{eq:momentum_general}
&P(Q{})= \pm [A(Q) + \sqrt{B(Q{})}]^{1/2}, \nonumber \\
&A(Q)= 1- Q^2 -2fQ, \nonumber\\
&B(Q{}) = 4fQ\left[\frac{4}{3}Q^2+fQ-\sgn(\delta\omega)\right]+4g,
\end{align}
as seen from the explicit expression (\ref{eq:Hamiltonian_RWA}) for $g(Q,P)$. 

The time on the classical trajectory can be continued to the complex plane in such a way that $Q(\tau;g)$ will remain real and will go to infinity. To this end, we start from the point on the real trajectory where $Q$ is maximal, $Q=Q_{\max}$ (the utmost right point on the red trajectory in Fig.~\ref{fig:schematic_contour_integral}). We  choose $\tau=0$ for $Q=Q_{\max}$ and then move along the imaginary axis on the  $\tau$-plane. As seen from Eq.~(\ref{eq:Hamilton_eqns}), in this case $Q(\tau;g)$ remains real, whereas $P(\tau;g)$ becomes purely imaginary. 

The real trajectory (\ref{eq:Hamilton_eqns}) spins clockwise, and therefore $\dot P(\tau;g)<0$ for $\tau = 0$ (i.e., for $Q=Q_{\max}$). Therefore the sign of Im~$P$ is opposite to the sign of Im~$\tau$, when we start incrementing Im~$\tau$. Since $P(\tau=0;g)=0$, function $P(\tau;g)$ is purely imaginary for small $|{\rm Im}~\tau|$,
\[{\rm sgn~Im}~P = -{\rm sgn~Im}~\tau.\]
From Eqs.~(\ref{eq:Hamilton_eqns}), $Q(\tau;g)$ remains real and $P(\tau;g)$ remains purely imaginary for purely imaginary $\tau$. Moreover, both $Q$ and $P$ increase in the absolute value as $|\tau|$ increases independent of the sign of Im~$\tau$. 

In the upper half-plane of the $\tau$-plane, the value of $\tau$ at which $Q\rightarrow \infty$ is given by $i\tau_{\infty}$,
\begin{align}
\label{eq:imaginary_time}
\tau_\infty =  \int_{Q_{\rm max}}^\infty dQ /| \dot Q|, %\nonumber \\
\quad \dot Q =  P(Q{})\sqrt{B(Q{})},
\end{align}
where we have used Eq.~(\ref{eq:momentum_general}) and chose Im~$P <0$, in agreement with Im~$\tau >0$; note that $B(Q{})>0$  and Re~$\dot Q =0$ for $Q>Q_{\max}$. %Since $\dot Q \propto Q^{5/2}$ for $Q\to \infty$, as seen from Eq.~(\ref{eq:momentum_general}), it is clear that $\tau=i\tau_\infty$ is the branching point of $Q(\tau;g)$.

When $Q$ goes to infinity, $|P|$ also goes to infinity and so does $|Q+iP|$ as well. Indeed, as seen from Eq.~(\ref{eq:momentum_general}), 
\begin{align}
\label{eq:P_asymptotic}
&P\approx \mp i(Q-2\sqrt{fQ/3}), \quad Q\rightarrow \infty.
\end{align}
From this expression, $Q+iP \propto Q$ or $\sqrt{Q}$ for $Q\to \infty$ depending on the sign of Im~$P$ or, equivalently, on the sign of -Im~$\tau$.

Once we know the locations of the singularities of the function $Q(\tau;g)+iP(\tau;g)$, we are ready to perform the integral in Eq.~(\ref{eq:matrix_elements}) in the limit of large $m$. For the case of $m<0 (m>0)$, we move the integration contour to the upper (lower) half-plane of the complex-$\tau$ plane; the case where the integral is shifted to the upper half-plane is shown in Fig.~\ref{fig:schematic_contour_integral}. With our choice of the time on the orbit,  the singularities of $Q(\tau;g)+iP(\tau;g)$ are on the imaginary-$\tau$ axis at $\tau = \pm i\tau_{\infty}$.

%We have chosen $\tau_\infty$ to be positive and the $\pm$ sign in front of $\tau_\infty$ corresponds to $\Im P <0$ and $>0$, respectively, as shown in Eq.~(\ref{eq:P_asymptotic}). 

To evaluate the contour integral, we need to know the function $Q+iP$ near the singularities. From Eqs.~(\ref{eq:imaginary_time}) and (\ref{eq:P_asymptotic}), for large $Q$ we have $\dot Q = P(Q{}) \sqrt{B(Q{})}\approx \mp i Q\sqrt{16fQ^3/3}$, where the upper and lower signs correspond to the upper and lower half-planes of the $\tau$-plane. Then for $|\tau-i\tau_\infty|\gg \tau_\infty$ 
\begin{align}
\label{eq:branching_up}
Q+iP &\approx 2Q\approx e^{i\pi/3}(3f/2)^{-1/3}(\tau-i\tau_{\infty})^{-2/3}, 
\end{align}
whereas for $|\tau+i\tau_\infty|\ll \tau_\infty$
\begin{align}
\label{eq:branching_down}
Q+iP &\approx e^{i\pi/6}(4f/9)^{1/3}(\tau+i\tau_{\infty})^{-1/3}.
\end{align}
The above equations show that the singularities at $\pm i\tau_{\infty}$ are branching points. We choose the branching cuts so that the contour integral does not cross them.
For the chosen contour, the integral of $\exp[-im\omega(g)\tau]\,[Q(\tau;g) + iP(\tau;g)]d\tau$ in Eq.~(\ref{eq:matrix_elements}) is reduced to the integral around the branch cut. Other contributions to the contour integral can be disregarded. Indeed, since $Q(\tau;g)$ and $P(\tau;g)$ are periodic with period $2\pi/\omega(g)$, the integrals along the vertical lines that go from $\pm \pi/\omega(g)$ in Fig.~\ref{fig:schematic_contour_integral} cancel each other. The integral along the horizontal sections of the contour with large $|{\rm Im}~\tau|$  is exponentially small for large $|m|$. 

The calculation of the integral around the branch cut is  straightforward with the account taken of Eqs.~(\ref{eq:branching_up}) and (\ref{eq:branching_down}).  It gives Eqs.~(\ref{eq:a_m<0})  and (\ref{eq:a_m>0}). The different dependence of the prefactors in these equations on $m$ for $m>0$ and $m<0$ comes from the different dependence of $a(\tau;g) \propto Q(\tau;g) + iP(\tau;g)$ on $\tau$ near $\pm i \tau_\infty$ in Eqs.~(\ref{eq:branching_up}) and (\ref{eq:branching_down}). 

Implied in the above calculation was an assumption that the singularities at $\pm i\tau_\infty$ are the closest singularities to the real-$\tau$ axis. Generally, there are other singularities of $Q(\tau;g)$ and $ P(\tau;g)$ for complex $\tau$. If we start from the leftmost point $Q_{\rm min}$  on a real trajectory in the $\nu=0$-well of $g(Q,P)$ (the red trajectory in Fig.~\ref{fig:schematic_contour_integral}) and move along the imaginary time axis, $Q$ diverges for Im~$\tau = \pm (\tau_{\rm tun}-\tau_{\infty})$. Here, $\tau_{\rm tun}\equiv \tau_{\rm tun}(g)$ is the imaginary part of the tunneling time to go from the classical real-time trajectory of the $\nu=0$ well of $g(Q,P)$ with a given $g$ to the classical real-time trajectory with the same $g$ in the $\nu=1$ or $\nu=2$ well; see Sec.~\ref{sec:tunneling} for details. However, an explicit calculation shows that  $\tau_\infty < \tau_{\rm tun} - \tau_\infty$, and therefore the singularities at  Im~$\tau = \pm (\tau_{\rm tun}-\tau_{\infty})$ do not affect the large-$|m|$ asymptotic of $a_m(g)$.  Starting from various points on the real-time $\nu=0$-trajectory and moving along the imaginary time axis, we checked numerically that the singularities at $\pm i \tau_{\infty}$ are indeed the closest to the real time axis.

%\begin{widetext}
\begin{figure}[ht]
\includegraphics[width=4.2cm]{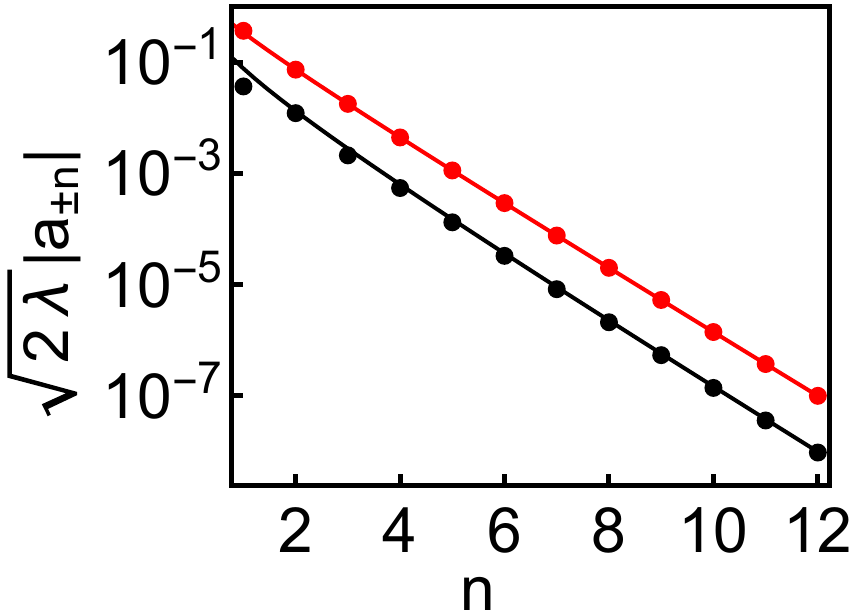}\hfill
\includegraphics[width=4.1cm]{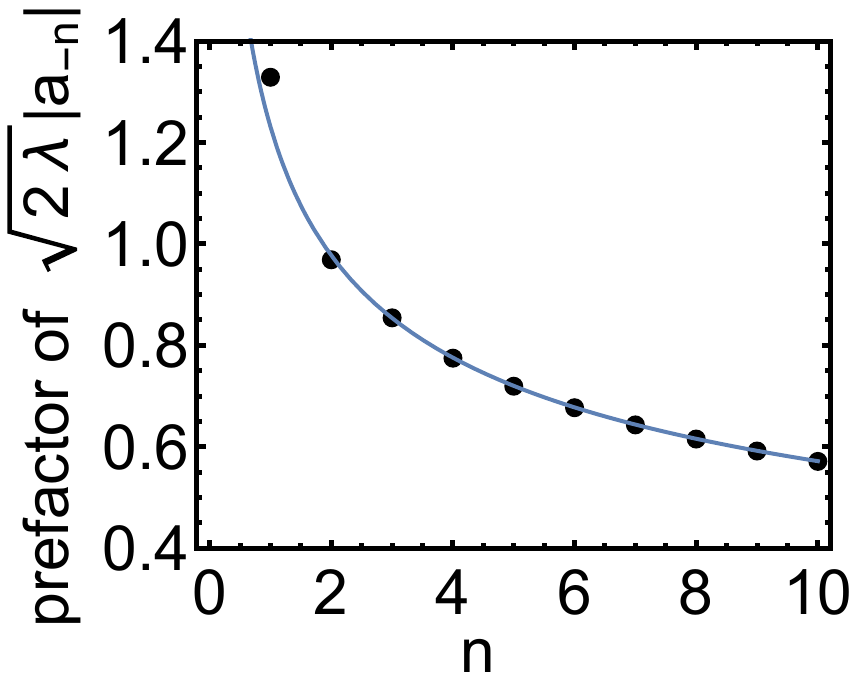}\hfill
\includegraphics[width=4.2cm]{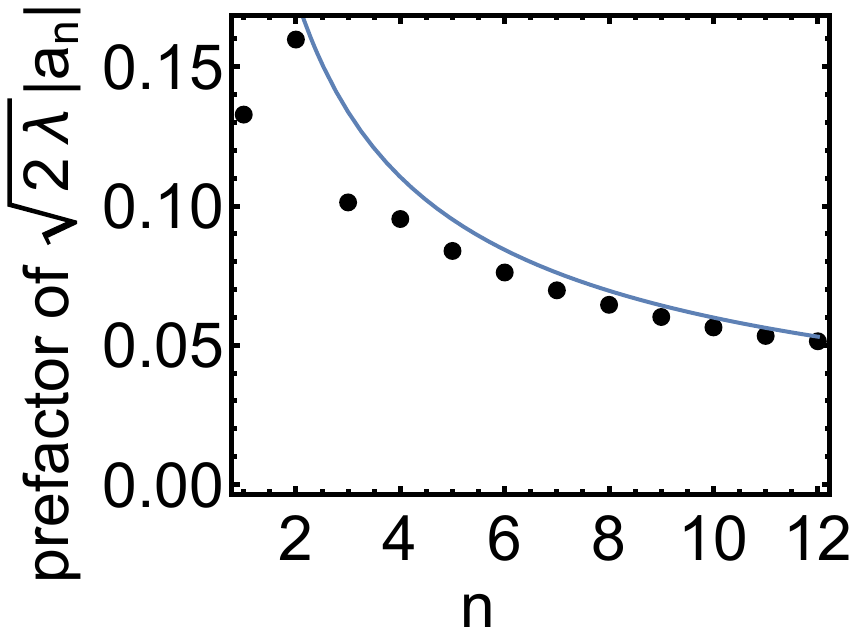}
\caption{Upper left panel: the absolute value of the semiclassical matrix elements $|a_{\pm n}|$ on the logarithmic scale. The red and black dots refer to $|a_{-n}|$ and $|a_n|$, respectively. The red and black solid lines are the corresponding asymptotic expressions~(\ref{eq:a_m<0}) and (\ref{eq:a_m>0}). Upper right and lower panels: the prefactor of the semiclassical matrix element $a_{\pm n}$. Dots and solid lines show the numerical and analytical results, respectively. The value of the scaled drive amplitude is $f = 0.5$, and the quasienergy of the orbit is $g = -0.1$. }
\label{fig:matrix_element}
\end{figure}
%\end{widetext}

We show in Fig.~\ref{fig:matrix_element} a comparison between the asymptotic expressions for the semiclassical matrix elements in Eq.~(\ref{eq:a_m<0}) and (\ref{eq:a_m>0}) and the results obtained using numerical integration of the Hamiltonian equations of motion inside a well of $g(Q,P)$. Both the decay exponents and the prefactors of the matrix elements agree well with the numerics. We note that the matrix elements always satisfy $|a_{-m}|>|a_m|$ for $m>0$. This means that the rate $W_{n\,n-|m|}$ of transitions down in quasienergy  is always larger than the rate $W_{n\,n+|m|}$ of transitions up, so that the stationary distribution over the intrawell states monotonically falls off with the increasing $g$.

The discussed here behavior of $Q+iP$ as a function of complex time for period tripling  is qualitatively different from the behavior of the analogous function for a parametric modulated oscillator~\cite{Marthaler2006} and for a nonlinear oscillator driven close to its eigenfrequency~\cite{Dykman2012,Guo2013}. This function has different forms for the parametric and resonantly driven oscillators, but in the both cases it is described by the Jacobi elliptic functions. It is double periodic in complex time, with two poles within the parallelogram of periods and with no branch points. This difference leads to a qualitative difference of the intrawell distribution and the probability of interstate switching, as discussed in the main text.

\section{The intrawell distribution in the classical limit}
\label{sec:classical_compare}
\begin{figure}
\includegraphics[width =4.2 cm]{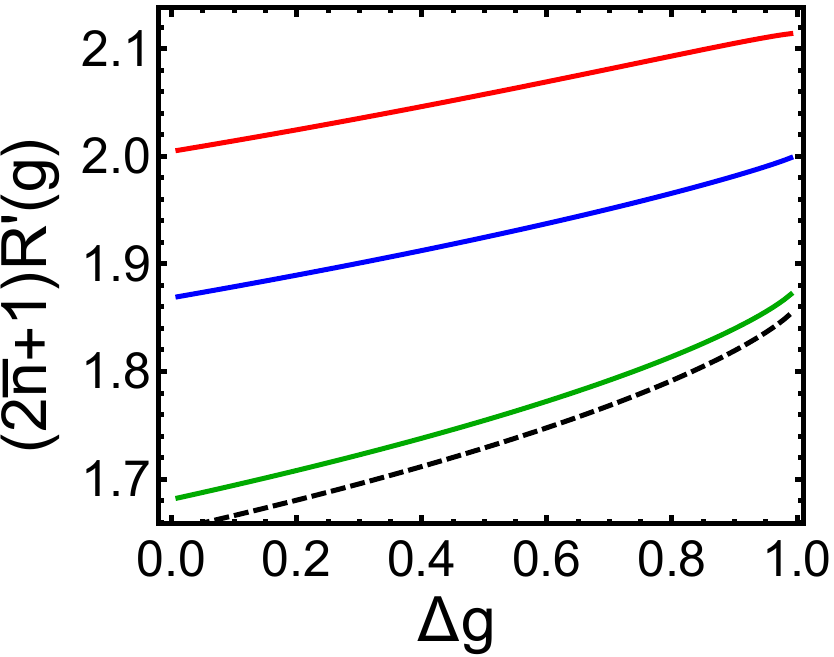} \hfill
\includegraphics[width = 4.2 cm]{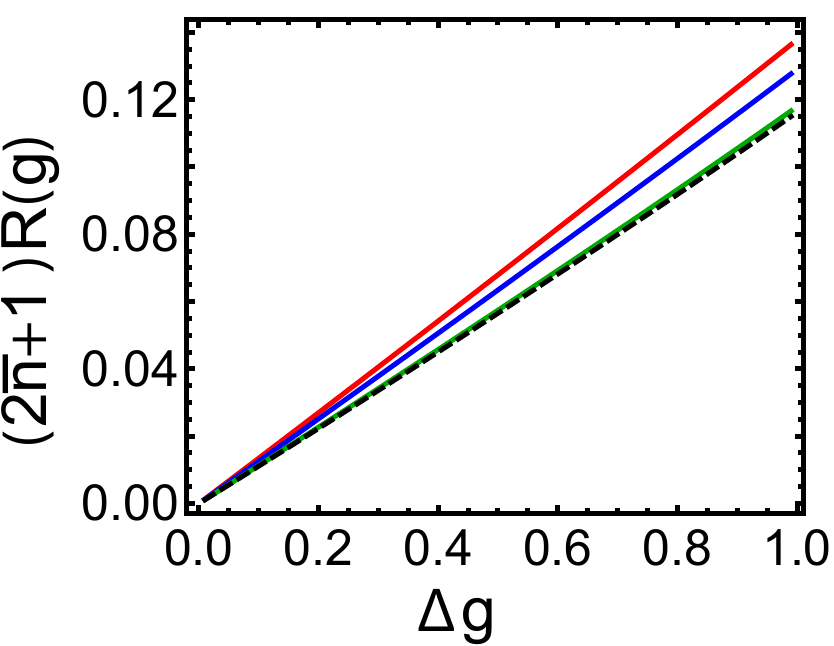}  \hfill
\includegraphics[width =4.2 cm]{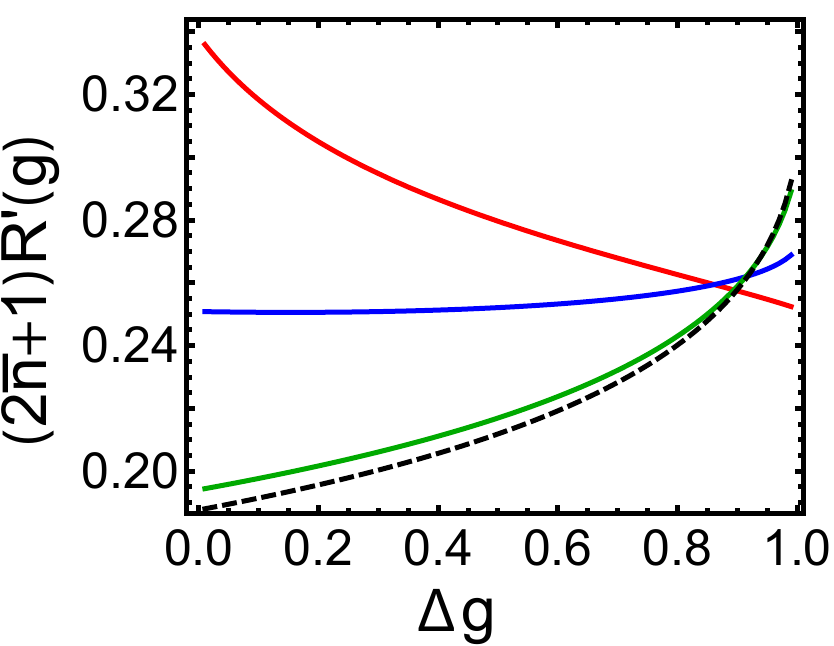} \hfill
\includegraphics[width = 4.2 cm]{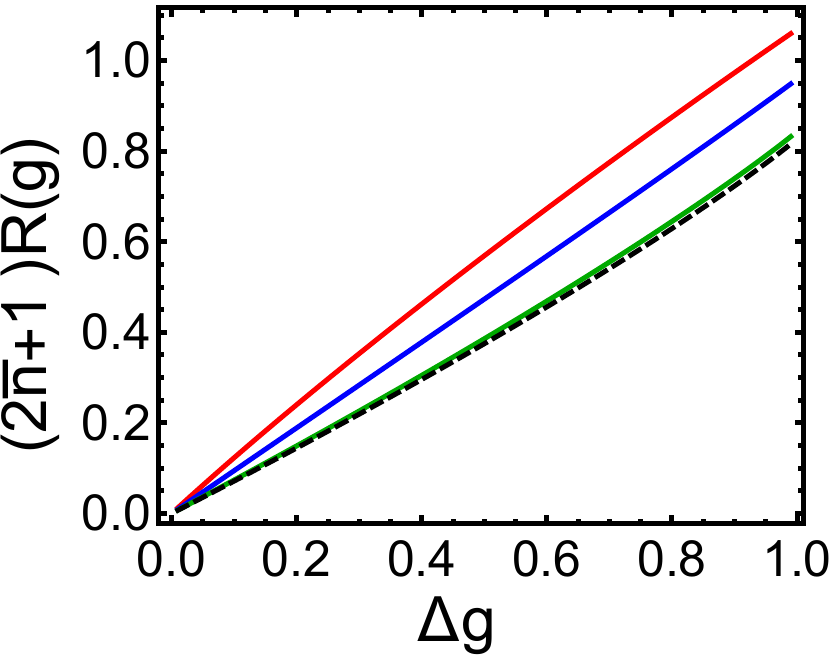} 
\caption{The dependence of the functions $R'(g)$ (left panels) and $R(g)$ (right panels) on the RWA energy $g$ for $\delta\omega>0$, $\Delta g = (g-g_s)/(g-g_{\rm min})$. The scaled drive amplitude is $f=0.1$ (upper panels) and $f = 2$ (lower panels). The red, blue, and green solid lines refer to $\bar n =0, 0.1, 1$, respectively. The dashed black line refers to the classical limit in Eq.~(\ref{eq:small_R_prime}). For the chosen values of the driving amplitude there holds the locality condition, and thus the distribution can be found in the eikonal approximation, as can be seen from Fig.~\ref{fig:gNL}.}
\label{fig:Rprime_classical}
\end{figure}

In this section, we compare the RWA-energy dependent  inverse effective temperatures $R'(g)$ and the scaled logarithms of the probability distribution $R(g)$ obtained by solving the full semiclassical balance equation (\ref{eq:equation_for_xi}) and by solving Eq.~(\ref{eq:small_R_prime}) that formally refers to the classical limit. The results are shown in Fig.~\ref{fig:Rprime_classical}.

For a weak drive (the upper left panel of Fig.~\ref{fig:Rprime_classical}), the behavior of $R'(g)$ is similar to that in the classical limit even at zero temperature. Here, the function $R'(g)$ monotonically increases from $g = g_{\rm min}$ to $g = g_s$ and the function $(2\bar n+1) R(g)$ weakly changes with temperature. In contrast, for a relatively large drive (the lower left panel of Fig.~\ref{fig:Rprime_classical}), $R'(g)$ at low temperature decreases as $g$ increases, which is significantly different from the behavior in the classical limit. As the temperature increases, both $R'(g)$ and $R(g)$ converge to the classical limit.  

\section{The imaginary tunneling time}
\label{sec:imaginary_tunneling_time}
The imaginary tunneling time $\tau_{\rm tun}(g)$ is the imaginary part of the time of moving, with a given $g$, along a Hamiltonian trajectory in the classically inaccessible region of the $(Q,P)$ plane between the wells of $g(Q,P)$. It is given by Eq.~(\ref{eq:tunnel_time_general}), which we repeat here to simplify the reading, 
\begin{align}
\label{eq:tunnel_time_general2}
\partial_g S_{\rm tun} \equiv \tau_{\rm tun}(g) & = {\rm Im}\,\int dQ/\partial_Pg(Q,P), \nonumber \\ 
\partial_P g(Q,P) & = P(Q{})\sqrt{B(Q{})}, 
\end{align}
where $P(Q{})\equiv P(Q,g)$ and $B(Q{})\equiv B(Q,g)$ are given in Eq.~(\ref{eq:momentum_general}). 
For concreteness, we will consider tunneling from the $\nu=0$ well of $g(Q,P)$, which has the minimum on the $P=0$-axis and is symmetric with respect to the reflection $P\to -P$. We assume that $\delta\omega>0$.

To calculate $\tau_{\rm tun}$ one has to establish the limits of the integral in Eq.~(\ref{eq:tunnel_time_general2}). For a given RWA energy $g$, these limits are determined by the boundaries of the classical real-time orbits $g(Q,P)=g$ shown in Fig.~\ref{fig:quasienergy_surface}~(c). We have to consider the orbits in the wells $\nu=0$ and $\nu=1$ or 2. We start with the  orbits in the $\nu=0$-well of $g(Q,P)$. Such orbits have two turning points $\dot Q\equiv P\sqrt{B(Q)}=0$ on the $P=0$ axis. These points, $Q_{\min}(g)$ and $Q_{\max}(g)$ are given by the equation $g(Q,0)=g$.

\begin{figure}[h]
\includegraphics[width = 8.5 cm]{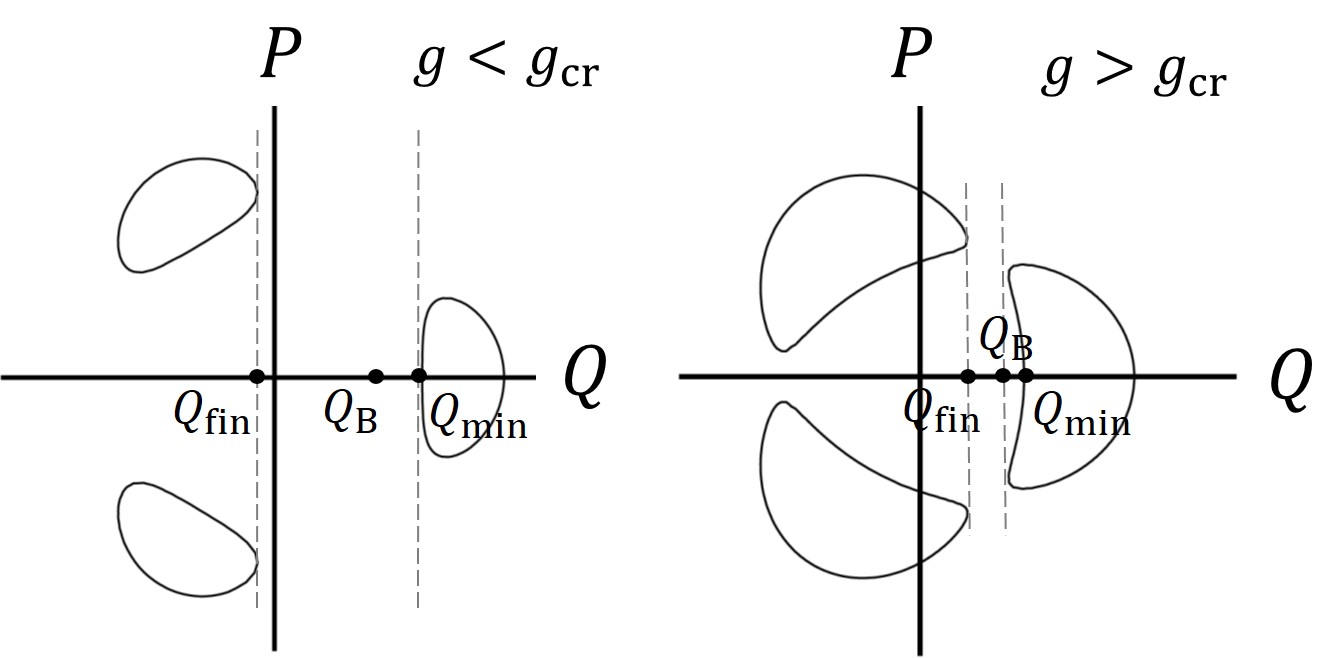}
\caption{An example of elliptical (left) and horseshoe-like (right) classical orbit when $g<g_{\rm cr}$ and $g>g_{\rm cr}$, respectively for $\delta\omega >0$.}
\label{fig:horse_shoe}
\end{figure}

The points $Q_{\min}(g)$ and $Q_{\max}(g)$ are the only turning points in the region of the RWA energy $g<g_{\rm cr}$, because $B(Q,g)>0$ in this region.  The shape of an orbit is close to elliptical, as seen in the left panel of Fig.~\ref{fig:horse_shoe}. The point $Q_{\min}(g)$ is the leftmost point on the orbit.

For $g_{\rm cr}<g<g_s$, function $B(Q,g)$ becomes zero on the classical orbit at a point $Q=Q_B(g)<Q_{\rm min}(g)$. At $Q=Q_B$ we have $\dot Q = 0$, and thus $|dP/dQ|\to \infty$, but $P\neq 0$. Respectively, the orbit becomes horse-shoe like 
as illustrated in the right panel of Fig.~\ref{fig:horse_shoe}. The point $Q_B(g)$ is the leftmost point on the orbit for $g>g_{\rm cr}$.

At $g=g_{\rm cr}$ the points $Q_B$ and $Q_{\min}$ merge, that is, the function $B(Q,g_{\rm cr})$ is zero at $Q=Q_{\min}(g_{\rm cr})$. From Eq.~(\ref{eq:momentum_general}) it follows that at this point $A(Q) =0$. These conditions give for $g_{\rm cr}$ and $Q_{\rm cr}\equiv Q_{\min}(g_{\rm cr})$ the expressions
\[g_{\rm cr} = g(Q_{\rm cr},0), \quad Q_{\rm cr} = -f+ (f^2+1)^{1/2}.\]

The integration over $Q$ in Eq.~(\ref{eq:tunnel_time_general2}) for tunneling from the $\nu=0$-well goes in the negative direction from the leftmost point on the classical orbit of this well to the rightmost point $Q_{\rm fin}$ of  the classical orbit of the well   $\nu=1$ or, equivalently, $\nu=2$. By construction, $\dot Q = 0$ at the rightmost point $Q_{\rm fin}$ of  the classical orbit of the well  $\nu=1$ (or $\nu=2$). Since the orbits of the wells $\nu=1$ and $\nu=2$ do not cross, $P\neq 0$ at this point. Therefore the point $Q_{\rm fin} \equiv Q_{\rm fin}(g)$ is given by the intermediate root of the equation $B(Q,g)=0$. 

The equation $B(Q,g)=0$ has three real roots. The smallest one gives the leftmost point on the orbits of the $\nu=1,2$ wells where $\dot Q=0$. The largest root is $Q_B$; for $g>g_{\rm cr}$ it lies on the classical orbit of the $\nu=0$ well, whereas for $g<g_{\rm cr}$ it lies in the classically forbidden region between the wells. In the region between $Q_{\rm fin}(g)$ and $Q_B(g)$ we have $B(Q,g)<0$, and thus the momentum $P(Q,g)$ is complex, as seen from Eq.~(\ref{eq:momentum_general}). 

Taking into account the difference between the regions $g<g_{\rm cr}$ and $g>g_{\rm cr}$, we write the expression for the imaginary tunneling time as
\begin{align}
\label{eq:tunnel_time_explicit}
\tau_{\rm tun}(g) =&  \int_{Q_{\rm fin}(g)}^{Q_B(g)}\frac{dQ\,{\rm Re}\, P}{|B|^{1/2}|P|^2}\nonumber\\
&+ \int_{Q_B(g)}^{Q_{\rm \rm min}(g)}\frac{dQ}{B^{1/2}{\rm Im}\,P} \Theta(g_{\rm cr}-g).
\end{align}
Here we have set $B(Q,g)^{1/2} = i|B(Q,g)|^{1/2}$ for $B(Q,g)<0$. Then, taking into account that Im~$P(Q,g)<0$ on the tail of the wave function of the $\nu=0$-well, we find from Eq.~(\ref{eq:momentum_general}) that, for $Q_{\rm fin}(g)<Q<Q_B(g)$ the momentum in the classically forbidden region takes the form 
\begin{align*}
P(Q,g) =& -[(A^2+|B|)^{1/2}+A]^{1/2}/\sqrt{2} \\
&-i[(A^2+|B|)^{1/2}-A]^{1/2} /\sqrt{2}.
\end{align*}

For $g<g_{\rm cr}$ and $Q_B(g)<Q< Q_{\min}(g)$, we have $A(Q)<0,B(Q,g)>0$, and $P(Q,g)$ is purely imaginary:
\[
P(Q,g) = -i |A + B^{1/2}(Q,g)|^{1/2}
\]
Clearly, $\tau_{\rm tun}(g)<0$. 

The integrand in Eq.~(\ref{eq:tunnel_time_explicit}) is singular at the saddle-point value of $g$ but the integral is finite. 
For $g=g_s$, the classical trajectory of the well $\nu=0$ has common points $(Q=-Q_s/2, P=\pm \sqrt{3}Q_s/2)$  with the classical trajectories of the wells $\nu=1,2$, where $Q_s$ is the position of the saddle point on the axis $P=0$; as seen from Eq.~(\ref{eq:Hamilton_eqns}),  $Q_s = [f-(f^2 +4)^{1/2}]/2$. Clearly, $S_{\rm tun}(g)\to 0$ for $g\to g_s$. From Eq.~(\ref{eq:tunnel_time_explicit}), near $g=g_s$ we have, to the leading order in $g_s-g$
\begin{align}
\label{tau_for_g_s}
\tau_{\rm tun}(g_s) = \pi\left[3f(f^2+4)^{1/2}Q_s^2\right]^{-1/2}.
\end{align}
%
%where $Q_s = (f- \sqrt{f^2 +4})/2$ is the position of the saddle point for $P=0$ (there are 3 saddle points, which can be obtained from $(Q=Q_s,P=0)$ by rotation in the $(Q,P)$-plane by $2\pi/3$).
One can show that $\tau_{\rm tun}(g)$ is linear in $g-g_s$ near $g_s$.

The integrands in Eq.~(\ref{eq:tunnel_time_explicit}) are also singular at $g_{\rm cr}$. In a narrow region around $Q=Q_B(g)$ the WKB approximation breaks down. However, the contribution of the corresponding region is small for small $\lambda$;  a somewhat cumbersome analysis shows that $\tau_{\rm tun}(g)$ as given by Eq.~(\ref{eq:tunnel_time_explicit}) is smooth and its first derivative is at least continuous at $g_{\rm cr}$.

%\bibliographystyle{apsrev4-1}
%\bibliography{references_v2}
%\bibliography{c:/Users/dykman/Dropbox/Refs/md10test}
%\bibliography{c:/Users/mark/Dropbox/Refs/md10test}

\begin{thebibliography}{45}%
\makeatletter
\providecommand \@ifxundefined [1]{%
 \@ifx{#1\undefined}
}%
\providecommand \@ifnum [1]{%
 \ifnum #1\expandafter \@firstoftwo
 \else \expandafter \@secondoftwo
 \fi
}%
\providecommand \@ifx [1]{%
 \ifx #1\expandafter \@firstoftwo
 \else \expandafter \@secondoftwo
 \fi
}%
\providecommand \natexlab [1]{#1}%
\providecommand \enquote  [1]{``#1''}%
\providecommand \bibnamefont  [1]{#1}%
\providecommand \bibfnamefont [1]{#1}%
\providecommand \citenamefont [1]{#1}%
\providecommand \href@noop [0]{\@secondoftwo}%
\providecommand \href [0]{\begingroup \@sanitize@url \@href}%
\providecommand \@href[1]{\@@startlink{#1}\@@href}%
\providecommand \@@href[1]{\endgroup#1\@@endlink}%
\providecommand \@sanitize@url [0]{\catcode `\\12\catcode `\$12\catcode
  `\&12\catcode `\#12\catcode `\^12\catcode `\_12\catcode `\%12\relax}%
\providecommand \@@startlink[1]{}%
\providecommand \@@endlink[0]{}%
\providecommand \url  [0]{\begingroup\@sanitize@url \@url }%
\providecommand \@url [1]{\endgroup\@href {#1}{\urlprefix }}%
\providecommand \urlprefix  [0]{URL }%
\providecommand \Eprint [0]{\href }%
\providecommand \doibase [0]{http://dx.doi.org/}%
\providecommand \selectlanguage [0]{\@gobble}%
\providecommand \bibinfo  [0]{\@secondoftwo}%
\providecommand \bibfield  [0]{\@secondoftwo}%
\providecommand \translation [1]{[#1]}%
\providecommand \BibitemOpen [0]{}%
\providecommand \bibitemStop [0]{}%
\providecommand \bibitemNoStop [0]{.\EOS\space}%
\providecommand \EOS [0]{\spacefactor3000\relax}%
\providecommand \BibitemShut  [1]{\csname bibitem#1\endcsname}%
\let\auto@bib@innerbib\@empty
%</preamble>
\bibitem [{\citenamefont {Einstein}(1917)}]{Einstein1917}%
  \BibitemOpen
  \bibfield  {author} {\bibinfo {author} {\bibfnamefont {A.}~\bibnamefont
  {Einstein}},\ }\href@noop {} {\bibfield  {journal} {\bibinfo  {journal}
  {Phys. Z.}\ }\textbf {\bibinfo {volume} {18}},\ \bibinfo {pages} {121}
  (\bibinfo {year} {1917})}\BibitemShut {NoStop}%
\bibitem [{\citenamefont {Peano}\ and\ \citenamefont
  {Thorwart}(2006)}]{Peano2006}%
  \BibitemOpen
  \bibfield  {author} {\bibinfo {author} {\bibfnamefont {V.}~\bibnamefont
  {Peano}}\ and\ \bibinfo {author} {\bibfnamefont {M.}~\bibnamefont
  {Thorwart}},\ }\href {\doibase 10.1016/j.chemphys.2005.06.047} {\bibfield
  {journal} {\bibinfo  {journal} {Chem. Phys.}\ }\textbf {\bibinfo {volume}
  {322}},\ \bibinfo {pages} {135} (\bibinfo {year} {2006})}\BibitemShut
  {NoStop}%
\bibitem [{\citenamefont {Martin}\ \emph {et~al.}(2017)\citenamefont {Martin},
  \citenamefont {Refael},\ and\ \citenamefont {Halperin}}]{Martin2017}%
  \BibitemOpen
  \bibfield  {author} {\bibinfo {author} {\bibfnamefont {I.}~\bibnamefont
  {Martin}}, \bibinfo {author} {\bibfnamefont {G.}~\bibnamefont {Refael}}, \
  and\ \bibinfo {author} {\bibfnamefont {B.}~\bibnamefont {Halperin}},\
  }\href@noop {} {\bibfield  {journal} {\bibinfo  {journal} {Phys. Rev. X}\
  }\textbf {\bibinfo {volume} {7}},\ \bibinfo {pages} {041008} (\bibinfo {year}
  {2017})}\BibitemShut {NoStop}%
\bibitem [{\citenamefont {Weinberg}\ \emph {et~al.}(2017)\citenamefont
  {Weinberg}, \citenamefont {Bukov}, \citenamefont {D'Alessio}, \citenamefont
  {Polkovnikov}, \citenamefont {Vajna},\ and\ \citenamefont
  {Kolodrubetz}}]{Weinberg2017}%
  \BibitemOpen
  \bibfield  {author} {\bibinfo {author} {\bibfnamefont {P.}~\bibnamefont
  {Weinberg}}, \bibinfo {author} {\bibfnamefont {M.}~\bibnamefont {Bukov}},
  \bibinfo {author} {\bibfnamefont {L.}~\bibnamefont {D'Alessio}}, \bibinfo
  {author} {\bibfnamefont {A.}~\bibnamefont {Polkovnikov}}, \bibinfo {author}
  {\bibfnamefont {S.}~\bibnamefont {Vajna}}, \ and\ \bibinfo {author}
  {\bibfnamefont {M.}~\bibnamefont {Kolodrubetz}},\ }\href@noop {} {\bibfield
  {journal} {\bibinfo  {journal} {Phys. Rep.}\ }\textbf {\bibinfo {volume}
  {688}},\ \bibinfo {pages} {1} (\bibinfo {year} {2017})}\BibitemShut {NoStop}%
\bibitem [{\citenamefont {Desbuquois}\ \emph {et~al.}(2017)\citenamefont
  {Desbuquois}, \citenamefont {Messer}, \citenamefont {G{\"o}rg}, \citenamefont
  {Sandholzer}, \citenamefont {Jotzu},\ and\ \citenamefont
  {Esslinger}}]{Desbuquois2017}%
  \BibitemOpen
  \bibfield  {author} {\bibinfo {author} {\bibfnamefont {R.}~\bibnamefont
  {Desbuquois}}, \bibinfo {author} {\bibfnamefont {M.}~\bibnamefont {Messer}},
  \bibinfo {author} {\bibfnamefont {F.}~\bibnamefont {G{\"o}rg}}, \bibinfo
  {author} {\bibfnamefont {K.}~\bibnamefont {Sandholzer}}, \bibinfo {author}
  {\bibfnamefont {G.}~\bibnamefont {Jotzu}}, \ and\ \bibinfo {author}
  {\bibfnamefont {T.}~\bibnamefont {Esslinger}},\ }\href@noop {} {\bibfield
  {journal} {\bibinfo  {journal} {ArXiv e-prints}\ } (\bibinfo {year}
  {2017})}\BibitemShut {NoStop}%
\bibitem [{\citenamefont {Schmidt}(2018)}]{Schmidt2018a}%
  \BibitemOpen
  \bibfield  {author} {\bibinfo {author} {\bibfnamefont {H.-J.}\ \bibnamefont
  {Schmidt}},\ }\href {\doibase 10.1515/zna-2018-0211} {\bibfield  {journal}
  {\bibinfo  {journal} {Zeitschrift f{\"u}r Naturforschung A}\ }\textbf
  {\bibinfo {volume} {73}},\ \bibinfo {pages} {705} (\bibinfo {year}
  {2018})}\BibitemShut {NoStop}%
\bibitem [{\citenamefont {Seetharam}\ \emph {et~al.}(2018)\citenamefont
  {Seetharam}, \citenamefont {Titum}, \citenamefont {Kolodrubetz},\ and\
  \citenamefont {Refael}}]{Seetharam2018a}%
  \BibitemOpen
  \bibfield  {author} {\bibinfo {author} {\bibfnamefont {K.}~\bibnamefont
  {Seetharam}}, \bibinfo {author} {\bibfnamefont {P.}~\bibnamefont {Titum}},
  \bibinfo {author} {\bibfnamefont {M.}~\bibnamefont {Kolodrubetz}}, \ and\
  \bibinfo {author} {\bibfnamefont {G.}~\bibnamefont {Refael}},\ }\href@noop {}
  {\bibfield  {journal} {\bibinfo  {journal} {Phys. Rev. B}\ }\textbf {\bibinfo
  {volume} {97}},\ \bibinfo {pages} {014311} (\bibinfo {year}
  {2018})}\BibitemShut {NoStop}%
\bibitem [{\citenamefont {Lohse}\ \emph {et~al.}(2018)\citenamefont {Lohse},
  \citenamefont {Schweizer}, \citenamefont {Price}, \citenamefont
  {Zilberberg},\ and\ \citenamefont {Bloch}}]{Lohse2018}%
  \BibitemOpen
  \bibfield  {author} {\bibinfo {author} {\bibfnamefont {M.}~\bibnamefont
  {Lohse}}, \bibinfo {author} {\bibfnamefont {C.}~\bibnamefont {Schweizer}},
  \bibinfo {author} {\bibfnamefont {H.~M.}\ \bibnamefont {Price}}, \bibinfo
  {author} {\bibfnamefont {O.}~\bibnamefont {Zilberberg}}, \ and\ \bibinfo
  {author} {\bibfnamefont {I.}~\bibnamefont {Bloch}},\ }\href {\doibase
  10.1038/nature25000} {\bibfield  {journal} {\bibinfo  {journal} {Nature}\
  }\textbf {\bibinfo {volume} {553}},\ \bibinfo {pages} {55} (\bibinfo {year}
  {2018})}\BibitemShut {NoStop}%
\bibitem [{\citenamefont {Shirley}(1965)}]{shirley1965}%
  \BibitemOpen
  \bibfield  {author} {\bibinfo {author} {\bibfnamefont {J.~H.}\ \bibnamefont
  {Shirley}},\ }\href {\doibase 10.1103/PhysRev.138.B979} {\bibfield  {journal}
  {\bibinfo  {journal} {Phys. Rev.}\ }\textbf {\bibinfo {volume} {138}},\
  \bibinfo {pages} {B979} (\bibinfo {year} {1965})}\BibitemShut {NoStop}%
\bibitem [{\citenamefont {Zel'dovich}(1967)}]{zeldovich1967}%
  \BibitemOpen
  \bibfield  {author} {\bibinfo {author} {\bibfnamefont {Y.~B.}\ \bibnamefont
  {Zel'dovich}},\ }\href@noop {} {\bibfield  {journal} {\bibinfo  {journal}
  {Sov. Phys. JETP}\ }\textbf {\bibinfo {volume} {24}},\ \bibinfo {pages}
  {1006} (\bibinfo {year} {1967})}\BibitemShut {NoStop}%
\bibitem [{\citenamefont {Ritus}(1967)}]{ritus1967}%
  \BibitemOpen
  \bibfield  {author} {\bibinfo {author} {\bibfnamefont {V.~I.}\ \bibnamefont
  {Ritus}},\ }\href@noop {} {\bibfield  {journal} {\bibinfo  {journal} {Sov.
  Phys. JETP}\ }\textbf {\bibinfo {volume} {24}},\ \bibinfo {pages} {1041}
  (\bibinfo {year} {1967})}\BibitemShut {NoStop}%
\bibitem [{\citenamefont {Kohn}(2001)}]{kohn2001}%
  \BibitemOpen
  \bibfield  {author} {\bibinfo {author} {\bibfnamefont {W.}~\bibnamefont
  {Kohn}},\ }\href {\doibase 10.1023/A:1010327828445} {\bibfield  {journal}
  {\bibinfo  {journal} {Journal of Statistical Physics}\ }\textbf {\bibinfo
  {volume} {103}},\ \bibinfo {pages} {417} (\bibinfo {year}
  {2001})}\BibitemShut {NoStop}%
\bibitem [{\citenamefont {Drummond}\ and\ \citenamefont
  {Walls}(1980)}]{Drummond1980}%
  \BibitemOpen
  \bibfield  {author} {\bibinfo {author} {\bibfnamefont {P.~D.}\ \bibnamefont
  {Drummond}}\ and\ \bibinfo {author} {\bibfnamefont {D.~F.}\ \bibnamefont
  {Walls}},\ }\href@noop {} {\bibfield  {journal} {\bibinfo  {journal} {J.
  Phys. A}\ }\textbf {\bibinfo {volume} {13}},\ \bibinfo {pages} {725}
  (\bibinfo {year} {1980})}\BibitemShut {NoStop}%
\bibitem [{\citenamefont {Kryuchkyan}\ and\ \citenamefont
  {Kheruntsyan}(1996)}]{Kryuchkyan1996}%
  \BibitemOpen
  \bibfield  {author} {\bibinfo {author} {\bibfnamefont {G.~Y.}\ \bibnamefont
  {Kryuchkyan}}\ and\ \bibinfo {author} {\bibfnamefont {K.~V.}\ \bibnamefont
  {Kheruntsyan}},\ }\href@noop {} {\bibfield  {journal} {\bibinfo  {journal}
  {Opt. Commun.}\ }\textbf {\bibinfo {volume} {127}},\ \bibinfo {pages} {230}
  (\bibinfo {year} {1996})}\BibitemShut {NoStop}%
\bibitem [{\citenamefont {Dykman}\ and\ \citenamefont
  {Krivoglaz}(1979)}]{Dykman1979}%
  \BibitemOpen
  \bibfield  {author} {\bibinfo {author} {\bibfnamefont {M.~I.}\ \bibnamefont
  {Dykman}}\ and\ \bibinfo {author} {\bibfnamefont {M.~A.}\ \bibnamefont
  {Krivoglaz}},\ }\href@noop {} {\bibfield  {journal} {\bibinfo  {journal} {Zh.
  Eksp. Teor. Fiz.}\ }\textbf {\bibinfo {volume} {77}},\ \bibinfo {pages} {60}
  (\bibinfo {year} {1979})}\BibitemShut {NoStop}%
\bibitem [{\citenamefont {Dykman}\ and\ \citenamefont
  {Smelyanskii}(1988)}]{Dykman1988a}%
  \BibitemOpen
  \bibfield  {author} {\bibinfo {author} {\bibfnamefont {M.~I.}\ \bibnamefont
  {Dykman}}\ and\ \bibinfo {author} {\bibfnamefont {V.~N.}\ \bibnamefont
  {Smelyanskii}},\ }\href@noop {} {\bibfield  {journal} {\bibinfo  {journal}
  {Zh. Eksp. Teor. Fiz.}\ }\textbf {\bibinfo {volume} {94}},\ \bibinfo {pages}
  {61} (\bibinfo {year} {1988})}\BibitemShut {NoStop}%
\bibitem [{\citenamefont {Marthaler}\ and\ \citenamefont
  {Dykman}(2006)}]{Marthaler2006}%
  \BibitemOpen
  \bibfield  {author} {\bibinfo {author} {\bibfnamefont {M.}~\bibnamefont
  {Marthaler}}\ and\ \bibinfo {author} {\bibfnamefont {M.~I.}\ \bibnamefont
  {Dykman}},\ }\href {\doibase 10.1103/PhysRevA.73.042108} {\bibfield
  {journal} {\bibinfo  {journal} {Physical Review A}\ }\textbf {\bibinfo
  {volume} {73}},\ \bibinfo {pages} {042108} (\bibinfo {year} {2006})}\BibitemShut {NoStop}%
\bibitem [{\citenamefont {Guo}\ \emph {et~al.}(2013{\natexlab{a}})\citenamefont
  {Guo}, \citenamefont {Marthaler},\ and\ \citenamefont
  {Sch{\"o}n}}]{Guo2013a}%
  \BibitemOpen
  \bibfield  {author} {\bibinfo {author} {\bibfnamefont {L.}~\bibnamefont
  {Guo}}, \bibinfo {author} {\bibfnamefont {M.}~\bibnamefont {Marthaler}}, \
  and\ \bibinfo {author} {\bibfnamefont {G.}~\bibnamefont {Sch{\"o}n}},\ }\href
  {\doibase 10.1103/PhysRevLett.111.205303} {\bibfield  {journal} {\bibinfo
  {journal} {Phys. Rev. Lett.}\ }\textbf {\bibinfo {volume} {111}},\ \bibinfo
  {pages} {205303} (\bibinfo {year} {2013}{\natexlab{a}})}\BibitemShut
  {NoStop}%
\bibitem [{\citenamefont {Zhang}\ \emph {et~al.}(2017)\citenamefont {Zhang},
  \citenamefont {Gosner}, \citenamefont {Girvin}, \citenamefont {Ankerhold},\
  and\ \citenamefont {Dykman}}]{Zhang2017}%
  \BibitemOpen
  \bibfield  {author} {\bibinfo {author} {\bibfnamefont {Y.}~\bibnamefont
  {Zhang}}, \bibinfo {author} {\bibfnamefont {J.}~\bibnamefont {Gosner}},
  \bibinfo {author} {\bibfnamefont {S.~M.}\ \bibnamefont {Girvin}}, \bibinfo
  {author} {\bibfnamefont {J.}~\bibnamefont {Ankerhold}}, \ and\ \bibinfo
  {author} {\bibfnamefont {M.~I.}\ \bibnamefont {Dykman}},\ }\href {\doibase
  10.1103/PhysRevA.96.052124} {\bibfield  {journal} {\bibinfo  {journal}
  {Physical Review A}\ }\textbf {\bibinfo {volume} {96}},\ \bibinfo {pages} {052124} (\bibinfo {year}
  {2017})}\BibitemShut {NoStop}%
\bibitem [{\citenamefont {Svensson}\ \emph {et~al.}(2017)\citenamefont
  {Svensson}, \citenamefont {Bengtsson}, \citenamefont {Krantz}, \citenamefont
  {Bylander}, \citenamefont {Shumeiko},\ and\ \citenamefont
  {Delsing}}]{Svensson2017a}%
  \BibitemOpen
  \bibfield  {author} {\bibinfo {author} {\bibfnamefont {I.-M.}\ \bibnamefont
  {Svensson}}, \bibinfo {author} {\bibfnamefont {A.}~\bibnamefont {Bengtsson}},
  \bibinfo {author} {\bibfnamefont {P.}~\bibnamefont {Krantz}}, \bibinfo
  {author} {\bibfnamefont {J.}~\bibnamefont {Bylander}}, \bibinfo {author}
  {\bibfnamefont {V.}~\bibnamefont {Shumeiko}}, \ and\ \bibinfo {author}
  {\bibfnamefont {P.}~\bibnamefont {Delsing}},\ }\href@noop {} {\bibfield
  {journal} {\bibinfo  {journal} {Phys. Rev. B}\ }\textbf {\bibinfo {volume}
  {96}},\ \bibinfo {pages} {174503} (\bibinfo {year} {2017})}\BibitemShut
  {NoStop}%
\bibitem [{\citenamefont {Peano}\ and\ \citenamefont
  {Dykman}(2014)}]{Peano2014}%
  \BibitemOpen
  \bibfield  {author} {\bibinfo {author} {\bibfnamefont {V.}~\bibnamefont
  {Peano}}\ and\ \bibinfo {author} {\bibfnamefont {M.~I.}\ \bibnamefont
  {Dykman}},\ }\href@noop {} {\bibfield  {journal} {\bibinfo  {journal} {NJP}\
  }\textbf {\bibinfo {volume} {16}},\ \bibinfo {pages} {015011} (\bibinfo
  {year} {2014})}\BibitemShut {NoStop}%
\bibitem [{\citenamefont {Guo}\ \emph {et~al.}(2013{\natexlab{b}})\citenamefont
  {Guo}, \citenamefont {Peano}, \citenamefont {Marthaler},\ and\ \citenamefont
  {Dykman}}]{Guo2013}%
  \BibitemOpen
  \bibfield  {author} {\bibinfo {author} {\bibfnamefont {L.}~\bibnamefont
  {Guo}}, \bibinfo {author} {\bibfnamefont {V.}~\bibnamefont {Peano}}, \bibinfo
  {author} {\bibfnamefont {M.}~\bibnamefont {Marthaler}}, \ and\ \bibinfo
  {author} {\bibfnamefont {M.~I.}\ \bibnamefont {Dykman}},\ }\href@noop {}
  {\bibfield  {journal} {\bibinfo  {journal} {Phys. Rev. A}\ }\textbf {\bibinfo
  {volume} {87}},\ \bibinfo {pages} {062117} (\bibinfo {year}
  {2013}{\natexlab{b}})}\BibitemShut {NoStop}%
\bibitem [{\citenamefont {Nayfeh}\ and\ \citenamefont
  {Mook}(2004)}]{Nayfeh2004}%
  \BibitemOpen
  \bibfield  {author} {\bibinfo {author} {\bibfnamefont {A.~H.}\ \bibnamefont
  {Nayfeh}}\ and\ \bibinfo {author} {\bibfnamefont {D.~T.}\ \bibnamefont
  {Mook}},\ }\href@noop {} {\emph {\bibinfo {title} {Nonlinear
  {{Oscillations}}}}}\ (\bibinfo  {publisher} {{Wiley-VCH}},\ \bibinfo {year}
  {Weinheim 2004})\BibitemShut {NoStop}%
\bibitem [{\citenamefont {Landau}\ and\ \citenamefont
  {Lifshitz}(1977)}]{landau1977}%
  \BibitemOpen
  \bibfield  {author} {\bibinfo {author} {\bibfnamefont {L.}~\bibnamefont
  {Landau}}\ and\ \bibinfo {author} {\bibfnamefont {E.}~\bibnamefont
  {Lifshitz}},\ }\href@noop {} {\emph {\bibinfo {title} {Quantum {{Mechanics}}:
  {{Non}}-{{Relativistic Theory}}}}}\ (\bibinfo  {publisher} {{Elsevier}},\
  \bibinfo {year} {1977})\BibitemShut {NoStop}%
\bibitem [{\citenamefont {Ziman}(1979)}]{Ziman1979}%
  \BibitemOpen
  \bibfield  {author} {\bibinfo {author} {\bibfnamefont {J.~M.}\ \bibnamefont
  {Ziman}},\ }\href@noop {} {\emph {\bibinfo {title} {Principles of the
  {{Theory}} of {{Solids}}}}},\ \bibinfo {edition} {2nd}\ ed.\ (\bibinfo
  {publisher} {{Cambridge University Press}},\ \bibinfo {address}
  {{Cambridge}},\ \bibinfo {year} {1979})\BibitemShut {NoStop}%
\bibitem [{\citenamefont {Landau}(1927)}]{Landau1927}%
  \BibitemOpen
  \bibfield  {author} {\bibinfo {author} {\bibfnamefont {L.~D.}\ \bibnamefont
  {Landau}},\ }\href@noop {} {\bibfield  {journal} {\bibinfo  {journal} {Z.
  Phys.}\ }\textbf {\bibinfo {volume} {45}},\ \bibinfo {pages} {430} (\bibinfo
  {year} {1927})}\BibitemShut {NoStop}%
\bibitem [{\citenamefont {Walls}\ and\ \citenamefont
  {Milburn}(2008)}]{Walls2008}%
  \BibitemOpen
  \bibfield  {author} {\bibinfo {author} {\bibfnamefont {D.~F.}\ \bibnamefont
  {Walls}}\ and\ \bibinfo {author} {\bibfnamefont {G.~J.}\ \bibnamefont
  {Milburn}},\ }\href@noop {} {\emph {\bibinfo {title} {Quantum {{Optics}}}}}\
  (\bibinfo  {publisher} {{Springer, Berlin}},\ \bibinfo {year}
  {2008})\BibitemShut {NoStop}%
\bibitem [{\citenamefont {Kubo}(1957)}]{Kubo1957}%
  \BibitemOpen
  \bibfield  {author} {\bibinfo {author} {\bibfnamefont {R.}~\bibnamefont
  {Kubo}},\ }\href {\doibase 10.1143/JPSJ.12.570} {\bibfield  {journal}
  {\bibinfo  {journal} {J. Phys. Soc. Jpn.}\ }\textbf {\bibinfo {volume}
  {12}},\ \bibinfo {pages} {570} (\bibinfo {year} {1957})}\BibitemShut
  {NoStop}%
\bibitem [{\citenamefont {Ong}\ \emph {et~al.}(2013)\citenamefont {Ong},
  \citenamefont {Boissonneault}, \citenamefont {Mallet}, \citenamefont
  {Doherty}, \citenamefont {Blais}, \citenamefont {Vion}, \citenamefont
  {Esteve},\ and\ \citenamefont {Bertet}}]{Ong2013}%
  \BibitemOpen
  \bibfield  {author} {\bibinfo {author} {\bibfnamefont {F.~R.}\ \bibnamefont
  {Ong}}, \bibinfo {author} {\bibfnamefont {M.}~\bibnamefont {Boissonneault}},
  \bibinfo {author} {\bibfnamefont {F.}~\bibnamefont {Mallet}}, \bibinfo
  {author} {\bibfnamefont {A.~C.}\ \bibnamefont {Doherty}}, \bibinfo {author}
  {\bibfnamefont {A.}~\bibnamefont {Blais}}, \bibinfo {author} {\bibfnamefont
  {D.}~\bibnamefont {Vion}}, \bibinfo {author} {\bibfnamefont {D.}~\bibnamefont
  {Esteve}}, \ and\ \bibinfo {author} {\bibfnamefont {P.}~\bibnamefont
  {Bertet}},\ }\href@noop {} {\bibfield  {journal} {\bibinfo  {journal} {Phys.
  Rev. Lett.}\ }\textbf {\bibinfo {volume} {110}},\ \bibinfo {pages} {047001}
  (\bibinfo {year} {2013})}\BibitemShut {NoStop}%
\bibitem [{\citenamefont {Goldstein}\ \emph {et~al.}(2001)\citenamefont
  {Goldstein}, \citenamefont {Poole},\ and\ \citenamefont
  {Safko}}]{goldstein2001}%
  \BibitemOpen
  \bibfield  {author} {\bibinfo {author} {\bibfnamefont {H.}~\bibnamefont
  {Goldstein}}, \bibinfo {author} {\bibfnamefont {C.~P.}\ \bibnamefont
  {Poole}}, \ and\ \bibinfo {author} {\bibfnamefont {J.~L.}\ \bibnamefont
  {Safko}},\ }\href@noop {} {\emph {\bibinfo {title} {Classical
  {{Mechanics}}}}},\ \bibinfo {edition} {3rd}\ ed.\ (\bibinfo  {publisher}
  {{Pearson}},\ \bibinfo {address} {{San Francisco}},\ \bibinfo {year}
  {2001})\BibitemShut {NoStop}%
\bibitem [{\citenamefont {Kagan}\ and\ \citenamefont
  {Leggett}(1992)}]{Kagan1992}%
  \BibitemOpen
  \bibinfo {editor} {\bibfnamefont {Y.}~\bibnamefont {Kagan}}\ and\ \bibinfo
  {editor} {\bibfnamefont {A.~J.}\ \bibnamefont {Leggett}},\ eds.,\ \href@noop
  {} {\emph {\bibinfo {title} {Quantum {{Tunneling}} in {{Condensed Media}}}}}\
  (\bibinfo  {publisher} {{North-Holland}},\ \bibinfo {year}
  {1992})\BibitemShut {NoStop}%
\bibitem [{\citenamefont {Larkin}\ and\ \citenamefont
  {Ovchinnikov}(1985)}]{Larkin1985}%
  \BibitemOpen
  \bibfield  {author} {\bibinfo {author} {\bibfnamefont {A.~I.}\ \bibnamefont
  {Larkin}}\ and\ \bibinfo {author} {\bibfnamefont {Y.~N.}\ \bibnamefont
  {Ovchinnikov}},\ }\href@noop {} {\bibfield  {journal} {\bibinfo  {journal}
  {J. Stat. Phys.}\ }\textbf {\bibinfo {volume} {41}},\ \bibinfo {pages} {425}
  (\bibinfo {year} {1985})}\BibitemShut {NoStop}%
\bibitem [{\citenamefont {Freidlin}\ and\ \citenamefont
  {Wentzell}(1998)}]{Freidlin1998}%
  \BibitemOpen
  \bibfield  {author} {\bibinfo {author} {\bibfnamefont {M.~I.}\ \bibnamefont
  {Freidlin}}\ and\ \bibinfo {author} {\bibfnamefont {A.~D.}\ \bibnamefont
  {Wentzell}},\ }\href@noop {} {\emph {\bibinfo {title} {Random
  {{Perturbations}} of {{Dynamical Systems}}}}},\ \bibinfo {edition} {2nd}\
  ed.\ (\bibinfo  {publisher} {{Springer-Verlag}},\ \bibinfo {address} {{New
  York}},\ \bibinfo {year} {1998})\BibitemShut {NoStop}%
\bibitem [{\citenamefont {Guckenheimer}\ and\ \citenamefont
  {Holmes}(1997)}]{Guckenheimer1997}%
  \BibitemOpen
  \bibfield  {author} {\bibinfo {author} {\bibfnamefont {J.}~\bibnamefont
  {Guckenheimer}}\ and\ \bibinfo {author} {\bibfnamefont {P.}~\bibnamefont
  {Holmes}},\ }\href@noop {} {\emph {\bibinfo {title} {Nonlinear
  {{Oscillators}}, {{Dynamical Systems}} and {{Bifurcations}} of {{Vector
  Fields}}}}}\ (\bibinfo  {publisher} {{Springer-Verlag}},\ \bibinfo {address}
  {{New York}},\ \bibinfo {year} {1997})\BibitemShut {NoStop}%
\bibitem [{\citenamefont {Dykman}(2007)}]{Dykman2007}%
  \BibitemOpen
  \bibfield  {author} {\bibinfo {author} {\bibfnamefont {M.~I.}\ \bibnamefont
  {Dykman}},\ }\href@noop {} {\bibfield  {journal} {\bibinfo  {journal} {Phys.
  Rev. E}\ }\textbf {\bibinfo {volume} {75}},\ \bibinfo {pages} {011101}
  (\bibinfo {year} {2007})}\BibitemShut {NoStop}%
\bibitem [{\citenamefont {Dykman}(2012)}]{Dykman2012}%
  \BibitemOpen
  \bibfield  {author} {\bibinfo {author} {\bibfnamefont {M.~I.}\ \bibnamefont
  {Dykman}},\ }in\ \href@noop {} {\emph {\bibinfo {booktitle} {Fluctuating
  Nonlinear Oscillators: From Nanomechanics to Quantum Superconducting
  Circuits}}},\ \bibinfo {editor} {edited by\ \bibinfo {editor} {\bibfnamefont
  {M.~I.}\ \bibnamefont {Dykman}}}\ (\bibinfo  {publisher} {{Oxford University
  Press}},\ \bibinfo {year} {2012})\BibitemShut {NoStop}%
\bibitem [{\citenamefont {Jordan}\ and\ \citenamefont
  {Smith}(2007)}]{jordan2007}%
  \BibitemOpen
  \bibfield  {author} {\bibinfo {author} {\bibfnamefont {D.~W.}\ \bibnamefont
  {Jordan}}\ and\ \bibinfo {author} {\bibfnamefont {P.}~\bibnamefont {Smith}},\
  }\href@noop {} {\emph {\bibinfo {title} {Nonlinear {{Ordinary Differential
  Equations}}, 4ed}}}\ (\bibinfo  {publisher} {{Oxford University Press}},\
  \bibinfo {year} {2007})\BibitemShut {NoStop}%
\bibitem [{\citenamefont {Kramers}(1940)}]{Kramers1940}%
  \BibitemOpen
  \bibfield  {author} {\bibinfo {author} {\bibfnamefont {H.}~\bibnamefont
  {Kramers}},\ }\href@noop {} {\bibfield  {journal} {\bibinfo  {journal}
  {Physica (Utrecht)}\ }\textbf {\bibinfo {volume} {7}},\ \bibinfo {pages}
  {284} (\bibinfo {year} {1940})}\BibitemShut {NoStop}%
\bibitem [{\citenamefont {Vijay}\ \emph {et~al.}(2009)\citenamefont {Vijay},
  \citenamefont {Devoret},\ and\ \citenamefont {Siddiqi}}]{Vijay2009}%
  \BibitemOpen
  \bibfield  {author} {\bibinfo {author} {\bibfnamefont {R.}~\bibnamefont
  {Vijay}}, \bibinfo {author} {\bibfnamefont {M.~H.}\ \bibnamefont {Devoret}},
  \ and\ \bibinfo {author} {\bibfnamefont {I.}~\bibnamefont {Siddiqi}},\
  }\href@noop {} {\bibfield  {journal} {\bibinfo  {journal} {Rev. Sci. Instr.}\
  }\textbf {\bibinfo {volume} {80}},\ \bibinfo {pages} {111101} (\bibinfo
  {year} {2009})}\BibitemShut {NoStop}%
\bibitem [{\citenamefont {Vijay}\ \emph {et~al.}(2012)\citenamefont {Vijay},
  \citenamefont {Murch},\ and\ \citenamefont {Siddiqi}}]{Vijay2012}%
  \BibitemOpen
  \bibfield  {author} {\bibinfo {author} {\bibfnamefont {R.}~\bibnamefont
  {Vijay}}, \bibinfo {author} {\bibfnamefont {K.~W.}\ \bibnamefont {Murch}}, \
  and\ \bibinfo {author} {\bibfnamefont {I.}~\bibnamefont {Siddiqi}},\ }in\
  \href@noop {} {\emph {\bibinfo {booktitle} {Fluctuating {{Nonlinear
  Oscillators}}}}},\ \bibinfo {editor} {edited by\ \bibinfo {editor}
  {\bibfnamefont {M.~I.}\ \bibnamefont {Dykman}}}\ (\bibinfo  {publisher}
  {{OUP, Oxford}},\ \bibinfo {year} {2012})\ pp.\ \bibinfo {pages}
  {362--389}\BibitemShut {NoStop}%
\bibitem [{\citenamefont {Kamenev}(2011)}]{Kamenev2011}%
  \BibitemOpen
  \bibfield  {author} {\bibinfo {author} {\bibfnamefont {A.}~\bibnamefont
  {Kamenev}},\ }\href@noop {} {\emph {\bibinfo {title} {Field {{Theory}} of
  {{Non}}-{{Equilibrium Systems}}}}}\ (\bibinfo  {publisher} {{Cambridge
  University Press, Cambridge}},\ \bibinfo {year} {2011})\BibitemShut {NoStop}%
\bibitem [{\citenamefont {Assaf}\ and\ \citenamefont
  {Meerson}(2017)}]{Assaf2017}%
  \BibitemOpen
  \bibfield  {author} {\bibinfo {author} {\bibfnamefont {M.}~\bibnamefont
  {Assaf}}\ and\ \bibinfo {author} {\bibfnamefont {B.}~\bibnamefont
  {Meerson}},\ }\href@noop {} {\bibfield  {journal} {\bibinfo  {journal} {J.
  Phys. A: Math. Theor.}\ }\textbf {\bibinfo {volume} {50}},\ \bibinfo {pages}
  {263001} (\bibinfo {year} {2017})}\BibitemShut {NoStop}%
\bibitem [{\citenamefont {Berry}(1989)}]{Berry1989}%
  \BibitemOpen
  \bibfield  {author} {\bibinfo {author} {\bibfnamefont {M.~V.}\ \bibnamefont
  {Berry}},\ }\href@noop {} {\bibfield  {journal} {\bibinfo  {journal} {Proc.
  The Royal Society Of London Series A-Math. Phys. And Engineering Sciences}\
  }\textbf {\bibinfo {volume} {422}},\ \bibinfo {pages} {7} (\bibinfo {year}
  {1989})}\BibitemShut {NoStop}%
\bibitem [{\citenamefont {Satzinger}\ \emph {et~al.}(2018)\citenamefont
  {Satzinger}, \citenamefont {Zhong}, \citenamefont {Chang}, \citenamefont
  {Peairs}, \citenamefont {Bienfait}, \citenamefont {Chou}, \citenamefont
  {Cleland}, \citenamefont {Conner}, \citenamefont {Dumur}, \citenamefont
  {Grebel}, \citenamefont {Gutierrez}, \citenamefont {November}, \citenamefont
  {Povey}, \citenamefont {Whiteley}, \citenamefont {Awschalom}, \citenamefont
  {Schuster},\ and\ \citenamefont {Cleland}}]{Satzinger2018}%
  \BibitemOpen
  \bibfield  {author} {\bibinfo {author} {\bibfnamefont {K.~J.}\ \bibnamefont
  {Satzinger}}, \bibinfo {author} {\bibfnamefont {Y.~P.}\ \bibnamefont
  {Zhong}}, \bibinfo {author} {\bibfnamefont {H.-S.}\ \bibnamefont {Chang}},
  \bibinfo {author} {\bibfnamefont {G.~A.}\ \bibnamefont {Peairs}}, \bibinfo
  {author} {\bibfnamefont {A.}~\bibnamefont {Bienfait}}, \bibinfo {author}
  {\bibfnamefont {M.-H.}\ \bibnamefont {Chou}}, \bibinfo {author}
  {\bibfnamefont {A.~Y.}\ \bibnamefont {Cleland}}, \bibinfo {author}
  {\bibfnamefont {C.~R.}\ \bibnamefont {Conner}}, \bibinfo {author}
  {\bibfnamefont {t.}~\bibnamefont {Dumur}}, \bibinfo {author} {\bibfnamefont
  {J.}~\bibnamefont {Grebel}}, \bibinfo {author} {\bibfnamefont
  {I.}~\bibnamefont {Gutierrez}}, \bibinfo {author} {\bibfnamefont {B.~H.}\
  \bibnamefont {November}}, \bibinfo {author} {\bibfnamefont {R.~G.}\
  \bibnamefont {Povey}}, \bibinfo {author} {\bibfnamefont {S.~J.}\ \bibnamefont
  {Whiteley}}, \bibinfo {author} {\bibfnamefont {D.~D.}\ \bibnamefont
  {Awschalom}}, \bibinfo {author} {\bibfnamefont {D.~I.}\ \bibnamefont
  {Schuster}}, \ and\ \bibinfo {author} {\bibfnamefont {A.~N.}\ \bibnamefont
  {Cleland}},\ }\href@noop {} {\bibfield  {journal} {\bibinfo  {journal}
  {Nature}\ }\textbf {\bibinfo {volume} {563}},\ \bibinfo {pages} {661}
  (\bibinfo {year} {2018})}\BibitemShut {NoStop}%
\bibitem [{\citenamefont {Chu}\ \emph {et~al.}(2018)\citenamefont {Chu},
  \citenamefont {Kharel}, \citenamefont {Yoon}, \citenamefont {Frunzio},
  \citenamefont {Rakich},\ and\ \citenamefont {Schoelkopf}}]{Chu2018}%
  \BibitemOpen
  \bibfield  {author} {\bibinfo {author} {\bibfnamefont {Y.}~\bibnamefont
  {Chu}}, \bibinfo {author} {\bibfnamefont {P.}~\bibnamefont {Kharel}},
  \bibinfo {author} {\bibfnamefont {T.}~\bibnamefont {Yoon}}, \bibinfo {author}
  {\bibfnamefont {L.}~\bibnamefont {Frunzio}}, \bibinfo {author} {\bibfnamefont
  {P.~T.}\ \bibnamefont {Rakich}}, \ and\ \bibinfo {author} {\bibfnamefont
  {R.~J.}\ \bibnamefont {Schoelkopf}},\ }\href@noop {} {\bibfield  {journal}
  {\bibinfo  {journal} {Nature}\ }\textbf {\bibinfo {volume} {563}},\ \bibinfo
  {pages} {666} (\bibinfo {year} {2018})}\BibitemShut {NoStop}%
\end{thebibliography}

%

\end{document}